\documentclass[oneside,10pt,final]{sty/ucthesis-CA2012}
\pdfoutput=1

\usepackage{fancyhdr}
\usepackage{hyperref}
\usepackage{amsmath, amssymb, graphicx}
\usepackage{xspace}
\usepackage{braket}
\usepackage{color}
\usepackage{setspace}
\usepackage{subcaption}

\RequirePackage[T1]{fontenc} 
\RequirePackage[tt=false, type1=true]{libertine} \RequirePackage[varqu]{zi4} 
\RequirePackage[libertine]{newtxmath}

\makeatletter
\let\normalsize\relax
\let\@currsize\normalsize
\makeatother

\newcommand{\be}{\begin{equation}}
\newcommand{\ee}{\end{equation}}

\begin{document}

\begin{frontmatter}

\title{Beyond Reality: Designing Personal Experiences and Interactive Narratives in AR Theater}

\author{You-Jin Kim}

\report{Dissertation} \degree{Doctor of Philosophy} \degreemonth{December} \degreeyear{2024}
\defensemonth{December} 
\defenseyear{2024}

\chairA{Professor Tobias Höllerer}  
\chairB{Professor Marko Peljhan} 
\othermemberA{Professor JoAnn Kuchera-Morin} 
\othermemberB{Professor Christina McCarthy} 
\numberofmembers{3} 

\field{Media Arts and Technology}
\campus{Santa Barbara}


	\maketitle
	\approvalpage
	\copyrightpage
	\begin{dedication}

\bigskip

${}$ \\

\bigskip

${}$ \\

\bigskip

${}$ \\

\bigskip

\begin{center}
\begin{Large}
This document is my PhD thesis, defended on December 13, 2024, for the Media Arts and Technology program at the University of California, Santa Barbara. I have uploaded it to arXiv to ensure it is easily accessible and with the sincere wish that more people get to experience the journey of discovery that I was fortunate to have during my PhD program. \end{Large}
\end{center}

\end{dedication} 
	\begin{acknowledgements}
During my time in graduate school, I received immense support from many wonderful people who have become integral to my life. I could not have completed this journey without the kindness, friendship, and support around me. They have become like family, which was not what I expected my PhD to be. My lifelong mission is to repay the kindness and support I received in Santa Barbara. While this section cannot fully capture how thankful I feel, I will attempt to encapsulate a part of it here. 

Since the beginning of my graduate school journey, I have been welcomed into the Four Eyes Lab, Media Arts and Technology program, and the Computer Science and Theater and Dance departments at UC Santa Barbara. Foremost, I would like to thank Tobias Höllerer for holding me to the highest standards of scientific rigor and for teaching me about research excellence. Thank you, Tobias, for seeing my potential and deciding to work with me even when I had nothing to show for it. I am grateful for your trust in me.

The remaining committee members also never spared their energy for me. Marko Peljhan, thank you for encouraging me to think big and pursue large-scale projects in AR theater and virtual production. Christina McCarthy, I appreciate your warm welcome into the Theater and Dance department. Your unwavering support for my projects was shown through the ways you always made time for me, your insightful questions that helped me refine my work, and for encouraging people you knew to complete our dance theater virtual production. Truly, you are the reason Dynamic Theater exists today. Weekly meetings with JoAnn Kuchera-Morin in office 2005 provided invaluable support and guidance that went far beyond academia. JoAnn, you encouraged me to envision what lies ahead and not be confined by the technology that exists today. Marko, Christina, and JoAnn, your support allowed me to pursue art and engineering freely and without you, I could not have completed this PhD.

Several other faculty at UCSB guided and shaped my journey. Matthew Turk eased the transition into graduate school by encouraging my unpolished ideas. My conversations with Marcos Novak on the surface of the moon (virtually) with classical music (Bach’s Cello Suite No.1) will remind me of discussions on the future of AI agents. Jennifer Jacobs taught me to be considerate and inclusive in research through her harmonious lab environment at ECL. I was inspired by the visionary Misha Sra, who always welcomed me at the Human-AI Integration Lab and treated me to coffee during my (many) visits. Yon Visell, your early guidance inspired me to pursue deeper research from the beginning of my PhD and showed me the world of haptics and soft robotics. Thanks to you, I always felt welcomed at RE Touch social events, as evidenced by how I made great friends in your lab.

When I was considering graduate programs, Mengyu Chen showed me that MAT students can thrive in both engineering and artistic pursuits without compromise. Mengyu is the reason I chose MAT and the reason why I strive to create experiences that bring people together. From him, I learned how artwork can offer insights into the world, whether as an artist or a scientist, while emphasizing the importance of good intentions and positive impact. I'll always cherish our memories of enjoying sushi and watching Terrace House during the summer when we both did internships!

I treasure the friendships built on the collective experiences of research, lab work, and graduate student life. I found community with Gregory Reardon, Yitian Shao, Rodger (Jieliang) Luo, and Myungin Lee when I first arrived at UCSB. Rodger helped me get settled in Santa Barbara. I enjoyed getting Greg’s unique perspective and encountering his ``tsundere'' charm during my many visits to the RE Touch lab. Yitian is a wonderful person to experience life with and I am always happy to see him whether it be in Dresden or Seattle. I often saw Myungin during the countless late nights as our offices were across the hall from one another. His presence was always a comfort, though the Halo nights and delicious Korean meals prepared at his home were a bonus. Thank you all for sharing meals, gaming, traveling, celebrating, and ultimately, for letting me share part of your lives. 

I also could not have done this without the steadfast support of my MAT peers. I am grateful for the many collaborations and shared experiences with Radha Kumaran. Radha's kind support and friendship enabled me to pursue as much research as I did. Despite saying each project was our last, I expect that there will be future collaborations. However, let's make sure to avoid outdoor studies! Ashley Del Valle-Morales’s care and friendship are a constant source of comfort; you make the whole department feel at home. Donna and I are always grateful for your presence, and I very much enjoy our singalongs to Disney music in the office at night. Alejandro Aponte, your generosity and friendship have been invaluable to me. As we are so close, I sometimes forget to express how much I appreciate your companionship. Your generosity towards others is a quality I aspire to emulate. I will miss coming into the office in the morning and finding the Legos on my desk rearranged into new stories crafted by you. Ana María Cárdenas, I am grateful for the countless coffees you provided to fuel my thesis writing and the evenings at Chili's we shared with Nathan and the group. Thank you for teaching me that sometimes kindness is all we need. 

I also extend gratitude to the many I have been lucky to take classes, work, and share space with. Yi Ding, whom I affectionately call ``sensei'' despite his insistence that we're just friends, I cherish all the celebrations we shared in Germany and look forward to meeting again in another city, another country, at a conference. Kangyou Yu, my WhatsApp friend, I hope we continue our conversations as promised!

Thank you to my ECL family (Mert Toka, Samuelle Bourgault, Devon Frost, Squid Wallace, and Emilie Yu), my MAT family (Nefeli Manoudaki, Iason Paterakis, Jenni Hutson, Diarmid Flatley, Marcel Rodriguez-Riccelli, and Emma Brown), RE Touch lab family (Bharat Chandrahas, Neeli Tummala, Mengjia Zhu, and Taku Hachisu), for always welcoming me. Further, I want to acknowledge Taivan Batjargal, Ellen Wu, Remy Wang, David Ren, Karan Ahuja, Tommy Sharkey, Tom Bullock, Jun Nishida, Shuying Yu, and the dance students in the Theater department for contributing invaluable memories to my graduate school experience. Andy Wilson and Eyal Ofek at Microsoft Research, Michael Coss at Nokia Bell Labs, and Barry Giesbrecht, thank you for the invaluable lessons and mentorship which continue to guide my research. 

I would also like to thank my mom and dad, who never ceased to support me in my endeavors, and my lifelong friend Q, who periodically visited me from Seoul. My animal friends Daisy, Daphne, Mochi, and Chiquis created warmth when the days were hard. Finally, my greatest appreciation goes to my partner, Donna. Although she studies the past and I study the future, we met in the present. I walked into an Asian Pacific Islander Graduate Student Alliance (APIGSA) meeting by accident, thinking it was for Asian international graduate students, during my first quarter at UCSB. The moment I entered, I realized I was in the wrong place, but as I was leaving, you stopped me and welcomed me into the group, which is when everything began. My life has gained so much meaning with you along this journey. Though Santa Barbara has given me so much, meeting you here is the most precious gift.

The completion of my PhD is as much a testament to the unconditional support I received from this loving community as it is to my own efforts. I am grateful for the incredible communities I encountered in Santa Barbara, which have created a nurturing environment for these six years. I will always fondly remember the people, the coastal winds, and the view of the coast from my office space in Elings 2005.

\end{acknowledgements} 
	 \begin{vitae}
\addcontentsline{toc}{chapter}{Curriculum Vitae}

\begin{vitaesection}{Education}

\item	Ph.D. in Media Arts and Technology (MAT), UC, Santa Barbara.
\item 	M.F.A. in Interactive Systems (EDA), Duke University.
\item 	B.F.A. in Game Development (ATS), School of the Art Institute of Chicago.

\end{vitaesection}

\textbf{Publications}

\begin{enumerate}

\item \textbf{You-Jin Kim}, Misha Sra, Tobias Höllerer. 2024. Audience Amplified: Virtual Audiences in Asynchronously Performed AR Theater. In \textit{2024 IEEE International Symposium on Mixed and Augmented Reality (ISMAR)}, 10 pages.

\item \textbf{You-Jin Kim}, Myungin Lee, Marko Peljhan, JoAnn Kuchera-Morin, Tobias Höllerer. 2024. Spatial Orchestra: Locomotion Music Instruments through Spatial Exploration. In \textit{Extended Abstracts of the CHI Conference on Human Factors in Computing Systems}, 5 pages.

\item Jamie Ngoc Dinh, \textbf{You-Jin Kim}, Myungin Lee. 2024. FractalBrain: A Neuro-interactive Virtual Reality Experience using Electroencephalogram (EEG) for Mindfulness. In \textit{Extended Abstracts of the CHI Conference on Human Factors in Computing Systems}, 5 pages.

\item \textbf{You-Jin Kim}, Andrew D. Wilson, Jennifer Jacobs, Tobias Höllerer. 2023. Reality Distortion Room: A Study of User Locomotion Responses to Spatial Augmented Reality Effects. In \textit{2023 IEEE International Symposium on Mixed and Augmented Reality (ISMAR)}, 10 pages.

\item \textbf{You-Jin Kim}, Joshua Lu, Tobias Höllerer. 2023. Investigating Search Among Physical and Virtual Objects Under Different Lighting Conditions. In \textit{Proceedings of the 29th ACM Symposium on Virtual Reality Software and Technology}, 11 pages.

\item Radha Kumaran*, \textbf{You-Jin Kim}*, Anne Milner, Tom Bullock, Barry Giesbrecht, Tobias Höllerer. 2023. The Impact of Navigation Aids on Search Performance and Object Recall in Wide-Area Augmented Reality. In \textit {Proceedings of the 2023 CHI Conference on Human Factors in Computing Systems}, 17 pages. 

\item \textbf{You-Jin Kim}*, Radha Kumaran*, Ehsan Sayyad, Anne Milner, Tom Bullock, Barry Giesbrecht, Tobias Höllerer. 2022. Investigating Search Among Physical and Virtual Objects Under Different Lighting Conditions. In \textit{IEEE Transactions on Visualization and Computer Graphics 28}, 11 pages.

\item Zhaozhong Wang, \textbf{You-Jin Kim}. 2022. Exploring Immersive Mixed Reality Simulations and Its Impact on Climate Change Awareness. In \textit{Asian Journal of Applied Science and Engineering 11}, 6 pages.

\vspace{0in} * indicates equal contribution. 
\end{enumerate}

\end{vitae}
	%
%

\begin{abstract}
\addcontentsline{toc}{chapter}{Abstract}

Augmented Reality (AR) technologies are redefining how we perceive and interact with the world by seamlessly integrating digital elements into our physical surroundings. These technologies offer personalized experiences and transform familiar spaces by layering new narratives onto the real world. 

Through increased levels of perceived agency and immersive environments, my work aims to merge the human elements of live theater with the dynamic potential of virtual entities and AI agents. This approach captures the subtlety and magic of storytelling, making theater experiences available anytime and anywhere. The system I am building introduces innovative methods for theatrical production in virtual settings, informed by my research and eight published works. These contributions highlight domain-specific insights that have shaped the design of an immersive AR Theater system.

My research in building a well-designed AR stage features avatars and interactive elements that allow users to engage with stories at their own pace, granting them full agency over their experience. However, to ensure a smooth and curated experience that aligns with the director or creator's vision, several factors must be considered, especially in open-world settings that depend on natural user movement. This requires the story to be conveyed in a controlled manner, while the interaction remains intuitive and natural for the user.

\end{abstract}

	\tableofcontents
\end{frontmatter}

\begin{mainmatter}

\pagestyle{fancy}
\renewcommand{\chaptermark}[1]{\markboth{{\sf #1 \hspace*{\fill} Chapter~\thechapter}}{} }
\renewcommand{\sectionmark}[1]{\markright{ {\sf Section~\thesection \hspace*{\fill} #1 }}}
\fancyhf{}

\makeatletter \if@twoside \fancyhead[LO]{\small \rightmark} \fancyhead[RE]{\small\leftmark} \else \fancyhead[LO]{\small\leftmark}
\fancyhead[RE]{\small\rightmark} \fi

\def\cleardoublepage{\clearpage\if@openright \ifodd\c@page\else
  \hbox{}
  \vspace*{\fill}
  \begin{center}
    This page intentionally left blank
  \end{center}
  \vspace{\fill}
  \thispagestyle{plain}
  \newpage
  \fi \fi}
\makeatother
\fancyfoot[c]{\textrm{\textup{\thepage}}} 
\fancyfoot[C]{\thepage}
\renewcommand{\headrulewidth}{0.4pt}

\fancypagestyle{plain} { \fancyhf{} \fancyfoot[C]{\thepage}
\renewcommand{\headrulewidth}{0pt}
\renewcommand{\footrulewidth}{0pt}}

\chapter{Introduction}

Technology has evolved since the pioneering work of A Touring Machine~\cite{feiner1997touring} in 1997, which opened the possibilities for augmented reality (AR) in large outdoor settings, and the innovations by Chock et al. (2002) in creating immersive theater experiences through physical movement and anchored content~\cite{cheok2002interactive}. It is becoming increasingly clear that the future of AR lies in ``always-on'' accessibility, which brings new challenges and opportunities. This shift necessitates focusing on user-centered design to create meaningful AR experiences that are accessible anytime and anywhere. As a result, this dissertation explores how stories can be told in AR, the considerations necessary for open environments, and how narratives can unfold through safe interaction, naturally extending the experience to virtual entities.

  My research demonstrates how AR can be tailored to meet users' demands to further enhance user interaction through natural locomotion and interactive narratives. Additionally, it advances artistic and technical strategies to craft personalized, immersive experiences that empower participants to explore narratives, dancers, and characters across augmented environments at their own pace. By following the overarching storyline set by the director or creator, unique spatial content, such as virtual bubbles and butterflies, can enrich the user's engagement and their perception of the ongoing dance and the theater stage. However, designing these open-space experiences requires careful consideration of user experience. Thus, it is essential to understand how stories unfold and how virtual productions should be presented. The immersive theater platform I propose expands interactive storytelling by utilizing physical locations and new ways for creators and directors to present personalized experiences.

\begin{section}{Motivation}
My motivation for this research is to explore and expand the possibilities of personalized storytelling and interactive narratives that users can experience and express in their physical environments. My goal is to create authentic, live theater experiences using AR, transforming traditional theater into a spatial, immersive journey.

The question of realism in art and media has been explored by several theorists over time. In the 1910s, German media theorist Walter Benjamin argued that even the most perfect reproduction of a work of art lacks its unique presence in time and space, which is crucial for its authenticity~\cite{haxthausen2004reproduction}. This idea was further expanded by French philosopher Jean Baudrillard in the 1970s, who suggested that our perception of reality has transformed so significantly that the distinction between originals, simulations, and simulacra is blurred~\cite{gane2000jean}. Building on Baudrillard's ideas, Philip Auslander in the 1990s argued that live and mediatized performances are perceived equally, with live performances becoming another form of reproducible text~\cite{auslander1997against}. Modern technologies, such as volumetric playback, offer new ways to capture and rewatch performances, changing our experience and perception of art.

As a theater enthusiast and multimedia designer, I was profoundly influenced by my experience with immersive theater, particularly \textit{Sleep No More} in New York City at the start of my PhD. This experience showed me how audience members could become active participants, moving through various stages and locations, and taking on an agency within the narrative. The production's interactive nature, where viewers wander freely through elaborately designed spaces and uncover different storylines at their own pace, inspired me. Witnessing how each person’s journey could be entirely unique revealed to me the potential for storytelling to be personalized and dynamic. This realization led me to see augmented reality as the perfect medium to capture and enhance these magical, immersive experiences. I recognized the production's careful user experience design, evident in the stage design and the natural progression of clues leading to the next part of the story, could be effectively translated into mixed reality experiences as well.

My dissertation addresses key components of designing AR immersive theater experiences by exploring the following research questions:

\begin{itemize}
    \item How will users engage with a narrative by walking around in an AR environment?
    \item How can AR stages be designed to let users experience different environments within the same physical space?
    \item How will users be guided to navigate and unfold the narrative in AR immersive storytelling?
    \item How can AI agents enhance the main narrative and user experience in AR?
    \item How can artistic performance be best translated into AR platforms? 
    \item How will AR immersive media and virtual production evolve in the future?
    
\end{itemize}

Through this research, I aim to drive the adoption of interactive narratives and immersive theater in AR, giving directors and artists more possibilities and control over the stories they express.

In immersive game environments, narratives are often guided through recorded animations, NPCs, and visual effects. My work places users in real-world locations, giving users control over how the story progresses as they explore the AR stage. I aim to shift the audience's role from passive observers to active participants, allowing them to engage with a personalized narrative that is unique and dynamic to their choices and preferences within an AR experience. My exploration combines scientific developments with artistic exploration and user experience design. By working with the latest AR glasses, anchoring systems, and digital twin technology, I aim to provide artists with a new space to curate user experiences that feel immediate and spontaneous, much like live theater or concerts. This approach offers more user choice within a curated narrative, transcending the limitations of traditional media by creating a more interactive and personalized theater experience.

\end{section}

\begin{section}{Why Augmented Reality Today}
Augmented Reality (AR) offers a transformative platform for capturing and sharing immersive storytelling; pushing the boundaries of human expression by blending augmented content with physical locations. AR, experienced from a user's point of view, allows for a tailored experience that extends human expression and interaction beyond traditional means. For instance, users can interact with virtual objects, fostering new forms of creativity and interaction. This expansion of the user's reality becomes personalized and adaptive to their environment. Subsequently, the field of interactive experiences has been actively designed and expanded in recent years, showcasing AR technology's capability to expand human creativity, imagination, and interaction.

Even with active research, however, the potential of AR remains largely unexplored. While researchers have been creating immersive augmented reality experiences for walking in large areas since the early 2000s, most immersive works we encounter today have not utilized the orientation and the placement of AR objects to align with physical objects or architecture in the physical environment. Furthermore, how these AR experiences should be conveyed or the order of experience for storytelling has not yet been designed for wide-area scale experiences that require users to walk around to interact, so no such extensive guidelines exist. With advancements in spatial mapping technology, standalone head-mounted displays (HMDs), and the ability to accurately anchor digital content to the physical environment, we now have the opportunity to present spatial content with greater control and precision. By carefully designing the user experience, we can accurately align augmented content with the user's vision and the real world, providing a true-to-life scale comparison between the digital and physical worlds.

With a deeper understanding of visual perception and cues, there are many potential ways to alter reality and influence user behavior or interaction using AR. Many possibilities arise for research expansion. First is presenting the same physical location in different ways. In other words, AR can allow users to feel as though they are in a completely new place, even when they are physically in the same location. This offers users a unique experience each time. Second, spatial content achieved through volumetric capture and accurate placement can be done with great precision over large areas. This allows pre-captured actors, animated avatars, and other elements to be viewed from multiple angles, presenting them from different vantage points to evoke completely different feelings, even if the content remains the same. Volumetric content is versatile, as it can be perceived differently with changes in lighting. An AR stage, like a theater stage with lighting, can create entirely different scenes and emotions for the audience. These assets can closely replicate and simulate the dynamics of live experiences. Combined with natural locomotion, this capability gives users an open-world experience where they can explore at their own pace, actively engaging with the environment and crafting unique stories.

Third, AI agents are another crucial component of AR  that demands further exploration. Recent advancements in AI technology, such as the introduction of ChatGPT, allow for deeper, truly intelligent conversations. Participants can utilize many senses we already use when interacting with humans, like gestures and behaviors, creating deeper connections within an AR experience. Additionally, AR technology has clear advantages over other 2D or screen-based platforms. These agents can be designed for specific tasks and learn directly from human motion tracking data, significantly enhancing user experience by providing tailored and efficient support to the main narrative.  Interacting with these entities through avatars from a human perspective creates a more intimate and personalized interaction.

The latest technology of standalone headsets now offers mixed reality experiences through optical see-through or passthrough displays, providing users with a wide range of options to experience augmented or mixed reality. While these technologies are widely available, each device has different physical interaction size limitations and recommended specifications, as detailed in their official documentation. Thus, choosing the right standalone headset is crucial. The Spatial Mesh, a 3D model acquired through natural locomotion and scanning parts of a space, generates part of the spatial mapping in both HoloLens 2~\cite{brown2022map} and Magic Leap 2~\cite{sagula2024spatial}, allowing for larger space interactions and expanding the potential for immersive experiences. Magic Leap 2 can handle spatial meshes of up to 250m²~\cite{sagula2024size}, while HoloLens 2 supports up to 125m² in my tests. However, issues like improper occlusion and content placement challenges persist. The Apple Vision Pro currently lacks walking capabilities~\cite{apple2024walking}, and the Meta Quest 3 faces tracking reliability issues beyond 15m²~\cite{oculus2021size}. Even with these technological advancements, there is a good reason to have a full layout of the space when designing a cohesive user experience. As different devices have varying capabilities, understanding these limitations is crucial for designing interactive narratives in AR. 

As technology evolves, future devices will likely support larger, outdoor spaces, emphasizing the importance of walking and natural movement. To fully utilize AR theater experiences, several challenges must be addressed. Rendering techniques need to be reimagined to blend physical and virtual elements seamlessly. As users will use this technology outdoors more often, understanding lighting conditions and investigating interactions between physical and virtual objects under different lighting conditions must be studied. Robust tracking and anchoring systems are necessary, especially in large spaces, to maintain immersive experiences. Ensuring proper transparency and occlusion of virtual objects depends on the orientation of the object, whether behind physical walls or in front of them. By addressing these challenges, AR can offer a curated yet open-world experience, where users have the freedom to explore and interact without explicit instructions. This vision of AR holds immense potential for creating truly immersive and personalized theater experiences.

To further illustrate these rapid developments in AR technology, here I summarize technical advancement in AR that allowed my research during my PhD:

\begin{itemize}
    \item Standalone AR headset innovations have led to devices such as Magic Leap 2, Microsoft HoloLens 2, Apple Vision Pro, and Meta Orion now being equipped with inside-out tracking technology and powerful processing capability. 

    \item Progress in display technologies has resulted in the advancements in optical see-through and passthrough displays. Experience through these displays allows users to see both virtual and physical objects. 

    \item Development in spatial mapping technologies has brought solutions like Spatial Mesh to understand physical layout for interaction. These offer accurate spatial mapping over large areas, enabling natural movement in larger spaces.

    \item Advances in 3D reconstruction and capture technologies have facilitated the creation of detailed 3D models and digital twins of real-world spaces for developers to utilize in designing user experiences. 

    \item Through research in computer vision and perception, significant enhancements in tracking and anchoring have been accomplished. These advancements contribute to the stability and accuracy of positioning virtual objects within wide areas. 

    \item Open-source projects such as Unity ML-Agents have been instrumental in creating simulation and training platforms~\cite{mlagent}. They offer environments that facilitate physical simulation-based training of intelligent agents.

\end{itemize}

\end{section}

\begin{section}{Building a Dynamic Theater System}
The immersive theater system presented in this dissertation utilizes AR technologies to transform traditional theater and dance performances into interactive, immersive experiences. The primary objective of this system is to expand user engagement in AR stages where physical layout becomes an important aspect of user experience, allowing audiences to experience spatial content in a way that feels both intimate and personalized. Furthermore, the system suggests how the act structure of traditional storytelling in theater can be adopted in immersive theater.

To demonstrate this concept, the immersive theater system is designed around personal devices and standalone devices, such as optical see-through (OST) devices like HoloLens 2, Magic Leap, and Meta Orion AR glasses. One of the key features of this system is its emphasis on natural locomotion and space utilization. Unlike traditional theater settings, where audiences are typically stationary, this system encourages users to move freely within the performance space. By doing so, it replicates the sensation of real-world exploration, enabling users to interact with the narrative at their own pace. This focus on natural movement addresses common challenges associated with virtual environments and highlights the benefits of location-based storytelling, where the physical space becomes an integral part of the immersive narrative experience.

The spatial content is strategically positioned to interact with the physical environment, creating a blended reality that fully immerses users in the performance. This approach not only enhances the level of immersion but also allows for greater interaction as users navigate the space, engaging with virtual content naturally and intuitively. The navigation guidance system I introduce artfully guides users through the performance space without distracting them from the main storyline. This enables users to experience floorwork moves, where pre-recorded dancers can initially approach their space or enter into the user's vision from hidden areas, such as behind a wall. This work captures human dancers' floorwork digitally and presents it artfully in virtual form as the director intended for everyone to enjoy. This shows how directors of virtual production can design user experience. 

Collaboration with musicians, thespians, and dancers is another crucial aspect of this system. Their involvement ensures that dance movements and the envisioned story atmosphere are effectively and coherently integrated into the AR environment and narrative. Musicians, thespians, and dancers come into the experience with a pre-established understanding of the use of a stage given their traditional theater training. By translating their art forms, body movements, and space utilization into AR platforms, this immersive theater system leverages their expertise to push the boundaries of traditional performance to create a new genre of theater that is dynamic and interactive beyond the limits of a traditional stage space. This collaboration further encourages creative exploration of space and the interplay between virtual and physical elements. 

Moreover, the system contributes to research in the field of Human-Computer Interaction (HCI) in mixed reality by conducting studies on user navigation and attention. Often, users walk and find prepared contents, which can divide their attention. The system accounts for their division of attention by preparing content in a predesignated area, allowing users to be fully immersed while remaining safe throughout the experience. The system not only informs the development of future systems by suggesting how spatial content is implemented and presented to the viewers but also offers guidance to artists and researchers interested in creating narrative-driven AR immersive narrative plays in various physical layout. 

In addition to presenting personalized experiences, the system also aims to recreate the social dynamics of live theater by incorporating AI-trained virtual audiences. These virtual audiences were trained for a specific physical layout to simulate the behavior of people walking and reacting within that space. The training data was collected from human users interacting in the same physical environment, engaging in activities such as moving around and consuming spatial content. These virtual characters simulate the presence of a live audience, enhancing the social aspect of the performance. This feature allows users to feel part of a collective event, even when experiencing the performance alone or asynchronously. By simulating the energy and presence of a live audience, the system fosters a sense of community and shared experience, which is often missing in narrative-driven extended reality (XR) virtual productions. My results from Audience Amplify~\cite{kim2024audience}, an expanded version of the initial project Dynamic Theater~\cite{kim2023dynamic}, suggest how these entities can be deployed to add a live theater-like social aspect to the experience, further advocating for my idea of the ``anytime, anywhere'' concept.

\begin{figure}[t]
\centering
\includegraphics[width= 1.0\textwidth]{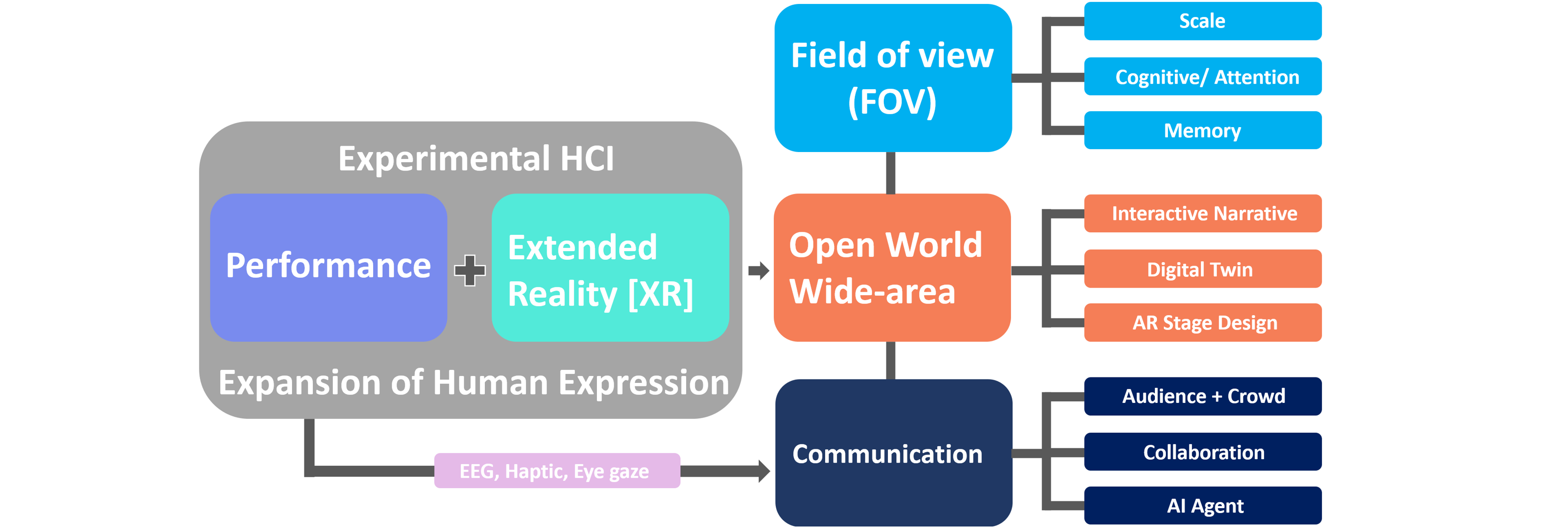}
\caption{An infographic describing the different areas of study included in this dissertation and their relationship to certain developments in the field of HCI. The subsequent chapters detail studies conducted to target areas of possible development in these categories. Related, EEG, haptic data, and eye gaze strengthen understandings for AI agents and audience participation. Combined, these studies contribute to greater corroboration between performance and extended reality.}
\label{fig:danceChart}
\end{figure}

By enabling performances to be experienced anytime and anywhere, while adapting to different physical layouts and including virtual audiences, the system offers a stark contrast to traditional live concerts or theater experiences, where attendance requires being on time and at a specific location. By employing experimental HCI techniques to expand human expression, my immersive theater system, Dynamic Theater, combines research on the field of view (FOV), open world/wide-area, and communication to enhance performance experiences in extended reality (Figure \ref{fig:danceChart}). Several studies, described in further detail in the forthcoming chapters, illustrate the use of these research areas and their contributions to immersive theater experiences. Dynamic Theater~\cite{kim2023dynamic} provides users with the flexibility to engage with content on their own terms, free from the constraints of location or timing. This adaptability aligns with current trends in virtual and social experiences, where users prefer to engage with content at their convenience, thereby broadening the potential audience for these performances.

The immersive theater system I introduce and present in this dissertation is designed to revolutionize the audience experience in theater and dance by leveraging modern technology to create dynamic, interactive, and socially engaging performances. This system explores new possibilities for storytelling and audience participation, providing an innovative and accessible approach to theater. By blending the physical and virtual worlds, it opens new avenues for artistic expression and engagement in the digital age, ensuring theater remains a vibrant and evolving art form. My goal is to capture the art form of human movement and its delicacy in mixed reality, demonstrating the platform's potential. It serves as a medium for artists to adapt their stories and theater plays, where imagination is the only limit. Additionally, the immersive theater platform I present offers a guide and starting point for artists to explore new forms of relationships within an immersive experience and design creative storytelling methods. It incorporates a guidance system to help the audience stay engaged with the narrative and suggests experimental ways to influence users in these AR stages, expanding their expression, such as playing music with virtual objects.

This user-centric approach is crucial for maintaining engagement and ensuring that the technology enhances, rather than distracts, from the performance. Without altering the human choreographic art form itself, my vision for immersive theater is to expand human creativity and expression and to preserve the art of human movement in the play, which the immersive theater system I present achieves. Dynamic Theater forecasts and suggests visions of how we will experience immersive performances across large physical locations, with dance performances at the center. 

\end{section}

\begin{section}{Dissertation Aims}
My dissertation explores how people consume spatial content and immersive interactive narratives from spatial computing platforms within AR. This involves fully utilizing the physical layout of space to guide users through storylines, ensuring they remain immersed without disruption, and experience the main narrative with personalized interactions. The result is a new interactive immersive theater system that can capture human physical movement, acting, and art forms, creating an immersive narrative. This opens possibilities for guided spatial computation experience design when consuming spatial content.

Furthermore, I aim to support creators in adopting this immersive storytelling platform. To support creativity, it is essential to consider the evolution of theater, from its ancient origins in the Theatre of Dionysus to the contemporary stage. This evolution has always been driven by an intrinsic need to enhance audience engagement and interaction. Over the years, dancers and choreographers such as Steve Paxton, Pina Bausch, Scott Graham, and Steven Hoggett have evolved, explored, and practiced artful approaches like footwork and floor work in theater entrance techniques~\cite{climenhaga2008pina, graham2012frantic, pallant2017contact}. These techniques are well-practiced in theater and dance today and are thoroughly documented.

 Working at the intersection of traditional theater production and virtual production, I see the implementation of this technology as a way to ensure that the art form of human performance is not altered. It should still be designed around the main story and performance while respecting the artist and art form. It is crucial that the expression of the human body as an art form is not lost during this transition or the capturing process into immersive media. My goal is to capture all these aspects of what makes physical performance great and present them digitally to users in a personalized manner.

Despite significant advancements in XR technology, current practices often underutilize the available physical space, thus limiting the potential for user engagement. This dissertation addresses this gap by integrating immersive dance performances with AR technology, transforming traditional theatrical experiences into dynamic, participatory performances that maximize audience interaction and engagement. I will demonstrate how interactive narratives in immersive environments, with content prepared within the spaces throughout, can offer more forms of narrative interaction and therefore a richer content consumption experience, akin to the excitement of attending a concert or theater. Additionally, deploying virtual audiences can enhance this experience even further. 

A significant aspect of this research is the optimization of space utilization in AR performances. The primary objective is to explore how AR technology can be employed to enhance user engagement by effectively utilizing the physical space. This involves investigating the spatial arrangement of digital content and the use of guidance systems to direct user movement within a designated area. By doing so, the research aims to create a more immersive and engaging experience for the audience, allowing them to interact with the performance without impacting the main content being prepared and placed for users to find and consume.
Navigation and interaction with the AR content are essential to achieving the research aims. These systems should enhance the immersive experience while ensuring user comfort and safety, allowing for natural locomotion and exploration. By leveraging AR technology and innovative navigation systems, this dissertation explores how AR can transform traditional theatrical experiences into dynamic, participatory performances that maximize audience interaction and engagement.

In the early stages of virtual production, the focus is on developing the conceptual design of the AR performance space. This involves collaborating with choreographers to map out dance movements and seamlessly integrate the play into the user experience design. This is followed by the development and testing of various guidance systems to direct user movement within the AR environment. By effectively utilizing physical space and integrating digital enhancements, this research aims to set a new standard for immersive theater experiences. The findings will provide valuable insights into the design of AR performances that maximize user engagement and interaction. The development of effective guidance systems for AR performances will contribute to the broader field of XR technology, enhancing user navigation and interaction in large-scale AR environments.

The development of AI-agents acting as virtual audience simulations is another significant aspect of this research. To replicate the social aspect of theater, the research will explore the use of AI-trained virtual audiences, designed to interact within the physical layout where users will experience and consume the virtual production. These digital crowds will be crafted to enhance the realism and immersion of the performance, creating a socially enriched experience even in the absence of live spectators. By incorporating virtual audiences as part of AR theater experiences, this dissertation project aims to create a more engaging and immersive experience for participants, allowing them to feel more connected to the performance and to each other. The implementation of AI-trained virtual audiences will demonstrate the potential of AI to enhance social experiences in digital performances, paving the way for more interactive and socially enriched XR applications. The research will also provide detailed design considerations for integrating physical and digital elements in AR performances, assisting future researchers and artists in creating cohesive and engaging immersive theater experiences.

In summary, this dissertation aims to transform traditional theater by integrating AR technology to create dynamic, participatory performances that maximize user engagement. It aims to provide insights into the following aspects:

\paragraph{Wide-Area AR}

\begin{itemize}

\item Present how to design a walkable, immersive AR stage for digital performances, allowing users to explore and be immersed in the narrative.

\item Present how to guide users toward the spatial content in the AR environment without rushing them or disrupting their immersive experience.

\item Present how manipulating the AR environment can predictably influence users' movements and interactions.

\item Present how to orchestrate an AR play in a large area transformed into AR stages, where users unfold the narrative by simply walking around.

\end{itemize}

\paragraph{Interactive Narrative}

\begin{itemize}

\item Present how to create a personalized experience where users set the pace of the story being told within an AR experience.

\item Present how to capture and encapsulate the artistry of human physical movements within AR platforms.

\item Present design guidelines and recommendations for creating and placing spatial content within AR environments for storytelling.

\item Present how virtual audiences and AI-agents can be deployed to generate a sense of social experience in an AR environment.

\end{itemize}

\paragraph{AR Theater}

\begin{itemize}

\item Present how traditional theater practice and movement can be translated into the AR platform.

\item Present how to create immersive virtual production in collaboration with choreographers and artists.

\item Present how to design a walkable, immersive AR stage for digital performances, allowing users to explore and be immersed in the narrative.

\item Present how users in an AR environment can express music through their motion and exploration of the environment.

\end{itemize}

By addressing the limitations of current XR practices and exploring innovative solutions for space utilization, navigation, and social interaction, this research seeks to enhance the immersive theater experience and set the stage for future advancements in this exciting field.

\end{section}

\begin{section}{Dissertation Overview}
This dissertation explores conceptual developments in dance performance, immersive experiences, immersive media, immersive narratives, and interactive narratives. The immersive theater system I present was built from numerous studies and experiences throughout my time in graduate school. As this dissertation does not discuss their developments chronologically, I have created a graphic (Figure \ref{fig:PHD_projects}) to illustrate their development in time and their contributions in either art or engineering realms.  

\begin{figure}[t]
\centering
\includegraphics[width= 0.9\textwidth]{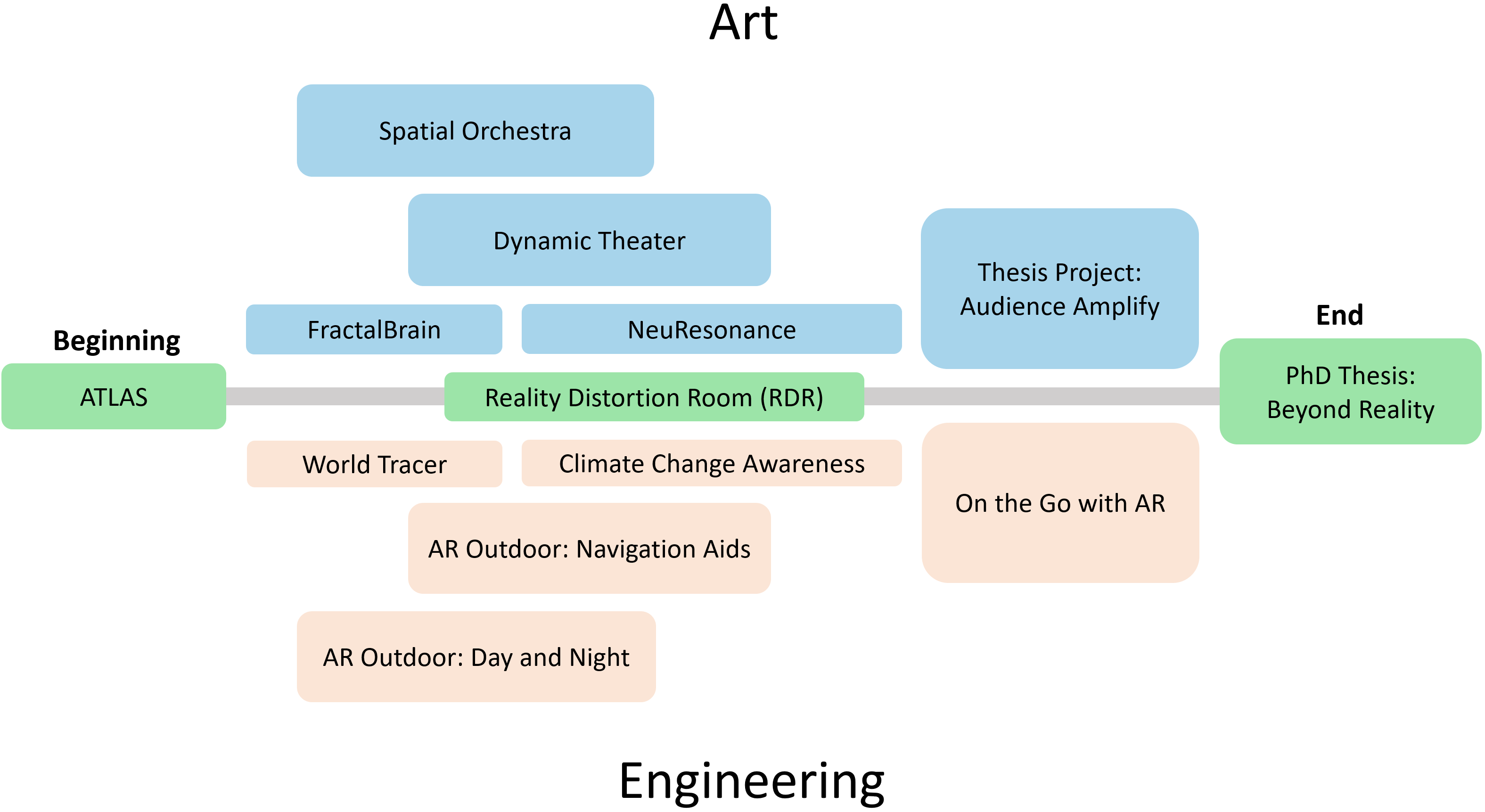}
\caption{This graphic illustrates the selected works featured in this dissertation as they were developed in graduate school. It also organizes the projects based on their association with more artistic works, such as Spatial Orchestra (Chapter 4), or more engineering-focused works, such as AR Outdoor Navigation Aids (Chapter 3).}
\label{fig:PHD_projects}
\end{figure}

The first chapter includes a comprehensive overview of the relevant research areas and recent adaptations, introducing several early works in theater and VR games, as well as the E.A.T studio performances from Bell Labs~\cite{nokia2024genesis}. These works have inspired my adoption of Dynamic Theater, as detailed in Chapter 2, which provides an in-depth conceptual investigation into digital twins, wide-area studies, and outdoor VR and AR theater. By examining inspirational works like the historically famous \textit{9 Evenings}~\cite{nokia2024genesis}, which featured collaborations between robots and performers in live events, I discuss how these ideas can be translated into capturing live events through multimedia and immersive media.

Chapter 3, ``Natural Locomotion in Wide-area Augmented Reality,'' expands on my E.A.T studio project, which I developed at Bell Labs, now part of the Nokia Corporation. The project, World Tracer, utilized depth sensors in floor robots to create a digital twin of a SLAM floor layout map. Expanding on what I learned, I made a digital twin model of an outdoor area on the University of California, Santa Barbara (UCSB) campus, measuring 1456 sq.m. (15,672 sq.ft.), using scanning, laser measurements, a Matterport camera, satellite images, and map data from Google Maps for proportion, alignment, and orientation. This acquired digital twin can assist wide-area interaction, where scale is very sensitive.

In Chapter 4, ``Movement and Perception in Augmented Reality,'' I explore how AR stages or AR objects can enhance or induce reactions, resulting in expanded interaction possibilities. AR objects serve as forms of artistic expression, where users play music with interactable virtual bubbles. Meanwhile, the AR stage utilizes scale and distorted visual cues to induce user movement patterns, exploring new ways of interaction and showing how this research journey in AR is just beginning.

In Chapter 5, titled ``Personalized Immersive Theater Experience,'' I introduce the concept of Dynamic Theater, envisioning a future where people can enjoy spatial content tailored to their preferences. This allows users to control the pace by moving around an augmented reality (AR) stage within the same indoor setting. Building on Dynamic Theater, the latter half of the chapter explores an extension that incorporates AI agents. These agents, trained in the digital twin of the physical space, simulate human behavior, enabling virtual audiences to partake in a concert-like experience alongside live participants.

I illustrate how my concept of wide-area, exploration-based storytelling can be adapted for Dynamic Theater. My system suggests design considerations essential for creating immersive media experiences, especially in outdoor or expansive environments where narratives unfold as users move through the space. This platform was also utilized in an immersive theater system in a class I taught at UCSB, where students developed their own plays, incorporating their chosen themes using the same platform employed in both Dynamic Theater and Audience Amplify.

The final chapter concludes the dissertation by summarizing the recommended considerations for designing immersive narrative systems, offering guidance for future directors and artists. It also highlights areas and topics for further research in this domain, identifying promising directions for future exploration in the field of AR research and for my own journey as a researcher in this field. The chapter discusses the future of immersive narratives in augmented reality and the design considerations that must be addressed. Additionally, I provide reflections and predictions on the evolving landscape of these immersive interactive narrative platforms as they continue to grow.

\end{section}

\begin{section}{Dissertation Contribution}
The goal of my dissertation, and the collective work leading up to it, is to showcase how to expand user interaction in immersive media and forecast how we will consume content in more interactive, replayable, and adaptive ways. I have included interactivity systems that expand user interaction possibilities in AR: one with the AR stage itself and virtual interactable objects to play music (Spatial Orchestra), and two immersive narrative projects where users experience dance theater (Dynamic Theater) and how engagements can increase in dance theater through the use of AI audiences (Audience Amplify). These projects, developed during my PhD program, aim to demonstrate the possibilities of immersive narratives, with the hope that they provide both conceptual and technical guidance to artists and researchers. 

This work presents more intimate and personalized experiences by utilizing the full physical layout in which the user is situated. I will introduce my immersive theater system, Dynamic Theater, where the user walks around and experiences a story as if a full stage set operator, lighting board operator, and theater team are presenting a personal show. This approach completely redefines how content is traditionally catered to users in theater or concert settings, which often accommodate hundreds of audience members. Additionally, the dissertation will demonstrate how the adaptation of AI agents in this AR space can enhance the user's immersive narrative experience. Furthermore, it explores the method of bringing a ``human aspect'' to the user's main story by introducing fellow audience members who walk along to see the show together. These virtual audiences are not part of the main show but are deployed to enhance the user's experience and main storyline.

User studies conducted on this immersive theater system have demonstrated that it: 1) offers a variety of methods for users to interact within augmented reality wide-area stages by utilizing a digital twin model, 2) illustrates how external factors, such as interactable virtual objects and dynamically altering AR stages, can be interactive and significantly influence user behavior, 3) enables users to explore and reveal spatial content within AR environments simply by walking around, and 4) creates new possibilities for specialized AR agents that are trained according to the physical layout to meaningfully enhance the overall user experience.

\begin{figure}[t]
\centering
\includegraphics[width= 1.0\textwidth]{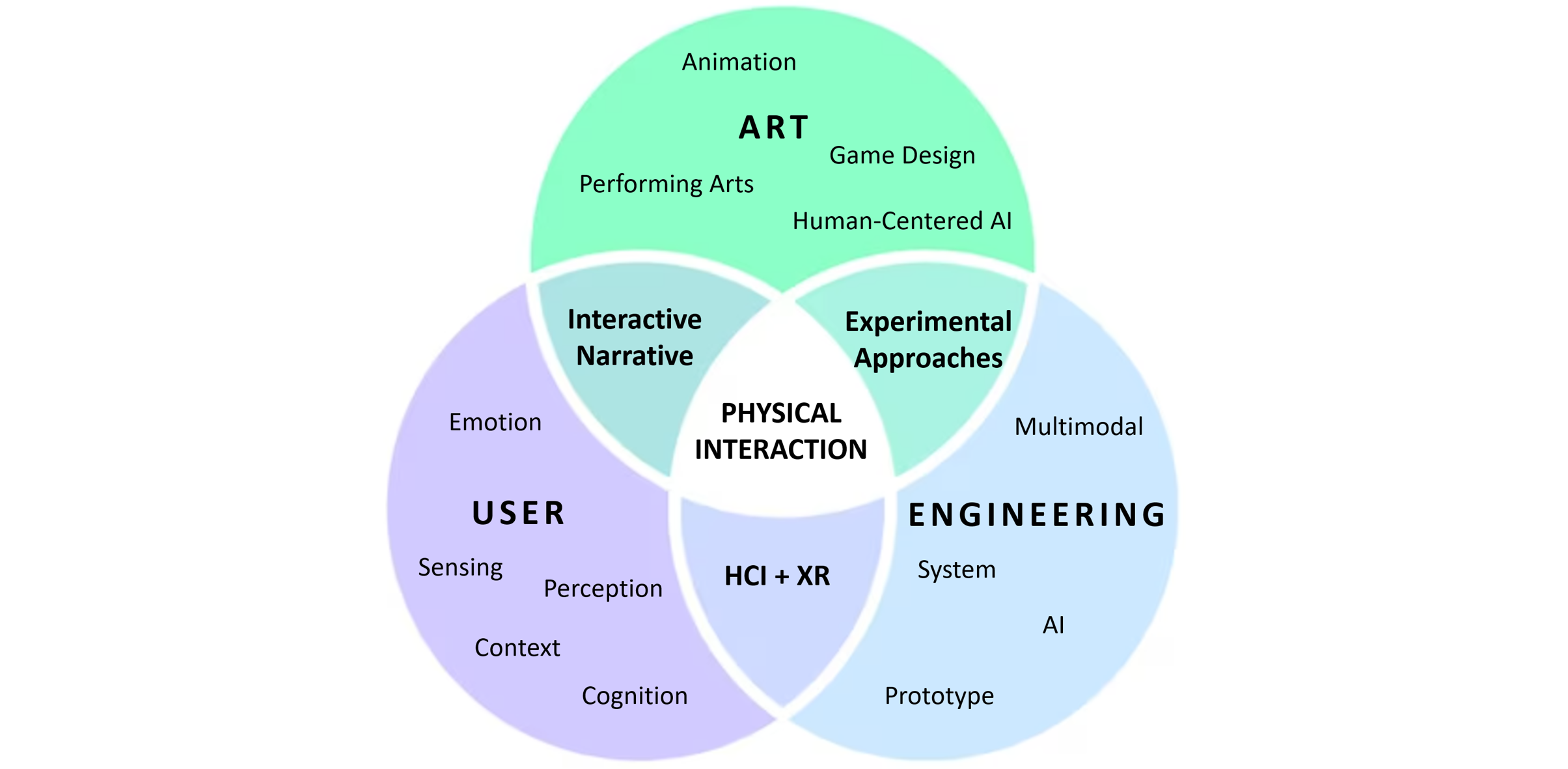}
\caption{This Venn Diagram illustrates the main fields of study I engaged throughout my research and the areas of overlap that I exploited in my projects. It is in these areas of overlap that this dissertation's vision comes to fruition and makes the systems I propose viable.}
\label{fig:vision}
\end{figure}

As I share this platform as an open-source repository, I expect to see more stories told in this way. This immersive theater platform presents many methods for visual cues to guide users to stay immersed and follow along with the story, aided by a gently floating butterfly to indicate where the next spatial content is prepared. This artistic approach to guiding users is my attempt to explore possible ways of creating spatial computing interactive narrative experiences. I aim to validate my ideas and approach through user studies so that when shared with the creative community, it presents a good starting point and toolset for them to adopt. Ultimately, my vision for this research is to highlight the possible intersections between art and engineering to highlight the experimental approaches that enhance audience and user participation (Figure \ref{fig:vision}). By emphasizing these areas of overlap, creators, both artistic and engineering alike, can find new avenues for innovation. With the AR theater system that I introduce, future creators and directors of immersive narratives can concentrate on the storytelling aspect, the most crucial component in the creation of AR theater. If they wish to introduce new visual cues in crafting user experience, they are encouraged to do so. The system's modular design makes adding such components intuitive, as it was specifically designed from the very beginning to allow for the easy integration of new elements.

\end{section}

\chapter{Background}

This dissertation builds on the methodologies of human-computer interaction and prior works in immersive and interactive narratives in mixed reality. To further illustrate the potential and current applications of these concepts in interactive narrative virtual production, it is valuable to analyze already existing examples that embody the principles and innovations discussed in this research. The works discussed in this background section discuss the developments of personalized experiences and live elements in immersive theater, the collaborations between human and technology in theater, and the technical developments in mixed reality, AR theater, and human-computer interaction. 

I begin by discussing the concept of personalized experiences with live elements in VR. Specifically, I focus on two games, Road 96~\cite{digixart2021road} and The Under Presents~\cite{game2019presents}, that created unique encounters that are difficult to replicate. This development of immersive digital theater, where viewers participated as characters stems back to location-specific theater from the 1980s. As this development significantly influences today's immersive theater, this opening discussion primarily focuses on traditional theater plays, though I briefly touch on VR works at the end, which might seem out of place here.

Next, I introduce \textit{9 Evenings}, a groundbreaking performance by the Experiments in Art and Technology (E.A.T.) studio at Bell Labs. This work demonstrated the integration of live components through collaboration between humans and technology~\cite{nokia2024genesis}. It showcased the dynamic interaction between audiences, artists, and robots, illustrating the depth of involvement and innovation in these performances. From this, I list various works that blend digital art and traditional performances and provide brief descriptions to illustrate where my work stands in relation to them.

\begin{figure}[t]
\centering
\includegraphics[width= 1.0\textwidth]{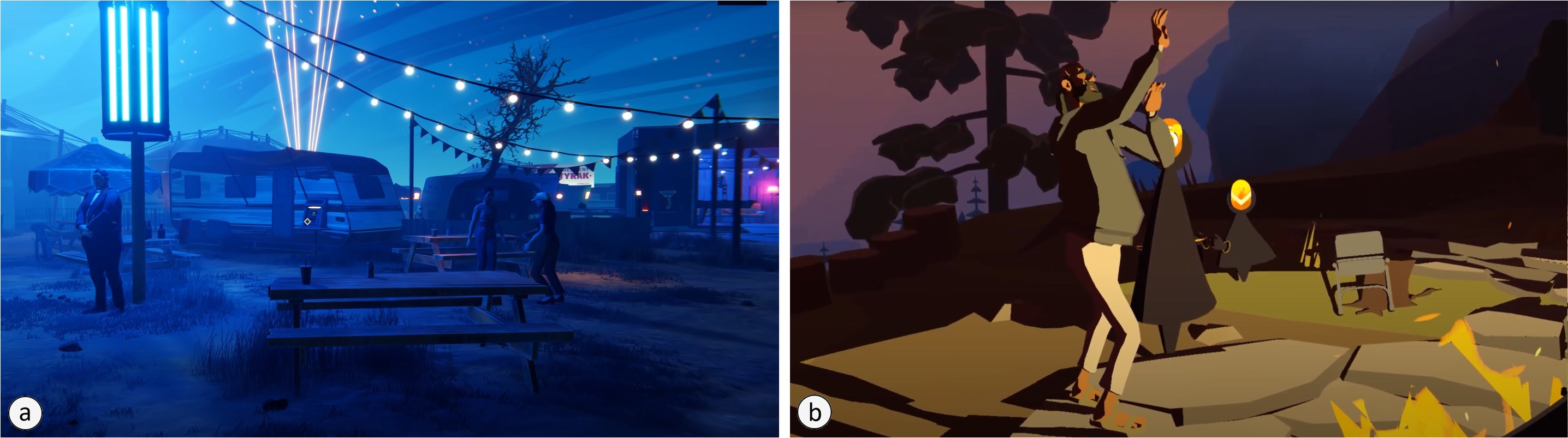}
\caption{(a) The popular VR game Road 96 was initially released in 2021; (b) The Under Presents released in 2019, included a live performance component.}
\label{fig:recentGames}
\end{figure}

\begin{figure}[h]
\centering
  \includegraphics[width= 1.0\textwidth]{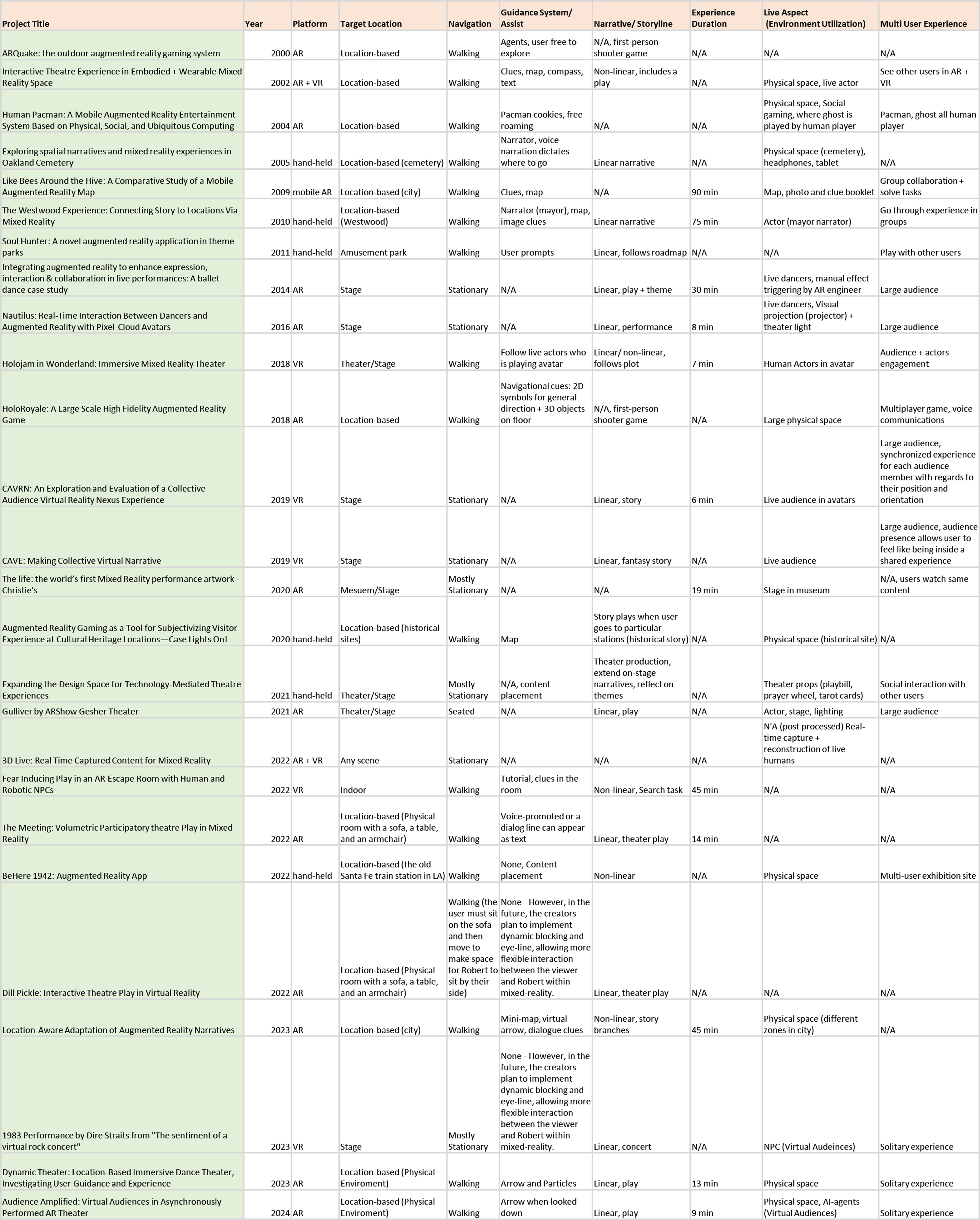}
\caption{Interactive narratives in mixed reality: An analysis of 26 projects}
\label{fig:AR Theater Literature}
\end{figure}

Following this, I delve into the technical advancements in computer science that have made these experiences possible. This includes topics such as interaction over wide areas in mixed reality, AR theater, and human-computer interaction (HCI) in dance. Subsequently, I present a chart of 26 projects (Figure \ref{fig:AR Theater Literature}), each featuring interactive narrative components in mixed reality and demonstrations of how spatial content can be experienced. These projects are analyzed based on elements such as the duration of the experience, storytelling style, navigation guidance, and platform type (handheld, VR, or AR). In addition to crediting the prior work that has inspired my research, this chart illustrates how interactive narratives in the mixed reality research domain have been actively practiced and explored. It shows trends in duration, interactive methods, and live theater components, among others. However, my focus is more on designing user flow and experience, the creation of these interactive works in close collaboration with artists and choreographers, the way stories are told, and the utilization of space in virtual productions.

Finally, I conclude by drawing connections to the aspects discussed in this related works section to what my AR system does and contains. This highlights how my research builds upon and contributes to the existing landscape of AR systems.

\begin{section}{Personalized Experiences \& Live Elements in Immersive Theater}

Today's immersive virtual productions capture the essence of live events that are directly influenced by user engagement and audience participation. One such example, the immersive VR adventure Road 96~\cite{digixart2021road} serves as a compelling case study for how dynamic narratives can be crafted and experienced in virtual environments. In short, the game play narrates several teenage hitchhikers attempting to flee the authoritarian nation of Petria without being arrested or killed. The inclusion of non-playable characters (NPCs) in Road 96 enhances the social and immersive aspects of the game, creating a rich, living world that evolves with player interaction. The random encounters with NPCs generate connection and meaningful interaction, yet, players who return to the game after completion may not experience the same story line. Road 96 illustrates how we can better understand how interactive storytelling elements and immersive theater techniques, such as providing players with unique and personalized narrative journeys, are being integrated into virtual reality. The game highlights the creative possibilities of virtual production and also demonstrates the evolving landscape of storytelling in digital media, where player choice and agency significantly impact the narrative experience. 

Furthering capturing the essence of live theater, the concept of integrating interactive elements to foster a sense of community and presence is exemplified in another groundbreaking project, The Under Presents\cite{game2019presents}. Under Presents is a multiplayer VR game and performance space that enables users to collectively view live or recorded performances, which encourages a shared virtual experience in a participatory theater setting. It includes live actors in the performance to help make the experience feel more real and, as one player described, ``...it's as if it was a real life experience which you can only experience once...''\cite{review2022}. Although it can be experienced entirely on one's own, these shared spaces where audiences build experiences together are on the rise. 

The rise of shared virtual experiences, as seen in projects like The Under Presents, reflects a broader trend towards participatory and immersive storytelling that transcends digital boundaries. This trend is deeply rooted in traditional theater practices, where immersive theater has evolved independently of digital media to transform audience engagement. By inviting audiences to become active participants in the narrative, immersive theater has created profound, multi-sensory experiences that resonate deeply with participants and push the boundaries of traditional theater. This evolution can be traced back to the cultural and political movements of the Baby Boomer generation, particularly from 1960 to 1980, where experimental theater forms encouraged interaction and community-building. This is exemplified by pioneering companies like The Living Theater and The Performance Group, or through the documentary~\cite{living1983theater}. These living theaters were instrumental to integrating actors with audiences, fostering a sense of community, and producing politically charged performances that encouraged participants to advocate for social change.

The rise of, site-specific theater in the 1980s sought to break free from the constraints of traditional stage performances. Notably Surface Tension (1981, at Berkley)~\cite{surface1981tension}, Storming Mont Albert (1982, at Melbourne's Yarra River)~\cite{storming1982mountalbert}, Full House (1989 at Saint Kilda Boarding House)~\cite{full1989house}, utilized ordinary spaces such as abandoned buildings, warehouses, and outdoor locations into immersive theatrical settings. This shift further blurring the lines between reality and performance and allowed audiences to engage with narratives in innovative and engaging environments~\cite{alively2000theater}.

A significant leap forward in immersive theater occurred in 2000 with the emergence of Punchdrunk, a company founded by Felix Barrett. Punchdrunk has become iconic in the field of immersive theater, synonymous with its unique style of performance. Their productions, such as \textit{Sleep No More}~\cite{sleepnomore2011}, emphasize exploration and audience agency, allowing participants to freely roam in elaborate theater spaces where actors engage in continuous action unaffected by audience interaction. This approach maintains a constant narrative flow while providing individualized experiences for each audience member. Emerging companies, like Third Rail Projects based in Brooklyn, continue to produce creative immersive theater experiences that involve more user attention and involvement in progressing with the narrative. The popularity of these experiences has grown, and in ways being recognized as mainstream media. However, only having a set number of people engaging in the experience at any given time remains a natural limitation. 

The onset of the COVID-19 pandemic posed significant challenges to live performances and necessitated innovation within the immersive theater community. As lockdowns restricted traditional theater experiences, creators turned to augmented reality (AR), virtual reality (VR), and digital platforms~\cite{game2019presents, thoravikumaravel2022dreamstream, schlagowski2023wish, li2022interactive}. These innovations included theater via Zoom and one-on-one immersive phone call experiences. The Candle House Collective~\cite{alternatereality2024}, for example, focuses exclusively on immersive phone call experiences, placing audiences at the heart of high-agency encounters. This theater company has received praise for its inventive programming, which seeks to displace audiences from their reality and engage their imaginations, offering scalable and sustainable theater experiences. More recently, productions like The Under Presents by Tender Claws blend VR and live performance~\cite{game2019presents}, enabling real-time interaction between audiences and performers. This hybrid experience bridges the gap between physical and digital realms, expanding the possibilities of immersive theater. Furthermore, VR became a prominent medium for meditation and exercise at home~\cite{FractalBrain}, providing users with innovative ways to enhance well-being and maintain physical fitness during lockdowns. Many companies turned to digital platforms to create virtual immersive experiences, reaching global audiences from the comfort of their homes. As immersive theater continues to evolve, its capacity to engage, inspire, and transform audiences ensures its enduring relevance in contemporary storytelling.

\end{section}

\begin{section}{Origins of the Interactions between Human and Technology in Theater}

Since the year 2000, the exploration of interactive narrative and artistic realism in the featured twenty-six projects lays the groundwork for understanding the subsequent innovations in immersive theater. These projects demonstrate the incorporation of cutting-edge technology and mixed reality through various devices, all while considering essential elements like story format, duration, and physical interaction. This analysis provides valuable insights into the evolving trends and specifications in immersive experiences. The projects also examine whether the experiences were designed for specific locations or targeted layouts. They assess the mobility aspect, whether the experience requires walking input, or if it is seated or stationary. Additionally, these interactive narratives are evaluated on whether they provided guidance or instructions as the story progressed, such as navigational aids or assistants. All these aspects are meticulously checked and compiled into a comprehensive chart to compare and identify trends or basic specifications of these projects.

The integration of interactive elements in these twenty-six projects traces back to earlier experimental efforts, particularly the pioneering work of Experiments in Art and Technology (E.A.T.) at Bell Labs. Notably, E.A.T.'s 1966 event, \textit{9 Evenings: Theater and Engineering}, was a groundbreaking collaborative platform where performers and technology came together to integrate synchronized communication and artistic elements into live performances and theater, aiming to redefine audience engagement through technology. One example of this was a live improvisation featuring a performer on a dance platform moved around by robots, presented in \textit{Solo} by Deborah Hay. This piece explored the degree of control and expression when working with long-range communication technology, highlighting how the performance could be both harmonious and individualistic at the same time~\cite{jeffers1979leaving}. By examining the pioneering efforts of E.A.T, we can see how these early collaborations between artists and technologists have had a lasting impact on aesthetic persuasion. Each performance piece presented by E.A.T continues to fascinate and inspire, highlighting the ongoing pursuit of collaboration between artists and research engineers to create performative pieces.

The recent innovations in immersive theater~\cite{game2023japan, game2023flow, game2020kizuna, game2020capsule}, accelerated by the challenges of the COVID-19 pandemic, echo the pioneering spirit of early experimental performances to underscore the importance of spatial and temporal connection in shaping audience perception and memory. This historical context enriches the understanding of how contemporary immersive experiences continue to evolve, emphasizing the importance of presence and shared space in storytelling—concepts that remain central to both traditional and modern performance practices~\cite{slater2016enhancing, grau2004virtual}.

Today's live performances, such as music concerts, feature large stages and distanced audiences. Members of the audience may find themselves relying on LED screens to see the details of the performance, such as the artist's facial expressions. Despite the detail and clarity of the screens, many people prefer to focus on the stage itself. This preference exists because looking directly at the stage helps orient viewers to the performer's location, allowing them to share the same space and time as the artist. In contrast, watching through a screen introduces a layer of separation between the audience and the performers. The way an event is remembered and perceived often depends on how individuals see themselves within the moment, as an extension of sharing space and time. To delve deeper into this notion, it is essential to focus on methods that capture and present a person in respect to the viewer's coordination through Mixed Reality, applying concepts such as place illusion, plausibility illusion, and plausibility paradox~\cite{hillis1999digital, vskola2020virtual}.

Scaling can be experienced through a blend of first-person and third-person perspectives, from experiencing an exponentially large San Francisco skyline while being a giant~\cite{amber2016city} to exploring the intricate underground world as a mouse~\cite{polyarc2018moss}. In this context, Place Illusion, as defined by Slater, goes beyond interactivity and appears realistic. While place illusion focuses on the visual reasoning of being in the projected environment, the plausibility illusion addresses how a depicted scenario aligns with reality. An important aspect of plausibility illusion is that the logic of the content does not have to be grounded in the extension of our reality; given the right conditions, a flying person, which defies real-world physics, can occur without breaking the immersive experience and narrative within the context of play. An excellent example of a first-person curated experience is the VR game Half-Life: Alyx~\cite{valve2020half}, where players experience the main character Alyx's perspectives, alternating between the present timeline in a first-person perspective and the past and future timelines in a third-person perspective. Even with constant perspective shifts and transitions into abstract spaces, entering a completely white space feels plausible in the context of Alyx's narrative experience.

While the notion of place illusion can be applied, there are fundamental differences between AR and VR experiences. In AR experiences, users are provided with Embodiment Illusion and Plausible Reality as these environments contain an appearance of truth with the user's physical body providing full agency. These elements of AR eliminate the process of reasoning with the user's awareness, convincingly integrating the virtual environment as reality.

This integration of virtual and physical realms in AR parallels the innovative approaches seen in early kinetic works, such as the E.A.T studio which utilized theatrical technology. Building on this historical foundation, the following section explores pioneering collaborations between artists and technologists that have significantly influenced artistic inspiration. By expanding our focus to mixed reality projects with narratives, we can better understand how 3D viewpoint systems and theater experiences intersect and offer insights into how components from this research can be adopted into an AR theater system. This analysis reveals how these elements translate into the overall user experience, bridging past innovations with future potential.
\end{section}

\begin{section}{Immersive Performance-Based Projects}

In this section, I explore ten immersive performance-based projects. These performances, conducted by performers or dancers in site-specific locations, utilize computational tools to create an immersive theatrical experience. Recent plays such as \textit{Natasha, Pierre and the Great Comet of 1812}~\cite{greatcomet2024}, \textit{Sleep No More}~\cite{sleepnomore2011}, and \textit{Then She Fell}~\cite{thenshefell2020}, which admits 15 audience members who participate in playing roles that advance the narrative, exemplify this approach by breaking the fourth wall. By eliminating the traditional use of the stage, these plays provide an immersive theatrical experience that completely engulfs the audience in the performance. This is achieved by using specific locations that allow audiences to converse with the actors and interact with their surroundings. In these projects, computational aid enhances the immersive quality, further blurring the lines between reality and performance.


\subsubsection{Boundary Functions by Theodore John Kaczynski}
\textit{Boundary Functions} (1998: Theodore John Kaczynski) is an interactive piece of art that separates viewers in the gallery by projecting lines from above onto the ground. When two or more users are present on the floor, a line is projected to divide the space between the users, cutting through the floor and altering the sense of physical space as they move. The division-making pattern is inspired by diagrams found in nature at all scales and is easily found in biology and nature surfaces. The invisible metaphorical distance and the physical space between the people are dynamically visualized in the floor diagram~\cite{kaczynski1967boundary}. 

\subsubsection{Pixel by Mourad Merzouki}
The dance piece \textit{Pixel} (2014: Mourad Merzouki, Adrien Mondot and Claire Bardainne) artistically integrates sensing technology and spatial mapping technology into performance dance arts. Eleven dancers perform with harmonious patterns and textures created by dots, flawlessly orchestrated through the projection. A collection of animated computer visuals that respond to the dancing movements are interacted with by the dancers in their performances. The natural-appearing synergy between the choreography and the visual overlay backdrop demonstrates how dance movement and expression can be amplified through the use of extended reality~\cite{pixel2022}. 

\subsubsection{Pattern Recognition by Alexander Whitley}
\textit{Pattern Recognition} (2016: Alexander Whitley and Memo Aktenis) is a performance for two dancers and light beams. The light show explores themes of learning memory using dance choreography assisted by a computer light system trained using machine learning and artificial intelligence. Whitley and Aktenis explore technology and design that seek to redefine the boundaries between computation response and human expression through pattern recognition. Both the human and the machine contexts work in collaboration, with one approach informing the other. The movement-responsive system tracks and observes dancers intelligently to mimic, and occasionally complete, the dancers' movements~\cite{patternrecognition2016}. 

\begin{figure}[t]
\centering
\includegraphics[width= 1.0\textwidth]{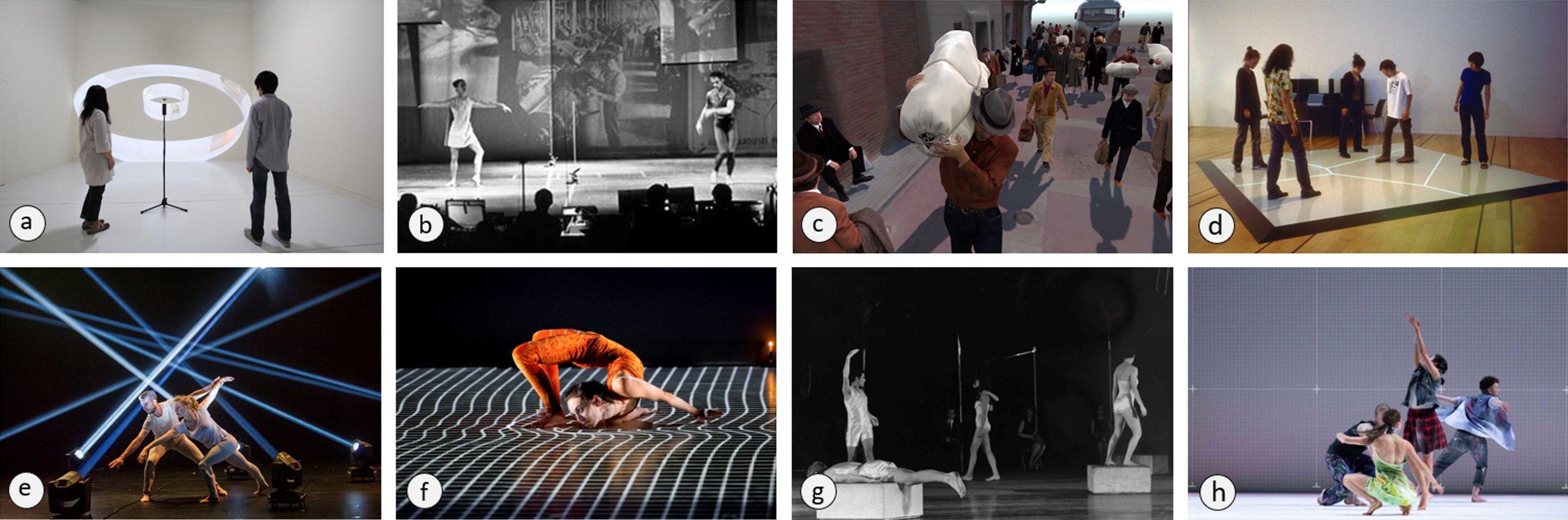}
\caption{Immersive Theater Works. (a) Morel’s Panorama by Masaki Fujihata. (b) Variations V by Merce Cunningham. (c) BeHere 1942 - Masaki Fujihata. (d) Boundary Functions by Theodore John Kaczynski. (e) Pattern Recognition by Alexander Whitley. (f) Pixel by Mourad Merzouki. (g) Solo by Deborah Hay. (h) Centaur by Pontus Lidberg.}
\label{fig:survey}
\end{figure}

\subsubsection{Centaur by Pontus Lidberg}
\textit{Centaur} (2020: Pontus Lidberg, Cecilie Waagner-Falkenstrm and Ryoji Ikeda), a dance play, was presented by Danish Dance Theatre using an interactive computational stage. The work explores a new kind of expressiveness using artificial intelligence technology that gives the performer's movement and the computation system equal weight. The interactive system was adopted to answer computer science concepts that elevate human qualities by combining artificial and human intelligence. The computational stage delivers a performance in response to each dancer's individual movement. The essential question that the creators explore is whether creative technology is a new form of art or an extension of the dancer’s movement, whether the projected environment is computer-generated art or a performance extension~\cite{centaur2024}.

\subsubsection{Solo by Deborah Hay}
\textit{Solo}, one of the performances of \textit{9 Evenings} (1966; Deborah Hay, Steve Paxton, and Robert Rauschenberg), was performed by 16 dancers and 8 radio-controlled moving platforms that moved about the stage. Following a single route that occasionally overlapped with other dancers, each dancer performed on platform rises that moved in slow motion while drifting across the stage. A radio conductor operates the moving platform to move in a pattern modeled like the motion of a cloud of atoms with an uncertain trajectory. \textit{Solo} is one of the earliest machine and human performances and a precursor to immersive theater plays~\cite{jeffers1979leaving}. 

\subsubsection{Variations V by Merce Cunningham}
Billy Kluver, photocells engineer at Bell Labs, used projections, displays, and appropriated Nam Jun Paik television graphics to create a dance performance. \textit{Variations V} (1966; Merce Cunningham, Billy Kluver, Nam June Paik and John Cage), a study of expanded cinema, incorporated a fifty-channel mixer and photocell beam to capture the intensely kinetic process of staging the show. The projected films and dancers' overlapping movements created a resonated pattern, allowing the dancers' motion to be entirely absorbed into the stage. Dancers moved freely around the stage, breaking the array of photocell beams to alternate synthesizers that produced sounds and graphic visuals for the projection~\cite{miller2001cage}.

\subsubsection{Morel’s Panorama by Masaki Fujihata}
Morel’s \textit{Panorama} (2003; Masaki Fujihata), a digital video installation, situated a custom-made panorama camera in the center of the gallery space. The camera captured 3D recreations of the gallery space with the viewers at the site of projection. Captured 3D recreations were then mapped onto a rendered cylindrical image projected on the walls of the exhibition space, updating every few minutes to highlight the latest condition of the area. The huge camera positioned in the room was concealed in the reproduced model of the gallery space as it progressively rotated in the projection. \textit{Panorama} camera technology widely adopted in surveillance systems used in prisons was used in the installation to reflect the gap between subjective experience and reproduced reality. The installation investigated how human senses can be recreated and digitally documented and how digitally captured and human senses can alter reality~\cite{fujihata2003morel}. 

\subsubsection{BeHere 1942 by Masaki Fujihata}
Masaki Fujihata’s BeHere 1942: A New Lens on Japanese American Incarceration tackles the chilling topic of the internment of Japanese Americans during World War II using augmented reality based on the historic photographs taken in the 1940s. A part of the Yanai Initiative for Globalizing Japanese Humanities, a joint UCLA and Waseda University project, the exhibit draws on the thousands of photographs taken by Dorthea Lange and Russell Lee to reimagine and reconstruct the process of incarceration. Participants occupy the role of a recently dispatched reporter sent to document the events surrounding Japanese American removal, which is experienced by aiming handheld devices (mostly tablets) at nearby buildings and spaces. Fujihata’s massive AR experience employed the use of actors in both Los Angeles and Tokyo to make the experience as realistic as possible. After careful consideration of the subject and the photograph to reconstruct the narrative, the actors would be recorded by volumetric video capture and integrated into the AR app. In this way, BeHere 1942 demonstrates the potential of non-headset-based augmented reality experiences~\cite{fujihata2022behere}. 

\subsubsection{The Life (2020) by Marina Abramović}
Claimed to be the “world’s first mixed reality performance artwork,” Marina Abramovic’s \textit{The Life }provided a 19-minute performance for participants to enjoy through the use of HoloLens 2 headsets. The performer, Abramovic herself, danced in real-time within a five-meter circle, though her physical body was elsewhere. The performance, which was captured using thirty-six cameras while Abramovic remained in front of a green screen, considered the physical orientation of the performer and participant. For participants wearing the headsets, dots lined the physical room where the performance artwork took place to allow for calibration of each user’s position during the experience~\cite{abramovic2019}. 

\subsubsection{Birder Memorial by Mark Skwarek}
The Border Art Memorial, an installation of the New York Museum of Modern Art (MoMA) opened in October 2010 to critical acclaim. The augmented reality public art project, officially titled \textit{Border Memorial: Frontera de los Muertos}, was dedicated to the thousands of migrants who have died in the US/Mexico borderlands. The project was part of a broader exhibit titled “We AR in MoMA,” which sought to discover how AR changes our understanding of public space and art. Using early handheld devices and smartphones, participants would see augmented artwork and installations based on their location in the museum’s exhibit thanks to GPS tracking and the application called Layar~\cite{memorial2012}.

Upon downloading Layar, participants could use their smartphones to scan areas along the border until a large virtually augmented object, in this case, a skeleton based on Oaxacan tradition called a calaca, would appear.  The application’s geolocation software, and previously collected GPS data on migrant deaths, allowed for the superimposition of these augmented calacas to symbolize the location where a migrant traveler had passed. In addition to raising awareness of the crisis at the border, which at the time of the exhibit remained largely outside the public consciousness, the Border Art Memorial demonstrated the use of location-specific augmented reality without the necessity of performance art. In many ways, the calacas operated as actors, in a nontraditional sense, by spinning and floating toward the sky upon the participant’s scanning. 

\subsubsection{All Kinds of Limbo by Raffy Bushman and Nubiya Brandon}
\textit{All Kinds of Limbo}, an XR production, allows for an immersive musical performance to be experienced in the comfort of the participant’s home. Launching in 2022, All Kinds of Limbo demonstrates the potential of collaborating tech and art, which in this case is seen in the partnership with Microsoft and writer/vocalist Nubiya Brandon, composer Raffy Bushman, and the performance of the United Kingdom’s Nu Shape Orchestra. What is unique about All Kinds of Limbo is the variety of engagements provided. The participant’s engagement with the performance piece depends entirely on the device they use. For example, participants using a smartphone or handheld device engage with an augmented performance projected on their home’s table or floor, while participants logging in on a laptop would use more tactile controls, similar to that found in a video game.  Arguably, the performance is best experienced through a virtual reality headset, which provides a way for the participants and performers to engage in the space together~\cite{bushman2019all}. 

\subsubsection{Mariza Dima’s Sutton House Stories}
The ongoing project, Sutton House Stories, presents an augmented reality experience of the Sutton House museum through the use of Hololens. The immersive experience allows participants to engage with four centuries of history through the Sutton House by traveling between rooms and interacting with the “complex relationships between culture, communication, learning and identity” (E. Cooper-Greenhill, 2007). The project employs actors, performers, designers, digital technicians, and historians to provide a sensorial, yet thorough, storytelling experience for museumgoers. Mariza Dimas, the project’s augmented reality mastermind, emphasizes sensorial engagement as an “embodied storytelling design,” a concept she calls \textit{affectual dramaturgy}. As she, and co-author Molly Maples, argue in a 2021 research study, affectual dramaturgy fuses “the users’ interactions with and through the AR technology with the historical characters, the story, the journey, and the site of the experience” (Maples, Dimas; 2021, 1). Using affectual dramaturgy allows the participant to consider multiple contexts — historical, environmental, emotional — to increase understanding of a historical site. While limited to the physical space of the Sutton House, the project demonstrates the potential of using an augmented reality program in museums and other cultural heritage sites~\cite{dima2022design}.

The use of site-specific augmented reality (AR) narratives is most effective when there is a strong connection between the spatial content and the physical location. A prime example of this is Masaki Fujihata’s BeHere 1942 experience, which allows users to engage with the historical context of Japanese American removal by occupying perspectives other than their own and interacting with a variety of predetermined characters. This application is designed to be experienced specifically at the Los Angeles Santa Fe train depot, the actual site of the historical event it represents. This site-specific nature adds significant meaning and authenticity to the experience, making it a must-try for anyone in the area. The immediacy and relevance of the location encourage more people to participate, as opposed to the possibility of postponing the experience.

However, because BeHere 1942 recreates a historical event, its narrative is largely predetermined, offering a curated experience that might be positioned in the middle of an interactivity scale. In contrast, Marina Abramović’s \textit{The Life} is a highly curated, one-person performance that is static, as the content is consumed from a fixed area and the presence of an audience does not alter the nature of the performance. In the chart~\ref{fig:location-narrative}, \textit{The Life} is positioned at the extreme end of the Location axis as static, indicating its location-specific nature. It is intended to be viewed from a designated area in the gallery, with each person experiencing the performance through HoloLens 2. This results in a highly location-specific and curated experience, as depicted in the chart~\ref{fig:location-narrative}.

\begin{figure}
\centering
\includegraphics[width=.9\textwidth]{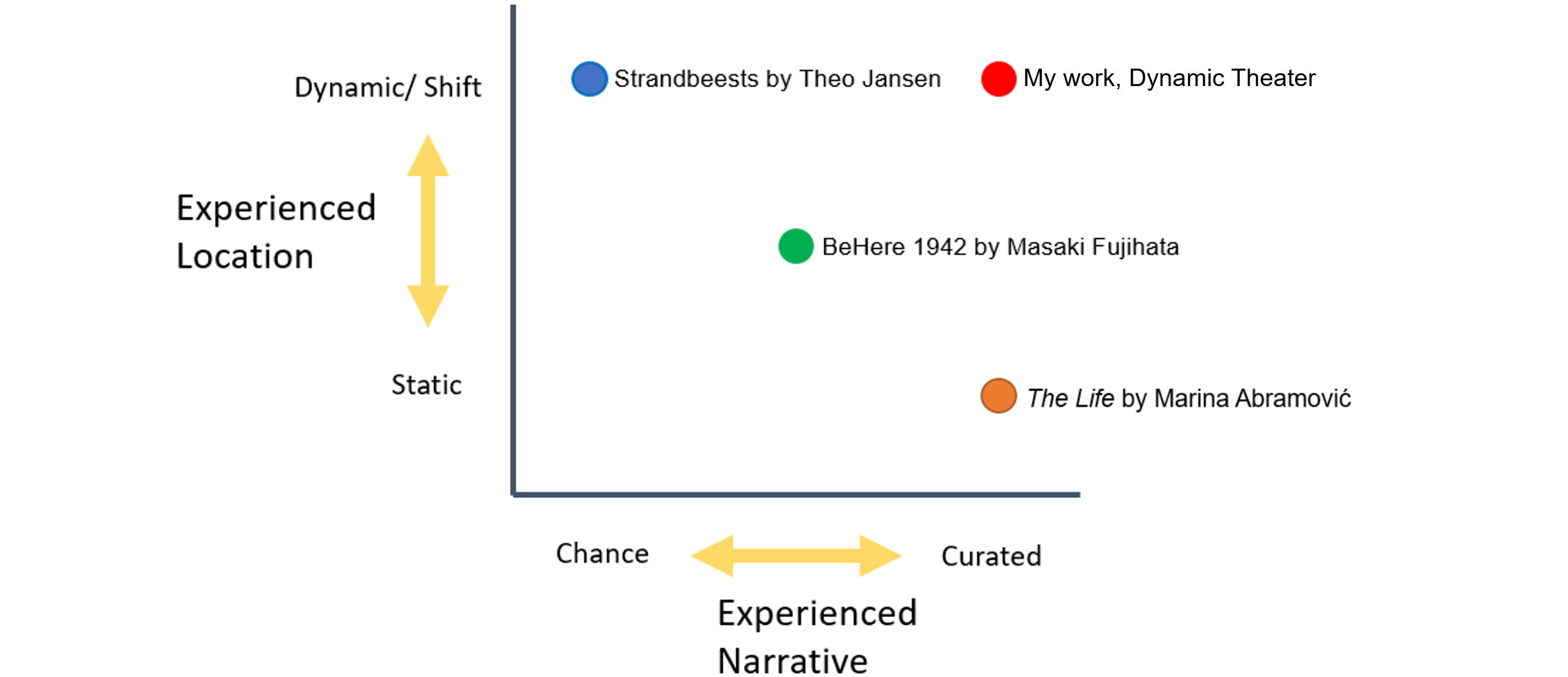}
\caption{The location and narrative graph illustrates the levels of flexibility in narrative and location, highlighting the potential for curating user experiences. This graph positions my work, Dynamic Theater, alongside the works of Theo Jansen, Marina Abramović, and Masaki Fujihata. It demonstrates that Dynamic Theater is capable of being both dynamic in location and providing a curated user experience.}
\label{fig:location-narrative}
\end{figure}

\end{section}

\section{Conclusion}

These studies demonstrate how technology can elevate human expression to be seen and experienced in digital environments, providing guidelines for implementation. The biggest challenges remain in stabilizing mixed reality platforms for interactive narratives, as the technology can be experienced in many ways. A key question is how to integrate human actors or agents into these spaces, allowing stories to be explored and unfolded through human navigation while maintaining flexibility to support diverse, natural interactions that enhance the user experience in AR theater. In essence, existing platforms and tools still have several milestones to achieve before they can fully meet creators’ needs in designing and building interactive immersive experiences. The ultimate goal is to present a platform or system for AR theater that can account for physical layout and utilize it effectively for storytelling. 

My research specifically aims to bridge the gap between the goals of interactive AR/VR storytelling and the utilization of physical spaces. It offers guidelines to artists on how to incorporate these elements into their storytelling. My AR Theater system aims to offer a foundational platform for storytellers, enabling them to translate their vision into a system that resembles the cue sheet of traditional theater. This helps them bring their creativity directly to the mixed reality platform. Additionally, my work lowers the technical barriers for creating interactive AR/VR scenes and stories, allowing creators with no programming experience to focus on storytelling and the effects they want to convey in a mixed reality system.

Unlike traditional theater experiences with linear narratives, my research offers a different approach where there are no explicit objectives, goals, or tasks assigned to the audience. Instead, the game logic and AR cue sheet I developed monitor user movement, progress, and pace, allowing viewers to unfold the story at their own rhythm. By providing guidelines on changing stages, creating new experiences within the same environment, and guiding users without overly instructing them, my system allows users to take ownership of the narrative. As users move around and reach certain zones, the story unfolds naturally—without strict rules on the order in which things should be seen. By pushing the boundaries of AR theater and offering personalized, location-based experiences, my research culls from the previously mentioned works to present an intricately crafted open-world theater experience that fully utilizes expansive spaces, allowing viewers to wander freely, get lost, or become completely immersed in their surroundings.

\chapter{Wide-Area Augmented Reality}

In this chapter, I present the foundational prototypes and projects critical to the development of my AR theater system. My experience as a research intern for Nokia Bell Labs, participation with the Santa Barbara Virtual Campus project, and a course I taught on Human-Computer Interaction with Augmented Reality, provided vital knowledge and experience to execute my own research. These projects, which were mostly conducted in 2021, explore the applications of large-scale, wide-area augmented reality settings. These projects explored methods to create accurate digital twin models, design AR stages, and simulate physical spaces to demonstrate how people navigate and interact within large spaces. Further, they illustrate the techniques for spatial anchoring augmented objects that can be used in headset systems. 

The chapter situates my first experience with spatial modeling, World Tracer, with research conducted at UCSB to develop digital-twin models. I begin by discussing my internship with Nokia Bell Labs' E.A.T studio, where I developed an intuitive tool for creating 3D environments using bitmap (PNG) representations of robot SLAM data. This tool allows users to trace walls and objects from SLAM data to construct a 3D model, which can be exported for physics simulations. This innovation enables the conversion of scattered floor layouts into a comprehensive representation of large spaces.

Building on my expertise in creating digital twins, the second part of the chapter focused on my development of a standardized pipeline for producing highly accurate 3D digital twins of wide-area outdoor spaces. This was implemented as part of the Santa Barbara Virtual Campus project, which empowered students to design AR experiences by contributing to a virtual representation of the UCSB campus. This work developed into three AR outdoor studies. The first, which focused on environmental lighting, explored the impact of mobile augmented reality on user interactions, behavior, and task performance in outdoor environments. A treasure hunt task using the Microsoft HoloLens 2 was employed to assess factors like lighting conditions and cognitive load.

Expanding from this project, we developed two additional prototypes. In the second outdoor study, which focused on the use of navigational aids, we evaluated three AR navigation aids to enhance search performance and user experience outdoors. Our research showed that in-world arrows were preferred, and users were less aware of physical objects compared to virtual ones. Finally, using the same platform, the third outdoor study emphasized understanding the consequences of climate change. The project increased user motivation for sustainable practices and showed significant potential for contextualizing climate change impacts. By using digital twin technology, we simulated natural disasters in familiar campus locations, providing a more profound understanding and impact. This project illustrates how to consider narratives and their impacts with AR systems.

\begin{section}{Early Work: World Tracer}
In the summer of 2021, I interned with the Future X Robotic team at Nokia Bell Labs' E.A.T Studio. Alongside Tommy Sharkey, a fellow PhD student from UC San Diego, we developed an intuitive tool for building 3D environments. Our project focused on optimizing floor robots at Nokia's Chennai factory in India, which manufactures network modems for 4G/5G networks. The factory employed five different types of Omron robots, each with varying specifications, such as speed and maximum load capacity, which made coordination complex. Additionally, the factory floor layout frequently changed due to shifting production demands, complicating robot navigation and necessitating a dynamic system for route optimization. Given the regular layout changes, hiring 3D modelers each time the factory layout changed was costly and inefficient. To address this, we leveraged the robots' onboard sensors and position data to collectively gather point cloud data, enabling the creation of simple 3D models in less than 10 minutes by untrained users and without needing special skills in 3D modeling.

\begin{figure}[t]
\centering
\includegraphics[width= 1.0\textwidth]{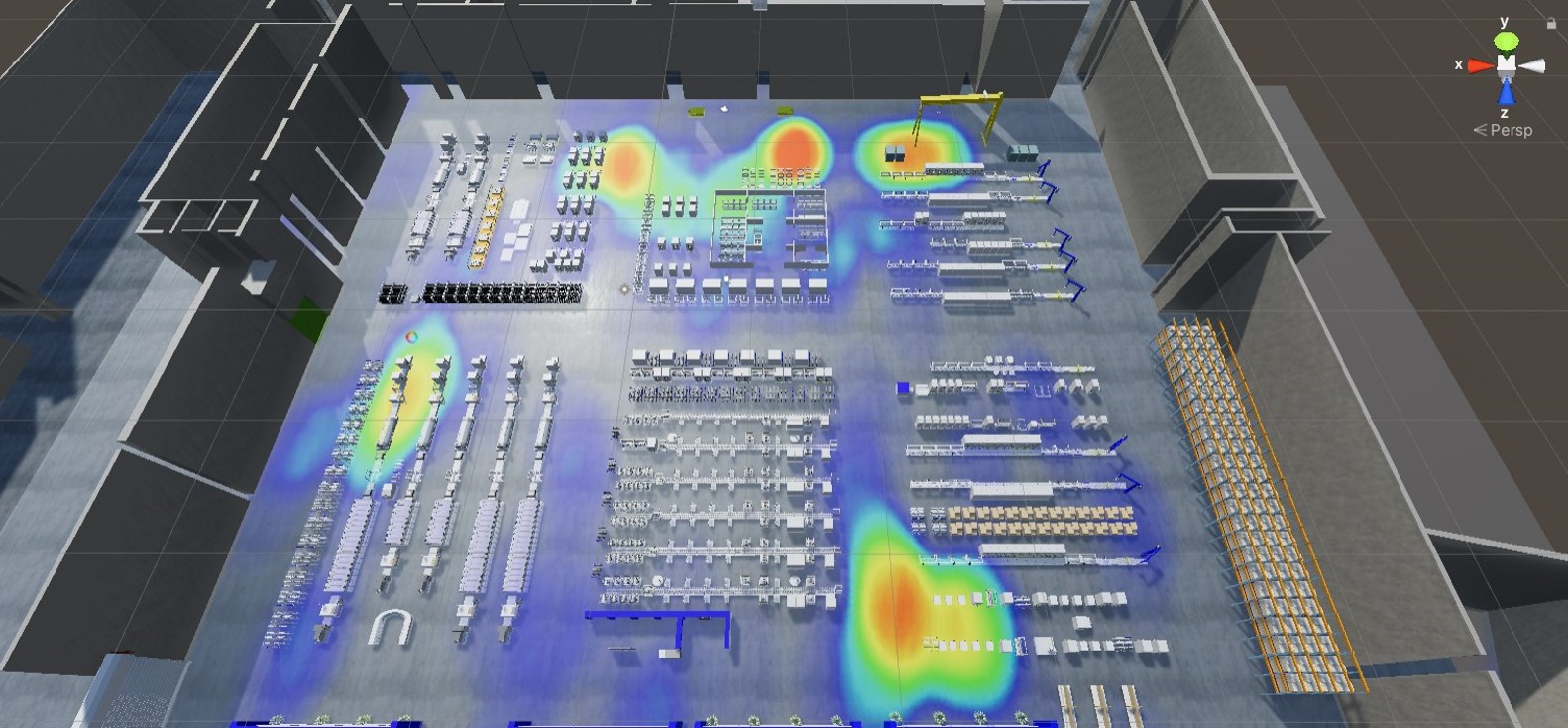}
\caption{A heatmap simulation of a robot deployed on the factory floor. By utilizing an accurate 3D model of the most up-to-date floor layout, generated through World Tracer, we can simulate the movement of Omron robots. This simulation helps identify areas with a higher likelihood of collisions. Based on these insights, we can adjust the properties or routes of the robots to maximize their efficiency.}
\label{fig:worldTracer}
\end{figure}

As a result, we developed World Tracer, a tool designed to facilitate the creation of 3D environments using bitmap (PNG) representations of robot SLAM (Simultaneous Localization and Mapping) data. It simplifies the conversion of 2D data into 3D models for simulations and other applications, as shown in Figure \ref{fig:worldTracer}. The key feature of World Tracer is its ability to quickly create a 3D floor model ready for simulation in just a few minutes, without requiring the operator to learn complex 3D modeling software. This allows operators to access the most up-to-date floor layout in OBJ 3D format with ease.

The process begins by importing SLAM data images into Unity, which serves as the development platform for World Tracer. Users start by cloning the repository, installing Unity 2020.3.12, and opening the project directory in Unity. They initialize the environment by pressing the play button. Once the images are imported, users navigate to the scene within Unity, integrate the images, and enter the file path. Users also adjust the pixel dimensions and scale (pixels per meter) to ensure the model accurately reflects the real-world environment. This scale setting is crucial for synchronizing the model’s dimensions.

World Tracer provides an intuitive interface with tools for tracing the outlines of walls and objects. Users select the desired tracing mode from a drop-down menu, choosing between modes, such as “Wall” and “Object.” The line tool is then used to trace features on the imported image. For walls, the tool extrudes traced lines into 3D forms, while objects require creating closed loops to form sealed 3D shapes. To aid modeling, World Tracer offers different perspectives, allowing users to switch between an extruded view to inspect 3D features and a bird's-eye view for ongoing tracing and detail addition. This flexibility ensures precise work and model adjustments as needed. For fine-tuning, World Tracer includes a correction tool for moving joints and lines in the model, allowing precise adjustments. When a joint is selected, it highlights it in yellow, indicating readiness for changes, such as repositioning walls. Once the model is complete, users can import additional 3D models into their layout or export their creations. The export function saves models in formats like OBJ or FBX, making them suitable for various applications, including physics simulations.

Additionally, World Tracer supports server integration that is useful for collaborative projects. Setting up the server involves installing Node.js, npm, and Assimp utilities. Once configured, the server can host World Tracer, allowing remote access and collaboration. Server configuration includes setting up hosted URLs for managing file storage and retrieval, enabling users to access and save their work online.

In short, World Tracer bridges the gap between 2D SLAM data and 3D modeling by providing an accessible and efficient way for developers and researchers to convert spatial data into detailed 3D environments. It is invaluable in spatial modeling and augmented reality applications. Drawing on this deep understanding of creating 3D digital twins of physical spaces, I focused on creating digital twins for experimental design which is crucial for my future work in AR applications spanning large areas, specifically through conducting AR studies at UCSB. 
\end{section}

\begin{section}{Digital-twin Model Development Pipeline}
As my understanding of creating digital worlds deepened, I aimed to develop a standardized pipeline that enables others to create highly accurate digital twin models for designing AR experiences. In this section, I outline the method used for creating a 3D digital twin of a large outdoor space as demonstrated in our outdoor studies~\cite{kim2021ucsb}. This method was shared during a workshop for the UCSB undergraduate CS group, as part of the Santa Barbara Virtual Campus project, which I lead. This initiative, driven by the undergraduate group Mentoring Unity for Virtual Reality Research, focuses on developing a detailed digital twin of the UCSB campus. The ultimate goal is to produce a complete digital twin of the campus with versatile 3D models optimized for various applications and research.

A key technical challenge the team addressed was managing large 3D model files on GitHub without relying on the Large File Storage (LFS) system. This involved strategically dividing and compressing assets to ensure they could be fully uploaded and easily shared, making them accessible to a wide audience. Additionally, the project included integrating a spatial anchoring system to accurately position virtual elements within the digital twin. The team researched optimal anchor point distribution across the campus map to ensure precise alignment and usability.

Through this project, I learned that detail and resolution are not everything. Simplified and easy-to-deploy physical layouts can be highly effective. It is important to highlight areas where user interaction will occur while providing a general understanding of the space. Simpler geometry, accurately aligned with high-definition photographic UV maps, enhances spatial understanding. A seminar workshop was conducted to educate participants on digital twin creation, empowering students to contribute to a comprehensive 3D Virtual Campus of UCSB for 3D or simulation research~\cite{kim2021ucsb}. 

I utilized a combination of tools to ensure an accurate and user-friendly pipeline for creating these 3D model digital twins. An iPad Pro (2020) with a LiDAR sensor was used to capture detailed 3D scans, providing precise depth information; however, any device equipped with LiDAR could perform this task. Satellite imagery was employed to align the model and ensure accurate scaling, while laser distance measures and the Matterport Pro MC200 3D Camera enhanced accuracy and detail. MeshLab was used to develop lightweight and detailed models suitable for standalone headsets displaying AR stages outdoors while serving as occlusion layers to ensure that physical walls could occlude AR objects. This approach enables the effective use of the digital twin in AR environments, which can be used to visualize the AR stage and design human interaction experiences.

\begin{figure}[t]
\centering
\includegraphics[width= 1.0\textwidth]{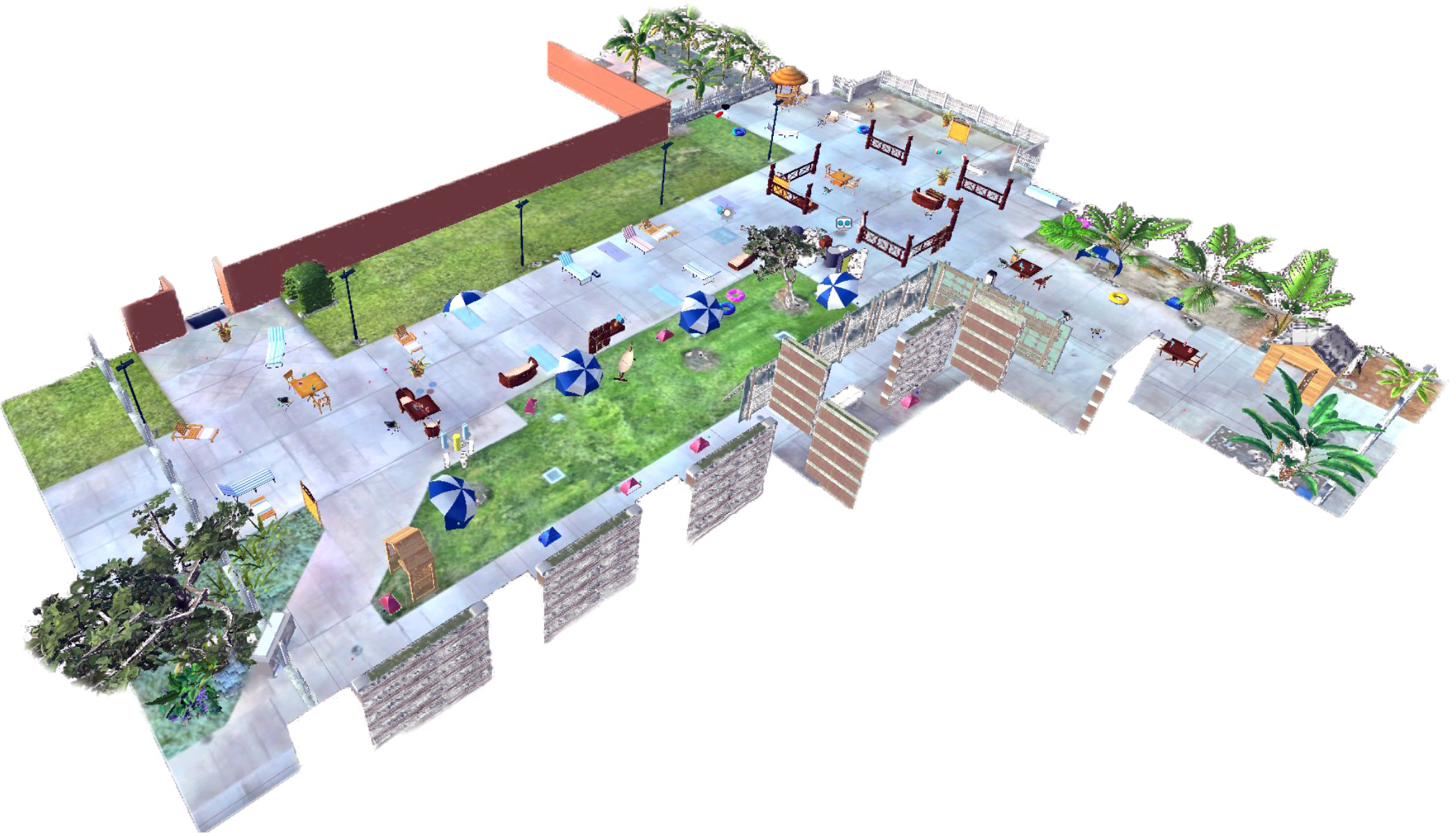}
\caption{Digital twin of UCSB's Kirby Crossing campus location used to design an AR stage}
\label{fig:digitalTwin}
\end{figure}

This process was used in the development of a digital twin for the UCSB campus, specifically the Kirby Crossing outdoor area next to the Kavli Institute for Theoretical Physics (Figure \ref{fig:digitalTwin}). The first step involves scanning the area using LiDAR-equipped devices, such as an iPad Pro with the Scanniverse app. This app, like others such as Polycam and Metascan, uses Light Detection and Ranging (LiDAR) to scan 3D scenes and export raw point cloud data can be used. After obtaining the point cloud data, which exceeded 10 GB for an area of 2,400 square meters, it was processed in MeshLab to compute normal and optimize the point cloud. The density of the point cloud is reduced using point cloud simplification, targeting a sample size of around 200,000 points to ensure optimization for AR headset performance. The optimized point cloud was then converted to a mesh using surface reconstruction techniques, such as Ball Pivoting in MeshLab.

Satellite imagery was used to correct any curvature or distortion by aligning the model accurately with the real-world physical environment in Blender. This ensured the scale and orientation of the digital twin 3D model were accurate. Higher-resolution satellite images from Bing Maps were also used for greater accuracy as those from Google Maps of UCSB's campus area were not as high quality at the time of creation in 2021. Ensuring correct scaling was crucial, so laser distance measures were employed to verify the dimensions of key features, such as the width of streets and distances between buildings. This step involved taking 12 different measurements around the area to achieve a high level of accuracy, reducing alignment discrepancies from 60 cm to around 5 cm.
I also accounted for any slight drifts in the models. Spatial anchors were used to stabilize the model when using headsets like the HoloLens. Once alignment was achieved, unique features for spatial anchors were identified which connected the digital data to the real world. It was crucial to maintain orientation and scale for alignment when new headsets retrieved this spatial anchor data from the server, as it allowed anyone to experience the model without additional setup. 

According to Microsoft Azure documentation~\cite{buck2022azure}, spatial anchors should be placed at easily identifiable visual cues, utilizing both point cloud and RGB data for tracking and referencing. In this project, six spatial anchors were strategically placed at unique locations such as signs and the school’s outdoor bulletin board. Detailed scans of these anchor areas were performed with the Matterport Pro MC200 to enhance alignment accuracy. This approach ensured the digital twin model had highly accurate surfaces where anchors were placed, allowing for precise alignment between the digital twin and the physical environment. This improved the synchronization of digital and physical spaces. Furthermore, small adjustments were made within the HoloLens AR environment by dragging and pinching the anchors to precisely align the virtual anchor points.

Once the digital twin was finalized, it was prepared for practical use. The design process involved carefully crafting the polygons to serve as visual cues, ensuring the digital twin effectively functions as an occlusion layer in augmented reality (AR) environments. When exporting to Unity, it was crucial to keep the 3D model under 100 MB, with each UV map also under 100 MB, to maintain compatibility with a GitHub repository without relying on Git Large File Storage (LFS). This facilitates ease of use for standalone headsets. The polygon count was reduced to maintain high-quality AR object placement, as the digital twin primarily acts as an occlusion layer rather than being directly displayed in AR. In the case of our Kirby Crossing model, the photo-realistic UV map was 258 MB, comprising one 3D model and two UV maps. Incorporating occlusion layers was essential to ensure virtual objects interact realistically with the physical environment. The goal was to balance detail and performance, making the digital twin effective for occlusion in AR applications and display simulations in Unity when designing user experiences and AR environments.

\end{section}

\begin{section}{AR Outdoor Study 1: Environmental Lighting} 

Using the UCSB digital twin model, this prototype project, which was developed in close collaboration with Radha Kumaran, explored the impact of environmental lighting conditions on search and classification tasks involving both physical and virtual objects within an augmented reality (AR) setting. Conducted outdoors, the study involved 48 participants tasked with finding and classifying virtual gems hidden among a mix of physical and virtual objects. The research aimed to understand how lighting conditions, specifically natural light in the evening and artificial light at night, affected user performance, comfort, fatigue, and cognitive load~\cite{kim2022investigating}.

\begin{figure}[t]
\centering
\includegraphics[width= 1.0\textwidth]{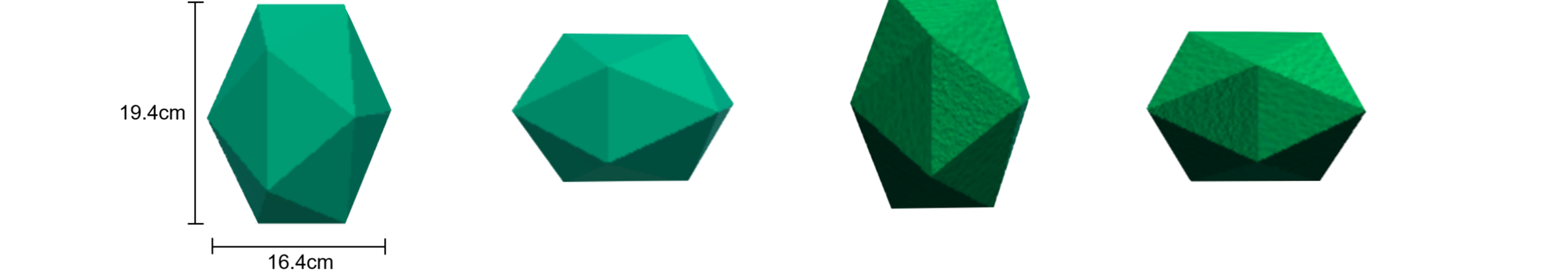}
\caption{Four different types of gems. From left to right: vertical and smooth, horizontal and smooth, vertical and rough, horizontal and rough.}
\label{fig:gems}
\end{figure}

Participants were randomly assigned to either the natural light condition or the night condition. The outdoor environment was set up to resemble a leisure-oriented scene, with objects like lawn chairs, umbrellas, and picnic tables. In the natural light condition, the study was conducted during the evening to avoid direct sunlight, ensuring consistent tracking and display operation. The night condition utilized artificial lighting from lamps. Participants were instructed to find and classify gems (Figure \ref{fig:gems}) that were distributed across three types of locations: physically occluded (hidden behind physical objects), virtually occluded (hidden behind virtual objects), and floating (not occluded by any objects).

\begin{figure}[t]
\centering
\includegraphics[width= 1.0\textwidth]{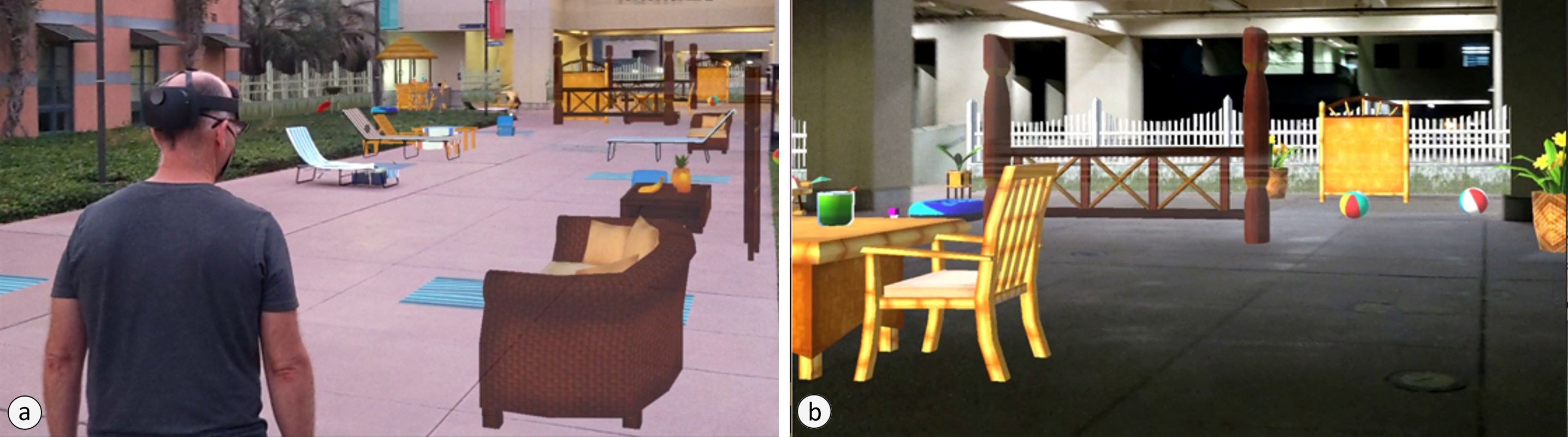}
\caption{Left: AR participant navigating part of the augmented outdoor environment for the daytime condition. Right: Part of the augmented environment during nighttime condition. Images captured via HoloLens-2's MR Capture. Relative brightness and opacity are not true to the experience.}
\label{fig:nightDay}
\end{figure} 

A dual-task condition was introduced in some trials, where participants engaged in an additional auditory task. This required them to respond to a target word within a sequence of spoken words, which added cognitive load to the primary gem search task. Data collection involved pre- and post-study questionnaires to capture subjective experiences regarding comfort, visibility, and fatigue, alongside performance metrics such as the number of gems found and task completion time.

The study was guided by hypotheses focused on performance, comfort, and cognitive load. Researchers hypothesized that participants would perform better under night conditions due to the increased visual salience of virtual objects against a darker background (Figure \ref{fig:nightDay}). It was also expected that participants would report higher comfort and lower fatigue levels at night, as these lighting conditions were presumed to be less demanding on their visual systems. Additionally, it was hypothesized that the dual-task condition would increase cognitive load, potentially degrading performance.

Results revealed that participants found more gems in natural light, with a statistically significant difference in performance between the conditions. Participants reported feeling more fatigued after the nighttime tasks. On the other hand, tasks were perceived as more demanding in natural light, which led to lower comfort ratings and trends toward increased fatigue.  However, only 22 out of 288 trials resulted in finding all gems, with a slight advantage during natural light trials. Visibility perceptions varied with lighting conditions; virtual objects were rated as more visible at night, while physical objects were clearer during the day.

The study’s findings highlighted the importance of considering environmental lighting in AR design, suggesting that while natural light might enhance task performance, it can also contribute to user fatigue and discomfort. These insights are essential for developing AR systems that are effective in dynamic real-world settings. The research contributes to understanding AR user experiences, particularly outdoors, and informs the design of future AR applications by emphasizing the need to optimize for various environmental conditions and user needs.
\end{section}

\begin{section}{AR Outdoor Study 2: Navigation Aids  } 

Continuing with my collaboration with Radha Kumaran, we explored the impact of three different AR navigation aids on search task performance and object recall in a wide-area outdoor setting using the Microsoft HoloLens 2 headset. This study was a natural progression from our first outdoor study, as users needed to search for gems over a large area. It made an important contribution to the mixed reality field by examining how users maneuver between physical and virtual environments to locate hidden gems. We tested navigational aids in a wide-area outdoor setting, where indications must be provided in all 360-degree directions, allowing users to decide the best path or order to find all the gems~\cite{kumaran2023impact}.

\begin{figure}[t]
\centering
\includegraphics[width= 1.0\textwidth]{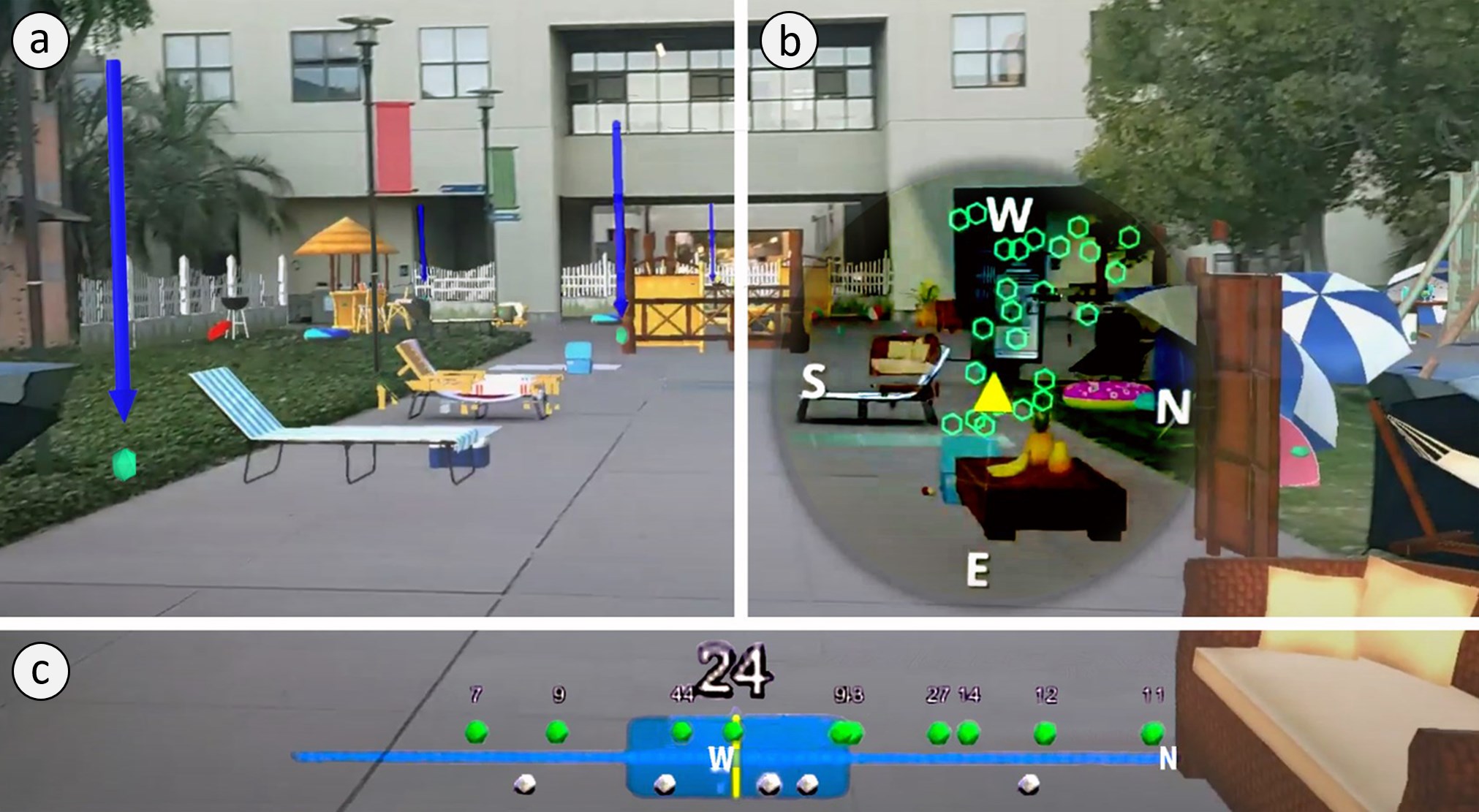}
\caption{A view of the experiment environment through the Hololens-2 headset, with examples of each of the three navigation aids (in practice, only one aid is visible at a time). (a) In-world arrows pointing to gems in the environment. (b) On-screen radar indicating positions of gems (green hexagons) relative to the participant (yellow arrow). (c) On-screen compass indicating relative positions of gems positioned in front of the participant (green gems with travel distance in meters displayed above each gem) and behind the participant (white gems). The blue rectangle indicates the forwards field of view and analogous zone behind.}
\label{fig:navigationAids}
\end{figure}

The study involved 24 participants who engaged in a visual search task where they needed to locate and classify virtual gems hidden among physical and virtual objects. The navigation aids tested included in-world vertical arrows pointing to gems, an on-screen radar, and an on-screen compass. The primary objectives were to assess how these aids influenced user performance, movement, eye-gaze behavior, and awareness of the physical environment. Participants were required to search for 24 virtual gems distributed throughout an outdoor area while wearing the HoloLens 2 (Figure \ref{fig:navigationAids}). The gems were classified based on orientation (vertical or horizontal) and texture (rough or smooth), necessitating a close proximity to each gem. To increase cognitive load and simulate real-world distractions, participants simultaneously performed an audio response task, where they listened for target words in a sequence and responded accordingly.

\begin{figure}[t]
\centering
\includegraphics[width= 1.0\textwidth]{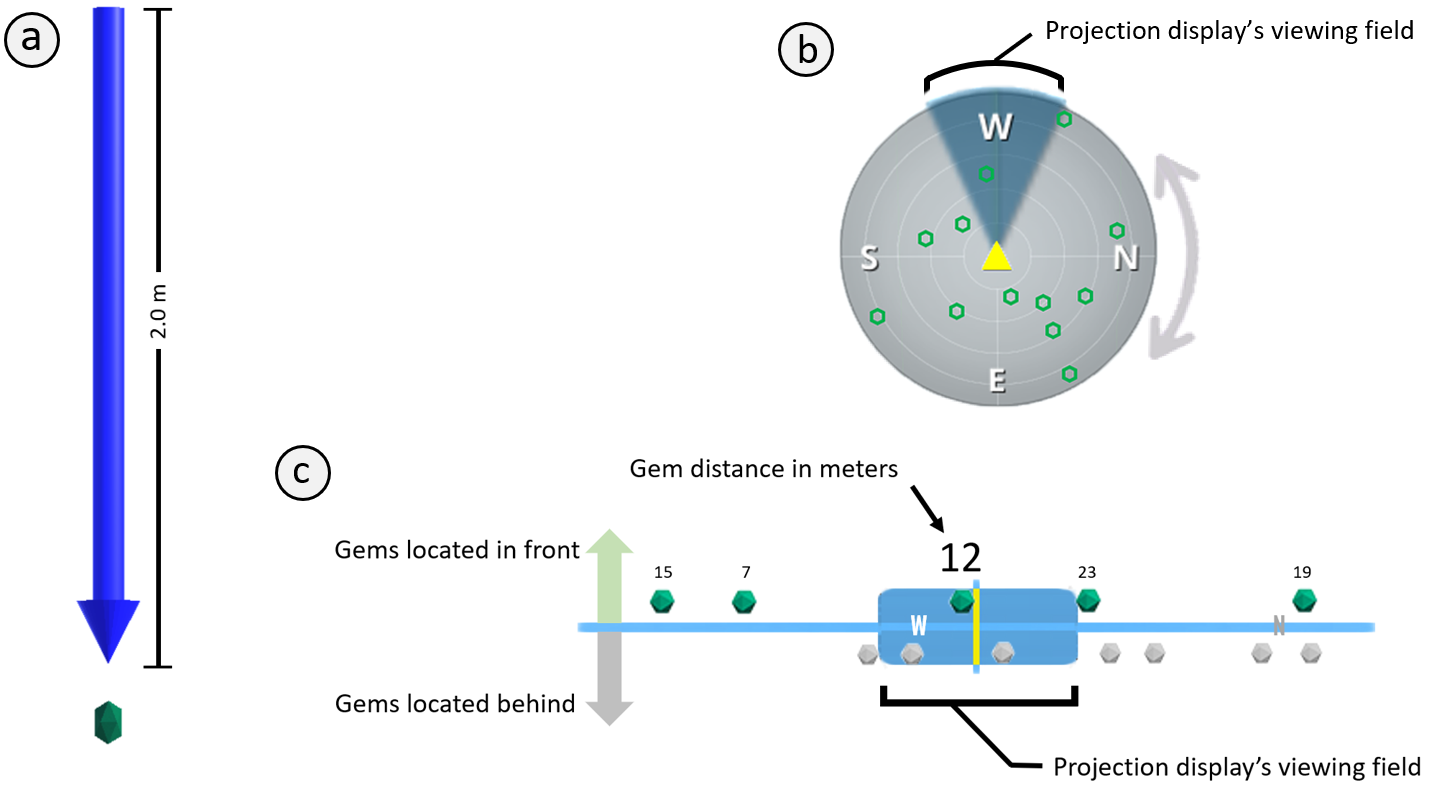}
\caption{Visuals of the three navigation aids the participants saw. Images not to scale. (a) In-world arrows, vertically above each gem. (b) On-screen radar, with a forward-up design and gems indicated by hollow green hexagons. The yellow triangle in the center indicates the position of the user in the space, and the blue cone represents the field of view. (c) On-screen horizontal compass, green hexagons above the line indicate gems in front of the user (with distance to each gem above the hexagon) and white hexagons below the line show gems behind the user.}
\label{fig:3nav}
\end{figure}

All three navigation aids (Figure \ref{fig:3nav}) enhanced search performance compared to the no-aid control condition. Participants completed the search tasks faster and with less head rotation when using the aids. Among the aids, in-world arrows were particularly effective in the initial phases of the search, resulting in quicker task completion and higher walking speeds compared to the on-screen aids. However, this advantage diminished during the later stages of the task when fewer gems remained to be found. Measuring cognitive load during the trials showed that the presence of navigation aids, especially the compass, increased cognitive load, as evidenced by poorer performance on the secondary audio task. This suggests that the compass introduced higher cognitive demands, potentially distracting participants from the auditory task.

Object recall and awareness results showed that participants exhibited lower recall accuracy for physical objects compared to virtual objects, indicating a significant shift in attention toward virtual elements. This was further supported by eye-tracking data, which revealed that participants focused more on the virtual field of view when using the head-stabilized aids (radar and compass), thus paying less attention to the physical environment. Users also strongly preferred the in-world arrows over the on-screen radar and compass. This preference aligned with the quantitative performance benefits observed with the arrows in the early stages of the search task.

Overall, the study provided valuable insights into the potential and limitations of AR navigation aids in enhancing user interaction with both virtual and physical environments. While navigation aids can significantly improve search efficiency, they may also lead to decreased awareness of the physical environment and increased cognitive load, particularly if the aids are complex or distracting. The results suggest that in-world navigation aids, despite their technical challenges in outdoor environments, might be more beneficial for tasks requiring spatial awareness and user engagement with the physical space.  Further, it underscored the importance of balancing virtual information with user attention to the physical world to optimize AR application design for real-world settings.
\end{section}

\begin{section}{AR Outdoor study 3: Climate Change Awareness} 

After completing two outdoor studies, I taught a summer session course titled 'Research in STEM: Human-Computer Interaction with Augmented Reality' as the instructor of record, supported by two teaching assistants: Stejara Dinulescu, a fellow MAT PhD student, and S. Shailja, an ECE PhD student. Following the course, I continued mentoring one of the students, Alex Wang, who was eager to further develop the AR application we had worked on in class. Together, we developed a Climate Change Awareness project that demonstrated the effects of climate change on familiar environments, such as the school campus. This system delivered realistic simulations of natural disasters on campus, allowing participants to see and experience potential impacts firsthand. By simulating natural disasters at a realistic scale within the physical environment, our goal was to create a powerful educational tool that enhanced students' understanding and awareness of climate change~\cite{wang2022exploring}.

The project, titled ``Exploring Immersive Mixed Reality Simulations and Its Impact on Climate Change Awareness,'' demonstrates how AR technology can be utilized to enhance public understanding and awareness of climate change. The research focuses on creating immersive experiences that effectively communicate the consequences of climate change, aiming to motivate individuals to adopt more environmentally friendly practices. This study highlights AR's potential as a powerful tool for climate change communication, finding that immersive simulations help users better grasp the urgency and complexity of climate change issues, ultimately encouraging changes in behavior and perception. The phrase ``seeing is believing'' captures this concept well, as AR uniquely offers the ability to simulate real-life disasters in the very places they could occur.

Despite the overwhelming evidence supporting climate change, there remains significant debate about its reality. Our goal was to provide a firsthand experience of how climate change can unfold, which is uniquely possible in an AR environment. By creating a digital twin, we could simulate scenarios such as how water would rise during a tsunami or flood, offering a tangible demonstration of potential outcomes. We developed the AR system for HoloLens 2 using a digital twin model of Kirby Crossing on the UCSB campus. This allowed us to craft and design simulations of floods and fires in that area, including scenarios where parts of buildings are damaged. We closely referenced real videos, reports, and the news to accurately simulate these events in Unity, utilizing a digital twin layer with UV textures of the environment to ensure realism. This setup enabled us to closely monitor how users experience these scenarios through the headset.

Two main scenarios were simulated: a flooding scenario depicting the potential impacts of extreme precipitation events and a wildfire scenario illustrating the increasing frequency and intensity of wildfires due to climate change. This is particularly relevant given California's recent fire seasons in 2022. An image of the simulation can be found in Figure \ref{fig:climate}.
The pilot study involved nine participants, most of whom had no prior experience with AR, although two had experience with VR. The study employed a counterbalanced design to compare the use of AR headsets with traditional methods, including paper and monitor simulations of natural disasters viewed on a 24-inch monitor. Data collection included pre- and post-experience questionnaires to assess participants' stress levels, inclination to adopt sustainable practices, and overall comfort in navigating the AR environment. Key metrics evaluated were user immersion, emotional responses, and any adverse effects such as nausea or discomfort during the AR experience, with exit interviews providing additional insights.

\begin{figure}[t]
\centering
\includegraphics[width= 1.0\textwidth]{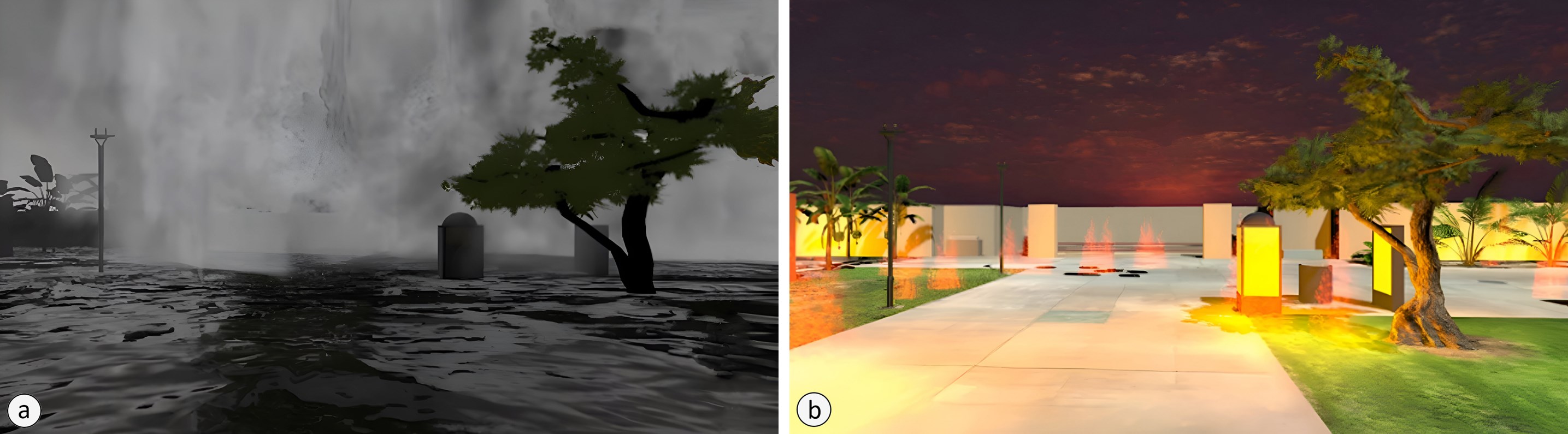}
\caption{Augmented reality simulation illustrating climate change impacts through natural disaster scenarios. The simulation includes: (a) a rain pooling and water-filling flood scenario, and (b) a fire scenario demonstrating debris breakdown.}
\label{fig:climate}
\end{figure}

The study revealed increased awareness and engagement among participants. The immersive nature of AR enabled users to visualize the potential impacts of climate change in a way that traditional media, such as television and newspapers, could not achieve. There was a positive correlation between AR use and the inclination to adopt environmentally friendly practices, with participants expressing a strong motivation to change their behaviors after the AR experience. The mixed reality experience elicited strong emotional responses, which was crucial for motivating behavioral change. Participants expressed feelings of concern and urgency regarding climate change after the simulations. The study also noted that AR provided a deeper level of immersion compared to other forms of dissemination, enhancing the overall impact of the message. Importantly, participants reported no significant adverse effects such as headaches, eye strain, or dizziness, indicating that the AR system is comparable in stability to traditional monitor experiences, especially in short sessions like ours, which lasted less than 10 minutes for both AR climate scenarios.

With a limited number of users, it was difficult to make an absolute decision that the AR experience was preferred, especially as it was expected that AR still had a novelty factor compared to a seated experience. While it cannot be directly compared, participants expressed the AR experience as better, because they were able to envision the incidents as happening in their present situations. More research must be done, but participants described the experience as more realistic and unlike anything they had encountered before. Seeing natural disasters unfold in front of them heightened their awareness and concern about the current reality of climate change, which was our objective: to show the unexaggerated potential outcomes of these events. The AR experience captured the scale and disastrous reality in a way that is challenging to achieve on a traditional screen.

Ultimately, the project highlights the transformative potential of augmented reality in raising awareness about climate change. By providing immersive and interactive experiences, AR can help individuals better understand the consequences of climate change and inspire them to take action. The positive feedback from participants underscores the importance of innovative communication strategies in addressing one of the most pressing challenges of our time. The implications of the findings suggest that AR technology could be a valuable tool for educational institutions, environmental organizations, and policymakers to communicate climate change issues more effectively. Future research directions include exploring more realistic and localized scenarios, incorporating real-time environmental scanning to create dynamic simulations, using geographical mapping to provide context-specific experiences, and enhancing the auditory experience with spatialized audio to further immerse users in the simulations.

\end{section}

\begin{section}{Conclusion} 

The three large, outdoor AR projects and prototypes I worked on in 2021 became the foundation of my AR theater system. While the first outdoor study prompted me to consider how an AR stage can be crafted and what level of AR coverage is necessary for users to feel immersed, the second study deepened my understanding of how navigation impacts user experience. The third study inspired me to consider innovative communication strategies and to take seriously the impact that our interactive narrative experience in AR can bring. They all encouraged me to explore large-scale experiences, experiment with fully augmented areas, and test more experimental navigational systems for interactive narratives. 

This scientific grounding gave me the confidence to be bolder and more experimental in my later artistic works and approaches. Ultimately, I wanted users experiencing my AR theater system to feel immersed and experience a sense of discovery and ownership in the narrative. Applying the knowledge from the previous projects, my AR theater system sought to develop more natural ways of interaction where information is not directly presented, where users discover the storyline with the guidance of natural elements and their own actions. 
\end{section}

\chapter{Interactivity and Perception in Augmented Reality}

This chapter explores two projects: Spatial Orchestra~\cite{kim2024orchestra} and the Reality Distortion Room (RDR)~\cite{kim2023reality} prototypes. Each project explores unique approaches to user interaction within augmented reality. Spatial Orchestra expands user expression by enabling music creation through body movement, while the Reality Distortion Room focuses on inducing natural user reactions and patterns of movement in a predictable way, through augmented visual cues projected in a Spatial Augmented Reality (SAR) setting. Both projects enrich user interactivity in augmented reality environments by effectively leveraging visual perception.

Spatial Orchestra allows users to interact with color-coded bubble objects in the AR environment to create musical notes, translating artistic expression through dance and enhancing the user's ability to express themselves in a mixed reality setting. In contrast, Reality Distortion Room demonstrates how external stage deformation and animation can predictably influence natural user locomotion. By controlling and deforming the augmented environment within an SAR system, it produces consistent and predictable user movement patterns. I have developed five distortion effects that can induce specific movements, such as directing users to a certain area or guiding them to the room's center.

While Spatial Orchestra empowers users to control their environment for creative expression, Reality Distortion Room demonstrates how the stage itself can predictably influence user behavior. Together, these prototypes highlight the vast potential for incorporating diverse elements into interactive narrative experiences using AR stages that are personalized to the space. These works underscore the need for further research, particularly in human-computer interaction within mixed reality, and the possibilities these interactive AR theater platforms hold. These AR stages not only serve as venues for user expression but also as environments that can significantly influence and enhance participants' experiences. 

Before discussing the two prototypes, I will briefly introduce a project I worked on during my first year in the PhD program (2018–2019) through the Magic Leap Independent Creator Program alongside fellow PhD student, Mengyu Chen. Project ATLAS, the title of our AR application, reimagined the Magic Leap experience by transforming a room into a submarine. This initial project inspired me to delve deeper into the possibilities within human-computer interaction and to push the boundaries of creativity and interactivity in augmented reality, expanding it as a medium for human expression and engagement. Through these three projects, I introduce the vast possibilities of AR and explore how to push its boundaries. 

\begin{section}{Early Work: ATLAS}

In 2018, a fellow PhD student, Mengyu Chen, and I enrolled in the Magic Leap Independent Creator Program. Together, we produced and developed a Magic Leap application, ATLAS, with technical assistance and device donations from Magic Leap. We showcased our ATLAS prototype at the UCSB Media Arts and Technology (MAT) End of Year Show on June 7, 2019. This event, named Media Art, Design, and Engineering (M.A.D.E.), highlighted graduate student projects that the MAT Program produces each year. It is a highly anticipated public event that concludes the academic year with exhibitions and technical demonstrations. This experience introduced me to the vast possibilities of augmented reality, prompting an exploration of its boundaries. It was also my first AR application development experience. 

ATLAS is an immersive augmented reality application developed using the Magic Leap Unity SDK on the Magic Leap One platform. It transforms users' living spaces into a virtual underwater world, allowing them to explore diverse habitats filled with aquatic life. We spent considerable time modeling this underwater world to be both interesting and calming, encouraging exploration and interaction while allowing users to experience relaxation, curiosity, and admiration simultaneously. In the experience, users embark on a collaborative adventure with other Magic Leap users to search for the mythical lost city of Atlas. Developed on version 0.97 of Lumin OS, the application supports multiplayer experiences for up to five users in the same room, facilitated by the Professional Developer package for multiplayer synchronization.

In the application, any room can become a submarine, where users can peer into ocean depths and navigate unknown territories. A 3D miniature navigation map mirrors the underwater world outside the submarine, providing spatial awareness and aiding exploration. The digital twin of the room is accurately created through Spatial Mapping, part of the Magic Leap ecosystem known as World Models. These include dense mesh data and planes, identifying walls, floors, and optimal surfaces for placing content. These elements enable occlusion and collision consistent with the user's environment, as well as the placement of virtual windows to view the ocean world and a digital cockpit for submarine control.
\begin{figure}[t]
\centering
\includegraphics[width= 1.0\textwidth]{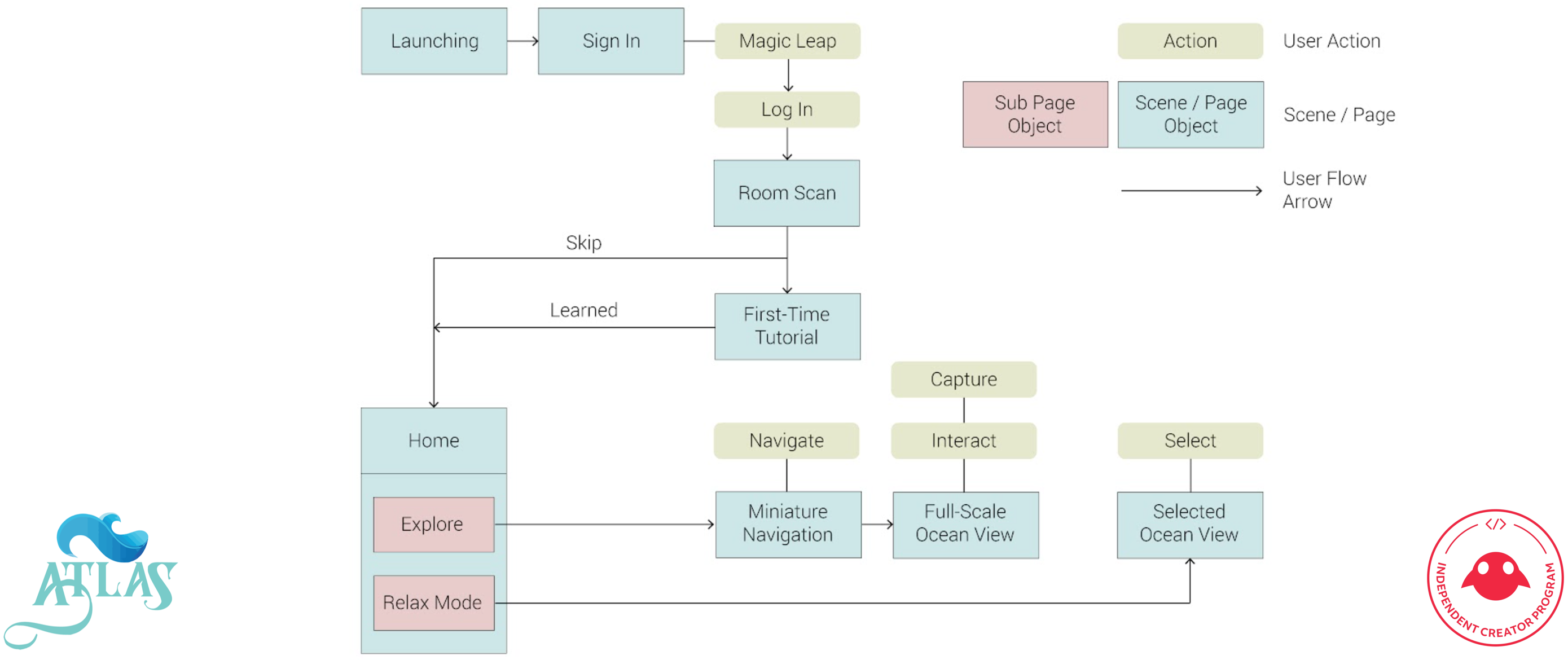}
\caption{The user experience pipeline for ATLAS demonstrating the flexibility of our gameplay modes. Users have the option to choose between ``Explore'' which offers game-like exploration, and ``Relax Mode,'' which allows them to simply enjoy the underwater scenery.}
\label{fig:ATLAS_chart}
\end{figure}

A unique feature of ATLAS is its manipulation of scale, allowing users to alter their perspective and engage with virtual content. Doing so leverages existing spatial relationships. Users can capture exotic marine life, displayed as miniatures within the submarine, contributing to their personal collection of discoveries. For this reason, we selected to begin the application with a tutorial, ``Follow the Starfish,'' to guide users through submarine operations. This serves as the first entry in the user's discovery log.

ATLAS fuses imaginative digital environments with natural ones, bridging technical considerations and artistic practice. By leveraging procedural generation, it creates near-infinite worlds that encourage exploration and discovery in both educational and entertaining ways. The core experience emphasizes social interaction and group problem-solving, as highlighted in the user pipeline chart (Figure~\ref{fig:ATLAS_chart}). Additionally, users can spend time in the submarine, observing the beautiful underwater world. The beauty of ATLAS lies in its flexibility; users can choose between ``Explore Mode'' for game-like exploration or ``Relax Mode'' to simply enjoy the underwater scenery from their living room submarine, interacting with sea creatures. The application also supports automatic exploration, allowing users to chat and relax as they watch nature unfold from within the submarine environment.

ATLAS exemplifies user interaction with virtual spaces by seamlessly blending imaginative digital worlds with real-world settings. This blending is achieved through dynamic elements, such as a whale gracefully swimming into view, and then naturally exiting without disrupting gameplay. This interaction conveys the sense of being underwater, even when users are in a static room where windows or walls simulate an ocean view, creating the illusion that the room itself is moving.

\end{section}
\begin{figure}[t]
\centering
\includegraphics[width= 1.0\textwidth]{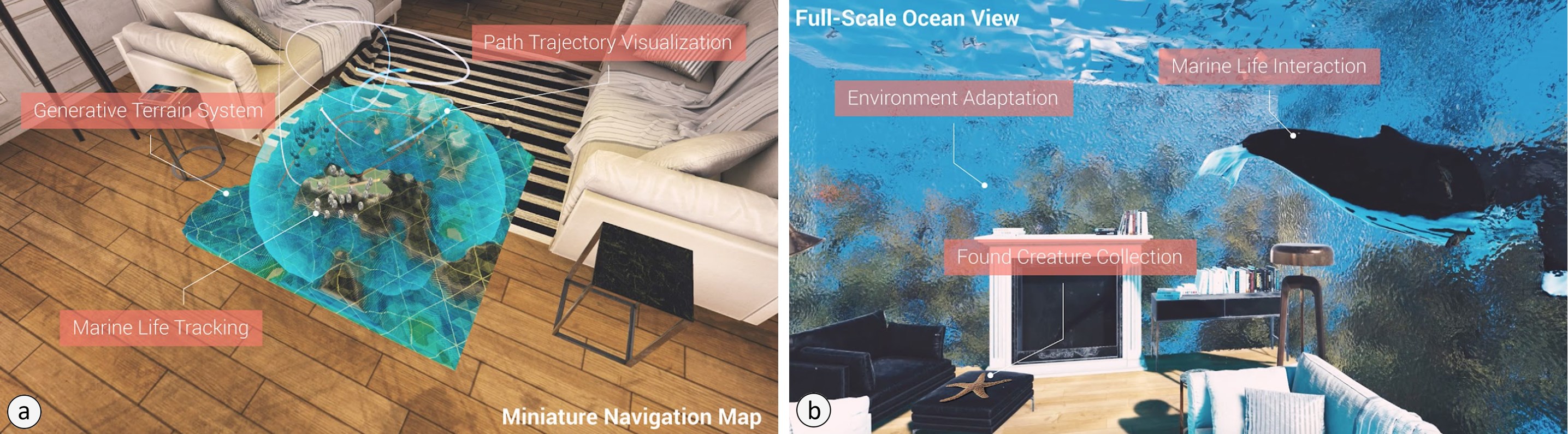}
\caption{(a) The navigational information system features a miniature navigation map prominently displayed in the center of the room. This setup allows users to easily see where their submarine is positioned within the virtual undersea world. (b) The walls transform into windows offering views of the undersea world, while physical elements such as the sofa and table remain visible. These elements are essential to maintaining a familiar living space for the user.}
\label{fig:ATLAS_image}
\end{figure}

Mengyu Chen and I explored various design choices for how much of the living room's window should be revealed to connect the audience with the outside (virtual) world. With the potential for entire walls to be made of glass, we aimed to design a experience that captured the transformation of a living room into a submarine. We experimented with different window designs and sizes to determine what elements were necessary to maintain a familiar living space for the user as illustrated in the Figure~\ref{fig:ATLAS_image}. These artistic choices leveraged AR technology to enhance the overall aesthetics and user experience, which offers possibilities not available in VR.

Developing in this environment demonstrated to me the next phase of mixed reality experiences that are interactive, personalized, and contain multi-narrative experiences. By seeking to redefine user engagement in virtual spaces by harnessing their creativity and imagination, the work done with ATLAS led to further exploration in augmented reality environments, such as the Spatial Orchestra and Reality Distortion Room. These concepts would eventually ground the concepts of Dynamic Theater, especially as they relate to unfolding experiences in multiple ways.

\begin{section}{Interactivity Project 1: Spatial Orchestra}
Spatial Orchestra was showcased at ``latent•ville,'' the MAT 2023 End of Year Show (EoYS) on Thursday, June 9th, 2023, at the 2024 EoYS during the soft AI+M event on June 5th, 2024, and once again at the ACM 2024 Conference on Human Factors in Computing Systems (CHI) Interactivity Session in Hawaii. Across these three demonstrations, Spatial Orchestra featured different variations, including bubble sizes, rendering materials, instruments, and AR stage designs. The version presented here, which I showcased at CHI 2024 in Hawaii, is the latest iteration.

Spatial Orchestra~\cite{kim2024orchestra} illustrates the ease of playing musical instruments using basic inputs such as natural locomotion, making it accessible to a wide range of users. Unlike traditional musical instruments, this system allows individuals of all skill levels to effortlessly create music by interacting with virtual bubbles. In this AR experience, users engage with ever-shifting sound bubbles by stepping into color-coded areas within a designated space, using a standalone AR headset. Each bubble corresponds to a cello note, emitting sound from its center, which allows users to experience spatial audio and express themselves musically, effectively transforming them into musicians (Figure~\ref{fig:so}).

This interactive element enables users to explore the intersection of spatial awareness and musical rhythm, extending to bodily expression through playful movements and dance-like gestures within the bubble-filled environment. This unique experience highlights the intricate relationship between spatial awareness and the art of musical performance.

\begin{figure}[t]
\centering
\includegraphics[width= 1.0\textwidth]{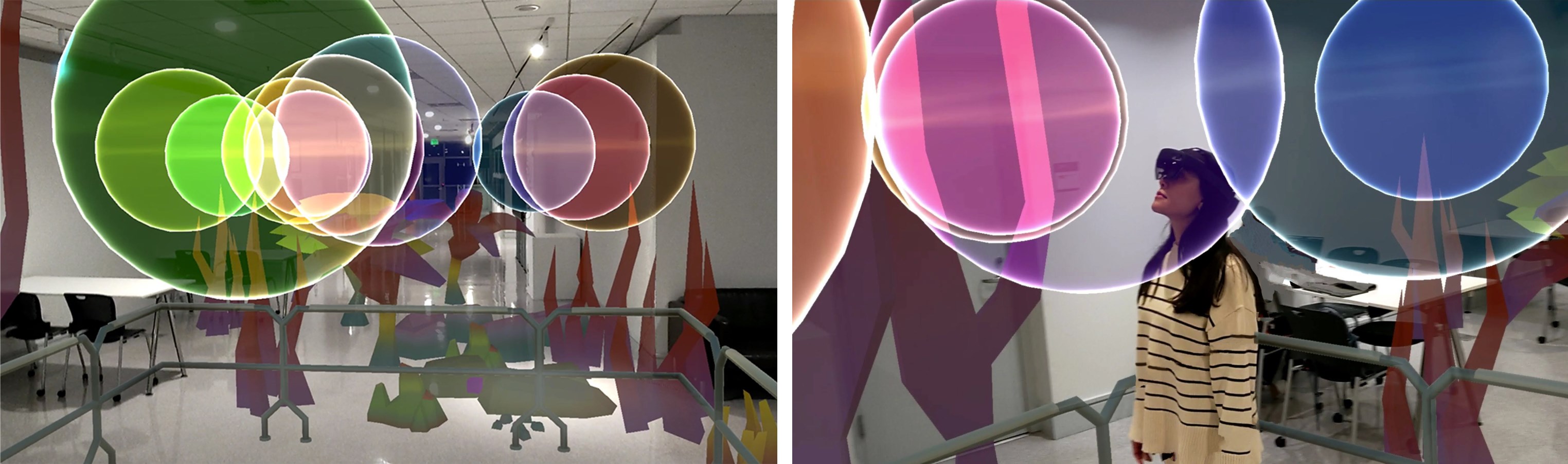}
\caption{Spatial Orchestra. Left: From the point of view of the user experiencing Spatial Orchestra, where they are in the augmented stage, freely moving around and interacting with ten virtual bubbles that are ever shifting and simulate cello chords. Right: A user immersed in the augmented environment. }
\label{fig:so}
\end{figure}

\subsection{Introduction}
As not everyone has the opportunity to acquire the skill of playing a musical instrument, the art of musical expression remains elusive. Within this context, I identified a novel application that sought to answer the question, ``What if individuals could create music using actions that are inherently familiar and well-practiced, utilizing the muscles and movements they employ in their daily lives?'' This led to the conceptualization of a musical instrument designed to be played entirely through locomotion and spatial awareness, engaging users in an intuitive interaction with gently drifting bubbles. Within this defined area, color-coded bubbles, each emitting a distinct cello note, float at the user's head level. Users can freely move around the play-space, engaging with these bubbles to compose music by entering them. Users are encouraged to navigate to areas where multiple bubbles are present, as they can create harmonies of mixed cello notes.

Spatial Orchestra utilizes research on natural locomotion within mixed reality in open spaces~\cite{cheok2002interactive, kim2023dynamic, kim2022investigating, kim2023reality}, focusing on spatial awareness facilitated by spatial audio~\cite{RiddershomBargum2023, Earnormous, DuoRhythmo, AuditoryImmersion} in recent years. My system leverages the spatial awareness that people already employ in their daily routines, eliminating the need for users to acquire traditional music notation or score-reading skills. Instead, users can engage with the floating bubbles in a manner analogous to real-life interactions. By merging spatial awareness with the act of navigating their environment, individuals can become adept at playing an AR musical instrument right from the start, offering a unique opportunity for everyone to become proficient music players. Moreover, users can further refine their skills with practice and experience. This means users can not only produce incidental sound but also develop their virtuosity to play sophisticated music. As proven by recent work, there are physical input-based and predictive approaches to musical embodiment~\cite{FormFollowsSound, maes2014action, maes2016sensorimotor, DuoRhythmo}. Spatial Orchestra fully engages this concept and utilizes musical embodiment and the role of prediction, designing interactions through user input and action-based effects on music perception.

\subsection{Related Work}
Advances in digital mediums facilitate new ways of musical expression. For example, the electronic medium gave freedom from the acoustic and physical design of the instrument, which allowed the development of wireless embodied interaction. While conventional instruments require years of training and mastery, this circumstance encouraged researchers to propose unique and easy ways to expand the users' creativity and think outside the box.

\subsubsection{Digital Instruments}
Gehlhaar's work~\cite{soundspace91} uses an ultrasonic echolocation system to specify the locomotion of participants within the space, and the computer synthesizes the sound using the information. Morreale's installation \cite{Morreale14} invites the audience to compose music by moving in the space using the distance between the participants. Lee's work \cite{lee21} utilizes machine learning-based gestural recognition using smartphones' accelerometers and gyroscopes to interact with simulated physics within an immersive projection space. In Junior's work~\cite{Junior21}, non-experts create coherent music through graphical elements in a virtual environment.

This circumstance has allowed a proactive music experience for conventional music consumers through more familiar mediations, including new instruments and games, while engaging in compositional techniques~\cite{winter2005interactive}. Rasamimanana's work \cite{Rasamimanana12} mapped the sports ball's movement and status into rhythmic sounds. These types of musical games utilize the concept of the trigger and playback-based composition methods.

\subsubsection{Musical Experience in Mixed Reality}
The advance of XR technology has revolutionized interaction by blending digital and real surroundings. One intriguing area of research involves the experimentation of Spatial Audio in mixed reality.

Schlagowski et al. (2023) explored the fusion of Spatial Audio with hand motion-controlled interfaces in VR, enabling users to collaboratively mix and play music in a mixed reality environment~\cite{schlagowski2023wish}. Another noteworthy project, DuoRhythmo by Riddershom and Bargum (2023), introduced a collaborative and accessible digital musical interface in mixed reality, focusing on designing a user-friendly experience~\cite{DuoRhythmo, RiddershomBargum2023}.

In education and entertainment, there have been efforts to leverage VR and Spatial Audio to enhance the way humans perceive and interact with games~\cite{Earnormous}. Additionally, projects like those enhancing auditory immersion in interactive VR environments demonstrate the diverse applications of Spatial Audio in immersive experiences~\cite{AuditoryImmersion}.

Turchet's research in 2021 highlighted a significant surge in the historical distribution of musical XR research over the past five years. The study aimed to define the musical XR field by analyzing 260 research characteristics~\cite{Luca21}. Bilow (2022) proposed an AR experience allowing participants to explore audiovisual elements through movements and to interact, using hand gestures~\cite{Bilbow22}. Furthermore, Wang (2022) conducted an empirical study comparing three audiovisual interface prototypes for head-mounted AR environments~\cite{wang22}.

These advancements collectively underscore the evolving landscape of XR technology, particularly in the integration of spatial audio for diverse applications ranging from collaborative music creation to educational experiences.

\subsection{Design of Spatial Orchestra}
To guide users to stay safety within the virtual environment, a virtual fence was employed within a 3.3 m by 3.3 m space. The fence was purposely stylized to differentiate it from other augmented objects in space. Ten bubbles, each measuring 80 cm in diameter, travel at a constant altitude level with the user's height, one meter every five seconds. Bubbles travel at random and bounce around the area while maintaining their altitude. Such behavior establishes a stochastic model like molecules in the closed space.

\begin{figure}[t]
\centering
\includegraphics[width= 1.0\textwidth]{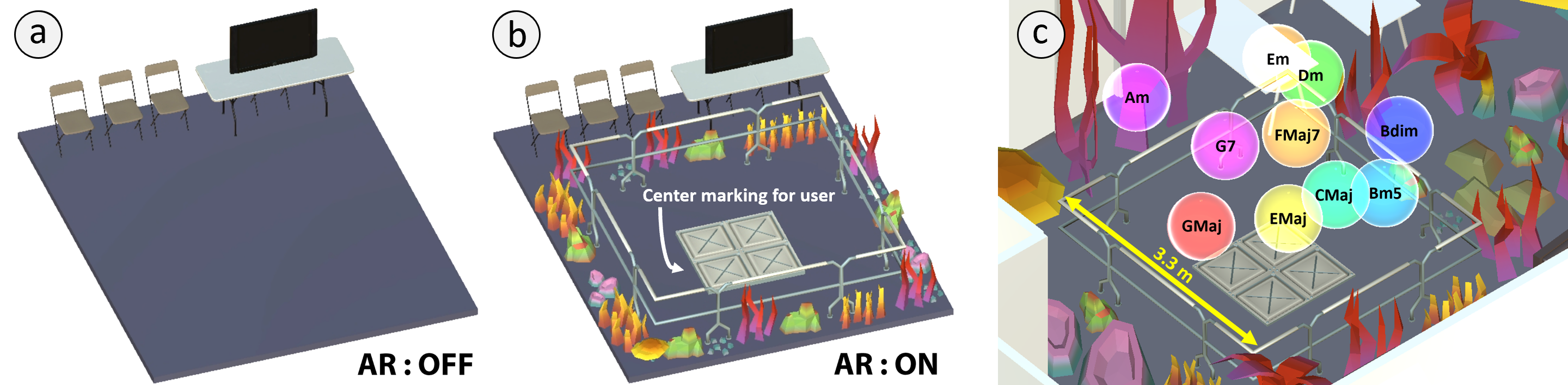}
\caption{An aerial perspective of the stage arrangement that the user explores. (a) The physical space used for the Spatial Orchestra. (b) The fencing and virtual stage are all augmented. (c) There are ten augmented virtual bubbles set up in the area. Each bubble emits unique cello notes that are color coded to enable user interaction.}
\label{fig:bubbles}
\end{figure}

Each bubble contains synthesized cello sounds that could be made from a single bow stroke. Sound omitted from the bubble's center can only be heard when the user's head is in the bubble. The colors of the ten bubbles represent the fundamental chords comprising: [EMaj, Em, FMaj7, GMaj, G7, Am, Bdim, Bm5, Cmaj, Dm].
These are deliberately selected to guide participants in generating a musical structure that can fold and unfold over time through variant chord progressions. While the bubbles travel like molecules in space, the user’s vibrant motion will increase the probability of triggering the cello strokes. Through the integration of melodic framework with their spatial and visual perception, users can craft imaginative multimodal experiences. (Figure~\ref{fig:bubbles}c).

\subsection{Experiencing Spatial Orchestra}

Users were instructed to stay within the virtual fence and to move slowly around the play-space. They were further informed that each bubble encountered would contain cello sounds. Color-coded bubbles matched specific notes encouraging users to interact with the bubbles individually or in clusters to compose music (Figure~\ref{fig:bubbles}c).

The size of the space, bubble dimensions, number of bubbles, and bubble's surface shaders were meticulously adjusted to ensure an optimal user experience. The goal was to create an environment full enough for creative music composition but not to the extent that users felt a lack of control.

\subsubsection{User Feedback}

I showcased the Spatial Orchestra during a university public event, allowing interested participants to sign up and experience it. Over the course of two days, 60 users had the opportunity to engage with the Spatial Orchestra.

\begin{figure}[t]
\centering
\includegraphics[width= 1.0\textwidth]{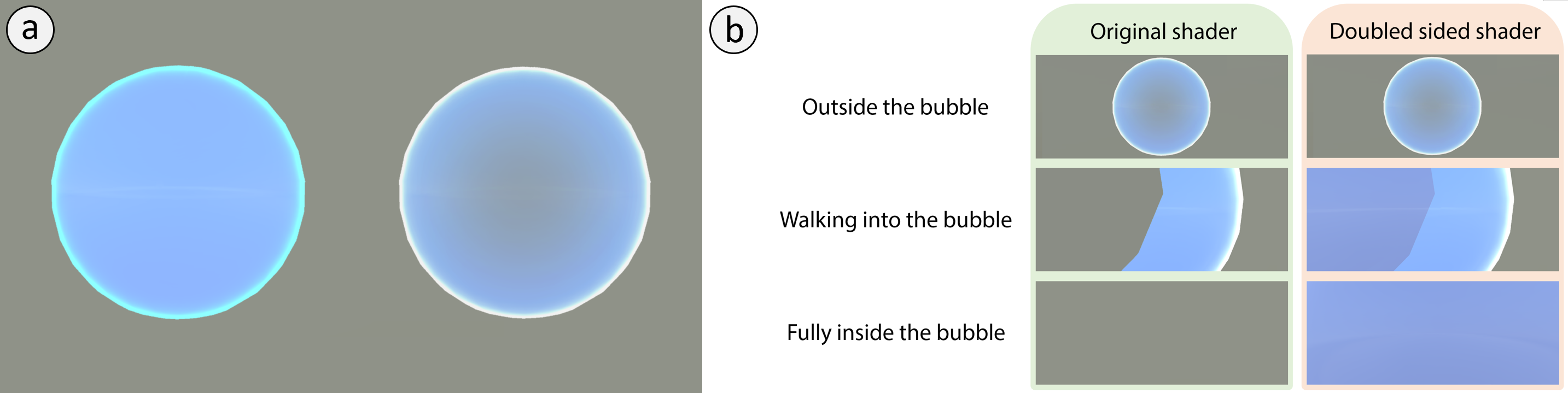}
\caption{(a) The bubble rendering model underwent improvements based on user feedback. Left: The original bubble featured a color highlight at the edge, representing the bubble's assigned color. Right: A new shader incorporated a white highlight for improved visibility of the bubble's boundaries. Adjustments were also made to the transparency and illumination levels, addressing user concerns about clarity when multiple bubbles are stacked and colors are mixed. (b) I addressed a critical issue where users struggled to identify if they were inside the bubble in the original shader. To solve this, I added a visual cue. Now, when users enter the bubble, the scene adopts a translucent color within the bubble's view. This was achieved through a double-sided shader, rendering both sides of the mesh.}
\label{fig:boundary}
\end{figure}

While the majority of feedback was positive, some users mentioned that they could not easily discern whether they were inside the virtual bubble or not, as the visual cue disappeared once they entered the bubble. To address this, I modified the rendering of the bubble texture to be visible from both inside and outside, providing users with a translucent color cue to indicate their position within the bubble (Figure~\ref{fig:boundary}b).

In response to user feedback, I also adjusted the transparency levels of the bubbles to enhance visibility, especially when multiple bubbles overlapped. This modification allowed users to better perceive individual bubble colors and easily identify and mix multiple cello notes in the same area. Experienced musicians told me that if they could see clearly when bubbles were stacked in a crowded area, they would be able to play music more effectively (Figure~\ref{fig:boundary}a).

Based on participant comments, I observed that many reported favorable experiences with high levels of engagement and felt immediately able to produce music while interacting with bubbles. Some participants even spent more than 20 minutes exploring and mastering the rhythm of the composition, exceeding my expectations for user engagement.

\begin{figure}[t]
\centering
\includegraphics[width= 1.0\textwidth]{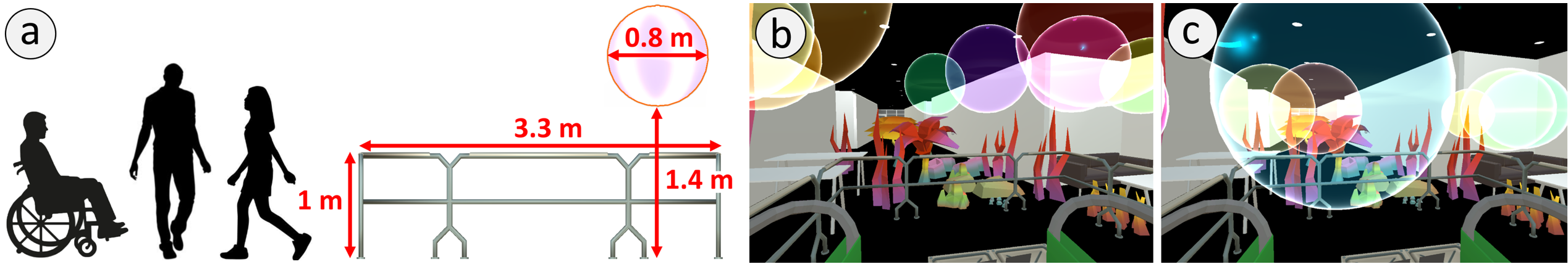}
\caption{(a) Compares the virtual bubbles' size to the virtual fence's height. Based on individual viewpoint heights, it may be challenging for some people to interact with bubbles directly. (b) View from the person in a wheelchair engaging with the bubbles. (c) I implemented an accessibility mode that allowed manual adjustment to the height of the bubbles. As you can see, the bubbles appear at the correct height when viewed from a wheelchair.}
\label{fig:accessibility}
\end{figure}

\subsubsection{Accessibility and Safety}

For users to understand their location in the play-space and proximity to the center, I included an augmented marker indicating the center. Colorful plants signified the boundary of the play-space and the proximity to the fence, adding a visual cue and safety buffer (Figure~\ref{fig:accessibility}).

Azure Spatial Anchors precisely aligned augmented objects during user motion~\cite{buck2022azure}. Multiple anchors ensured alignment throughout the experience; if one failed, others maintained the objects' locations. Only one anchor failed in a continuous five-hour test. The participant remained unaware, highlighting the system's reliability. Lastly, to accommodate wheelchair users and children, I also implemented Accessibility Mode, which allowed me to manually adjust the height of the bubbles that float.

\subsection{Conclusion}
In conclusion, Spatial Orchestra is a spatial musical instrument that utilizes natural locomotion. Playing this instrument is not only an immersive experience but also an expression of physical movement. Utilizing a stand-alone augmented reality headset, I showcased a musical instrument that leverages walking as a means of interaction. It is designed for anyone to learn quickly and play effortlessly, even without prior musical training.

During a university event, participants discovered various patterns and techniques to play music. They adjusted the rhythm by entering bubbles more or less frequently, showcasing their virtuosity through empirical user tests. Some users also found creative ways to produce specific sounds by combining bubbles or waiting before entering them.

Users enjoyed the distinctive experience of interacting with bubbles to create sounds. Despite the challenges of generating precise notes and rhythms, the simplicity of using natural inputs like walking offers users an accessible and enjoyable way to play music.

\end{section}

\begin{section}{Interactivity Project 2: Reality Distortion Room}
This project was completed in collaboration with Andy Wilson, a partner researcher at Microsoft Research, during my 12-week research internship at Microsoft Research. I was part of the Extended Perception, Interaction, and Cognition (EPIC) Research Group, and I continued working on it upon returning to UCSB. The project was presented at the International Symposium on Mixed and Augmented Reality (ISMAR) in October 2023.

The Reality Distortion Room (RDR)~\cite{kim2023reality} is a proof-of-concept augmented reality system that uses projection mapping and unencumbered interaction with the Microsoft RoomAlive system to study how users respond to visual effects that appear to transform the physical room they are in. This study introduces five effects that alter the appearance of a physical room to subtly encourage user movement. Our experiment examines users’ reactions to different distortion and augmentation effects in a standard living room. These effects are projected as wall grids, furniture holograms, and small particles in the air, making the room seem elongated, wrapped, shifted, elevated, and enlarged. The study results provide valuable insights into how AR experiences can be implemented in confined spaces by offering an initial understanding of how users can be gently encouraged to move throughout a room.

A growing number of virtual experiences take the user’s physical environment into account, which leads to an expansion of potential and possibilities in immersive home entertainment. Many gamers consuming entertainment in their homes are increasingly turning to immersive experience technology such as virtual reality (VR) and augmented reality (AR) for an extended reality or presence platform experience~\cite{home2022schell, world2022oculus}.

Some VR work specifically addresses the question of supporting navigation in large VR environments while relying on real walking in smaller physical environments. Redirected walking offers natural locomotion with correct proprioceptive, kinesthetic, and vestibular stimulation, but it requires sizable actual tracking spaces~\cite{nilsson201815}. Interaction-based redirected walking uses techniques such as warping~\cite{dong2017smooth, williams2021arc} and sensory technologies~\cite{sun2018virtual} perceptual illusion~\cite{sra2018vmotion}, or space mapping~\cite{sun2016mapping, hartmann2019realitycheck}. These techniques can operate in a smaller space; however, the experience is regularly interrupted to correct the user's position when the user approaches the limit of the available walking space.

\begin{figure}[t]
\centering
\includegraphics[width=1.0\textwidth]{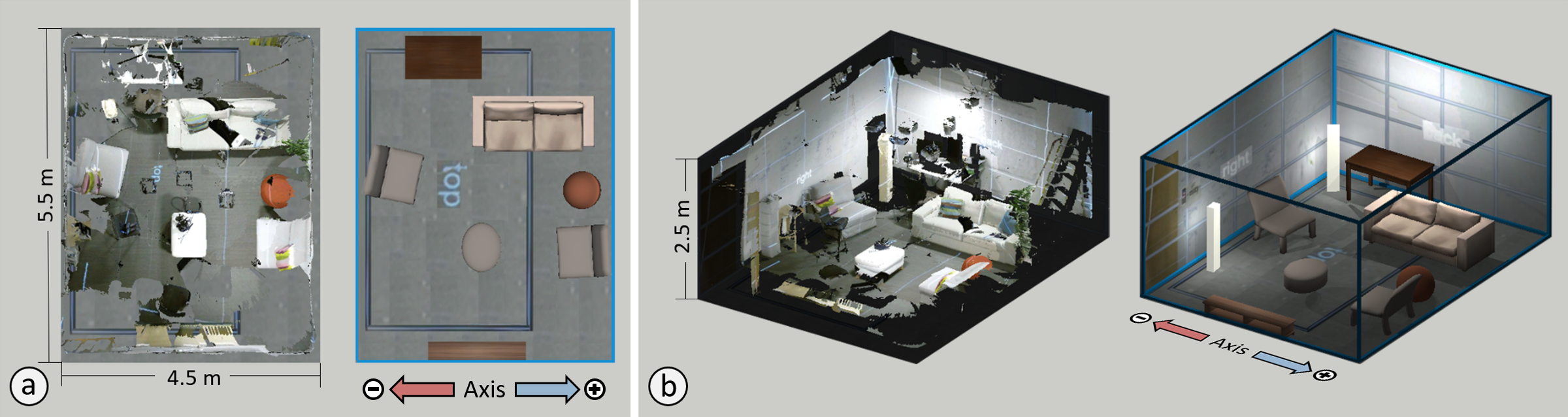}
\caption{(a) Top view of the 4.5 m x 5.5 m room with furniture where the study and distortion effects were conducted. Left: Floor layout as scanned by Kinect v2 sensor cameras.  Right: Digital Double 3D model of room. (b)	Left: Using SLAM (Simultaneous Localization and Mapping), a live 3D reconstruction of the room from a side angle. Right: Digital Double 3D model of the room from the same angle.}
\label{fig:pic3}
\end{figure}

VR routinely utilizes techniques such as room marking systems to assist users in navigating safely within the room when using VR. Room setup features introduced in SteamVR and Meta Quest SDK allow users to mark out surfaces of the physical layout of the home to better avoid collisions when the user is immersed in virtual reality~\cite{steam2022valve, sdk2022oculus}. These methods, such as collision bounds and Chaperone, cannot be directly applied to AR applications as the physical layout is present at all times in the platform experience. While redirected walking in mixed reality using VR headset and passive haptics~\cite{kohli2005combining, suma2013redirected} has been examined, our research marks a step towards redirected walking in visual AR. 

Motion parallax and perspective-correct rendering of computer graphics content allow augmented experiences that are different from the physical layout the viewer is in~\cite{gibson1979theory, bruder2012tuning}. In this work, the possibility of subtly manipulating a user’s natural locomotion via visual motion effects is additionally explored.

Reality Distortion Room presents five room distortion treatments in augmented reality that employ the user's visual perception and spatial understanding to subtly manipulate their position via natural locomotion. Using wireframe overlay effects, mapped walls and furniture generate an omnidirectional room-scale display that renders a 3D reconstruction and extension of the user's physical space. Out of the five distortion treatments, three treatments are designed to impact the user's directional motion pattern (movement along an axis, refer to Figure~\ref{fig:pic3}a for axis directions), whereas the other two treatments are designed to impact the user's motion to and from the room center (distance-to-center). 

\begin{figure}[t]
\centering
\includegraphics[width=1.0\textwidth]{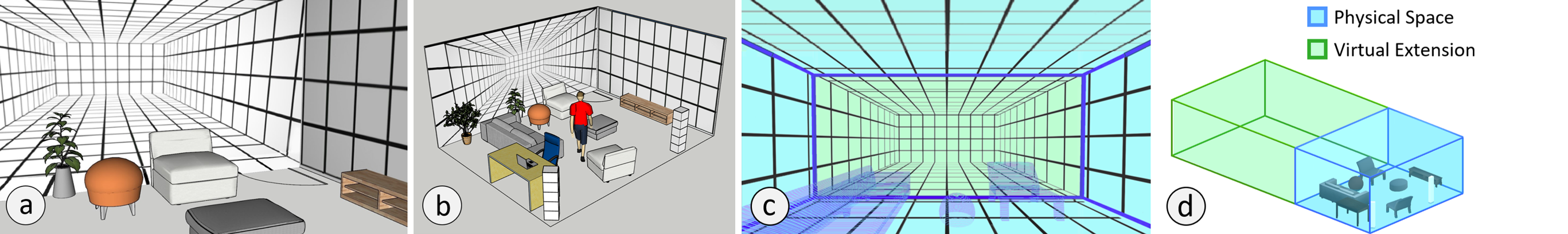}
\caption{(a) First-person point of view (POV) from the perspective of a user viewing a virtually extended space in the room. (b) Overhead view of the room environment where the user experiences the distortion effects. (c) First-person POV as seen via head-tracking and perspective correction, where the green area designates the space that is extended through projection using the Elongation Distortion. (d) Overhead view of the physical space of the room compared to its virtual extension during the Elongation Distortion.}
\label{fig:pic2}
\end{figure}

The study examines natural locomotion and visual perception. RealityShader was selected to simulate distortion treatments because it seamlessly blended projected and physical environments while the physical layout remained undistorted~\cite{youtube2021realityshader}. AR headsets with see-through waveguide displays, such as HoloLens 2 or Magic Leap 2, are equipped with relatively small field-of-view displays, which limits immersion. Video pass-through MR, such as the Varjo XR-3 or Meta Quest Pro, is becoming more commonplace but still has fidelity problems. Thus, I opted to run this study in spatial augmented reality, using spatial projection as seen in Figure~\ref{fig:rdr}a. RealityShader's projected augmented reality system~\cite{jones2014roomalive} captured and reacted to the user’s locomotion response. The RealityShader system allows the assessment of visual distortion effects while enabling users to walk in a fully surrounded projected space and untethered to a physical device as seen in Figure~\ref{fig:pic2}b. I performed several additional treatments, including augmenting the space with randomly floating particles and overlaying furniture outlines as seen in Figure~\ref{fig:rdr}.

The system encourages users to move in certain ways within the room. In applying the distortion treatments in an immersive AR experience, the study found that the Reality Distortion Room impacts users’ movement and reactions in specific ways, without explicitly telling them to do so. This study provides the first empirical evidence that developers can influence the motion of users by warping and modulating the AR space, which suggests a potential mechanism to be used in the eventual realization of redirected walking in an AR environment. There is great application in this system for helping users make modest positioning adjustments, especially when standard tools for adjustments, such as sound systems, are being used. Reality Distortion Room makes the following contributions: 

\begin{itemize} 
    \item Design and pilot testing of different distortion geometries for the system, the Reality Distortion Room.
    \item A user study (n=20) demonstrating the effectiveness of systematically influencing a user's natural locomotion. 
    \item  Analysis of study results, demonstrating user locomotion responses to generic room shape changes in a projected augmented environment. In particular, a directional effect and a center distancing effect are demonstrated as a reaction to the geometric deformation of the environment (distortion treatment) alone.

\end{itemize}

\subsection{Related Work}

The Reality Distortion Room platform offers a virtual world that alters and extends existing physical space. It is inspired by previously published scholarship that illustrates how to make use of physical layout to enhance the experience in a virtual world.

Fuchs envisioned the potential ways to utilize the combination of virtual and real worlds, implementing the CAVE system in the Office of the Future project~\cite{raskar1998office}. 
The spectrum of the AR-VR continuum was previously explored from the human-computer interaction aspect~\cite{milgram1994taxonomy, sayyad2020walking, benko2014dyadic, benko2015fovear, shapira2016reality}. Each method utilizes space in different ways when users are exploring virtual environments. For example, one such method detaches the user entirely from the physical space, making the movement frictionless and stationary using a treadmill~\cite{darken1997omnidirectional,iwata2005circulafloor,iwata2006powered,median2008virtusphere}. Other methods include adjusting sensitivity of input and output of the tracked movement between virtual and real to redirect the user~\cite{langbehn2018redirected, sun2016mapping} or providing a dynamic haptic environment where the physical layout adopts to a virtual one~\cite{iwata2005circulafloor, cheng2015turkdeck, suzuki2020roomshift}. In addition, previous work investigating ``vection'', or ``illusions of self-motion'' study how to convincingly simulate human locomotion in virtual environments without having to allow for full physical movement of the user~\cite{riecke2012selfmotion, riecke2013perceptual}. Reality Distortion Room, on the other hand, looks at the problem from the perspective of inducing the full physical movement of the user without explicit guidance systems or instructions. Therefore, the system was designed to induce movement patterns in user locomotion without informing them about the desired movement pattern, meaning participants were only instructed to move around freely in the environment. Inspired by these earlier studies, this work embraces the situated physical reality and our senses within it in a mixed reality experience. Through a proof of concept and user study, I demonstrate how to subtly manipulate user locomotion in AR space. 

\subsubsection{Experiencing Large Space}
Experiencing a larger environment than the one in which one exists can be both physically and conceptually disengaging for users~\cite{hartmann2019realitycheck}. To overcome these issues of disengagement, creative measures are taken to redirect attention and enable redirected walking~\cite{razzaque2001redirected, sra2018vmotion, sun2016mapping, peck2010improved} through procedurally generated virtual space from 3D reconstructed physical space~\cite{sra2018oasis, sayyad2020walking}. Event-based methods, such as dynamic saccadic redirection, have showcased a way to reduce the space required for immersive experiences~\cite{sun2018virtual}. Remixed Reality~\cite{lindlbauer2018remixed} explores the direct manipulation of the environment by offering users different and larger room layout options from their perspective. To further extend virtual space through augmentation and remote user presence, Room2Room~\cite{pejsa2016room2room} presented a prototype implementation for rearranging and extending the virtual layout in consideration of physical space through augmentation. Additionally, Reality Check~\cite{hartmann2019realitycheck} demonstrated how a virtual game environment can be combined with real-time 3D reconstruction of the room, resulting in a presence platform experience.

\subsubsection{Room-scale Interaction}
The physical constraints of a room can make it difficult to fully experience the six degrees of freedom in VR. However, extensive research led to the exploration of redirected walking~\cite{razzaque2001redirected, nilsson201815, steinicke2010estimation}, a method that subtly adjusts the user's direction without them noticing, allowing the user to be immersed in a large virtual environment within limited physical layout they are operating in to provide a seamless experience~\cite{williams2021arc, wilson2018object, langbehn2018evaluation}.

Recent works~\cite{jones2013illumiroom, jones2014roomalive, benko2014dyadic} adopt a SAR system to embrace limited room space and scale and deliver a customized layout. This platform suggests strategies that use a room for an access point, port, and physical interactive space while expanding the interactivity well beyond the room scale.

In recent years, many room-scale VR games use the layout of the room both as part of the virtual environment layout and use the full floor space as an interactive space~\cite{rune2020eye, curiousvr2022custom, tea2022void}. For example, the VR game Custom Home Mapper: Castle Defender (2020) uses the player’s room layout to generate the top floor of the tower balcony. Eye of the Temple (2020) constructs a temple maze from surroundings for the player, and Tea for God (2021) converts the player’s room to a bunker with windows on each wall. Room-scale interactability is appealing as it can be catered to current VR users, which is why research targeting room-scale mixed reality persists~\cite{langbehn2018evaluation, sra2018vmotion, gal2014flare}. My system also caters to the standard room-scale layout as seen in Figure~\ref{fig:pic3}.

\subsubsection{3D Reconstruction for Physical Spaces}
The wide accessibility of depth cameras has produced much remarkable research around 3D reconstruction using these technologies~\cite{newcombe2011kinectfusion, peasley2013replacing} such as adding IMU sensors~\cite{giancola2018integration} to examine outdoor 3D navigation~\cite{qayyum2013kinect}. Robust 3D reconstruction in a combination of object detection research~\cite{lai2013rgbd, alexandre20123d, li2020volumetric, meka2020deep} provides insight into always-on display AR experiences~\cite{nassani2015tag, ibrahim2018arbis}. HoloLens and Magic Leap, both augmented reality (AR) devices, utilize spatial mapping~\cite{magic2022what, zeller2022spatial}, scan and examine the physical layout of the user's surroundings to find the optimal location to place virtual contents~\cite{kumaran2023impact, kim2022investigating}. Projects like FLARE, SnapToReality, VRoamer and Dynamic Theater exhibit the potential these intelligent virtual content projections have to enrich immersive mixed reality experiences~\cite{gal2014flare, nuernberger2016snaptoreality, cheng2019vroamer, kim2023dynamic}. While current spatial mapping for VR-AR systems does well with scanning surroundings for planar surfaces, finding usable space to project virtual objects, and designing a virtual environment, they are ultimately restricted to the physical layout.

\begin{figure}[t]
\centering
\includegraphics[width= 1.0\textwidth]{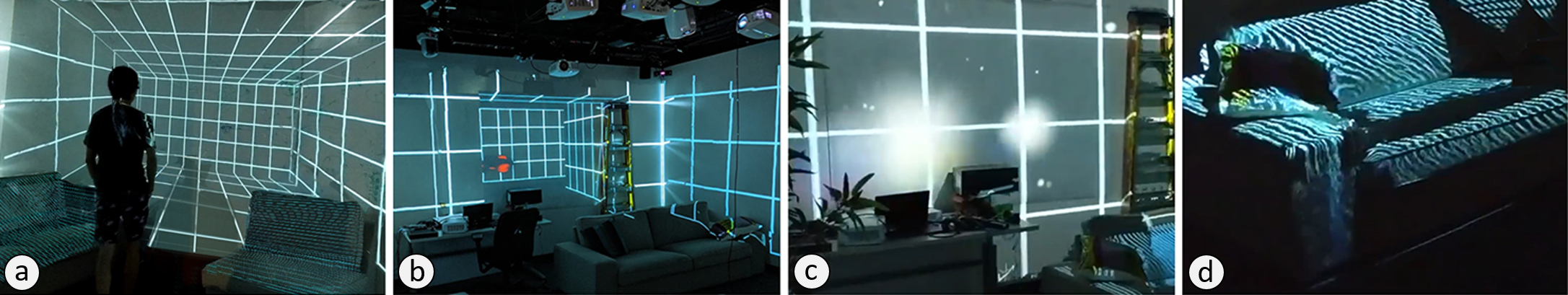}
\caption{The Spatial Augmented Reality (SAR) environment is augmented by four depth cameras and five short throw overhead projectors. (a) User looking at a wall as the room's geometric shape transforms. (b) First-person point of view (POV) from the perspective of a user viewing a virtually extended space in the room. (c) To improve users' spatial awareness, numerous floating particles were rendered to their POV in the environment. (d) In order to ensure user safety in a relatively dark area, texture pattern overlays are projected onto physical furniture to make it look like an illuminated hologram-like item.}
\label{fig:rdr}
\end{figure}

The benefits and impacts of a wide field of view beyond this vision are clear~\cite{ren2016evaluating}. Reality Distortion Room's projection mapping system utilizes a 3D reconstructed model of the room to create a view from the virtual environment. That view is then projected onto real physical surfaces such as the walls and furniture in the room, simulating the perspective of a co-located virtual world~\cite{hartmann2019realitycheck} using the real world as a baseline.
In conducting the user study in a full-surround spatial augmented reality system, it was necessary to augment the entire human field of view and beyond, to create the most convincing user experience through the transformation of their surroundings.

I designed a set of virtual environments that are aligned or partially aligned with the physical room that is deformed, extended, and subtracted from the perspective of the user's eyes as demonstrated in Remixed Reality~\cite{lindlbauer2018remixed}.
In my system, the user sees the physical world at all times as seen in Figure~\ref{fig:rdr}b. Building the experience around the room layout, including the furniture, is done by overlaying it with thin wireframe renderings. This technique highlights the furniture in dark environments, providing safer navigation for the user. My primary goal was to examine the user's natural locomotion in a room filled with everyday objects while simulating the distortion treatment.

\subsection{Reality Distortion Room}
Reality Distortion Room uses the RoomAlive infrastructure~\cite{benko2014dyadic, jones2014roomalive} to deliver a full surrounded augmented reality experience. The room geometry is scanned and loaded into the game environment and Unity workspace where sets of geometric space transformation (distortion treatment) are deployed. Virtual models are placed into the scene in relation to the physical room to project reconstructed geometry. This allows my system to extend the virtual world from the real world, as if the room is transforming, by rendering the physical world within the scene. To simulate the changing geometric environment through the projected walls of the physical room, the synchronization between the user's head position, digital twin, and the real world is crucial. The live reconstruction of the physical environment is done using four RGB-D cameras (Microsoft Kinect v2) that are placed in each corner of the room. The real-time geometric representation of the world is then directly placed in the virtual game environment that warps and enlarges while tracking the head position to reflect the user's perspective.

\subsubsection{System Infrastructure} \label{sec:system}
The AR projected room of approximately 4.5 × 5.5 meters incorporates four Microsoft Kinect v2 depth cameras in each corner of the ceiling line. I placed the furniture objects as typically arranged in a living room, making a close representation of the living room where the user would use my system. Five wide field-of-view projectors render a 360$^{\circ}$ Spatial AR system, projecting to surfaces within the room and fully utilizing the objects inside the room including the furniture and moving objects. Virtual objects are presented from the user's viewpoint as found in RoomAlive system~\cite{jones2014roomalive}. The distorting 3D geometry model is placed and oriented in the physical room and the scene is constructed based on the viewpoint of the user. As the projection is rendered in a view-dependent manner, the participant is free to walk around the room without a headset. As the user in the room is not tethered to anything they are encouraged to walk around and conduct true normal locomotion. Since Reality Distortion Room features full surround visual coverage of the virtual environment, projected onto the physical environment, what the user will see is the room they are standing in, transformed into a different shape. While the distortion treatment is underway, the user's stereopsis will not align with the intended visual of a deformed room. I made sure that all my distortion treatments returned back to the default physical room layout with the standard projection mapping applied. Coming back to a condition where the virtual environment aligns with the physical arrangement is crucial for the user's affordance as this process blends real and virtual. For the virtual room to be precisely aligned with the physical counterpart, the system goes through a calibration process. Static 3D geometry that includes stationary and moving objects is reconstructed with data collected from scanning a cloud of 3D points. With these baseline dimensions, projected content may be precisely aligned with the physical layout. Details on this calibration process can be found in the RoomAlive paper~\cite{jones2014roomalive}.

\subsubsection{Distortion Treatment Components}
Distortion treatment alters the geometric perception of the physical layout. Various transformations and distortion treatments of the environment were implemented to determine their effectiveness in generating consistent locomotion transitions. I examined 10 treatment designs: elongation, warp, shift, elevation, enlarge, enlarge (even larger), rotation, twist, furniture rotation, and furniture shift. Highlighted grid panels, where each tile is 65cm x 65cm, overlay the entire room assisting the user in understanding the transforming geometry while showing the scale transformation. I also added particles in the space to demonstrate accurate reflections of motion parallax as users moved, as well as to assist with seeing added or subtracted space. This easily allows better spatial awareness, while inducing more movement without presenting one target. Each particle is 1.92cm in radius and floats at a speed of 1 cm per second in a random direction (Figure~\ref{fig:rdr}c). Approximately 712 particles float around in every 10-meter cubed space. Every trial tested featured particles in space except for one trial, which was baseline without distortion.

To design three distortion treatments that stimulate the occupant's movement along an axis identified in Figure~\ref{fig:pic3}), I imagined how I would move and respond to a transforming space around me. I identified where people would be most likely to walk toward to secure their safety or view when the room began distorting. The same concept was applied to the two distortion treatments that influenced the user’s distance to the center of the room. When designing the floor layout of the distortion treatment I made sure the virtual space did not transform any smaller than the physical walkable space, presenting the full room layout for users to walk around without being able to step out from the virtually rendered space. The design of the system allows users to interact as they would in an indoor space while experiencing a much bigger environment.

Each distortion treatment is divided into two phases: apply and return. The first Apply segments represent where the distortion treatment begins, and Return entails the latter segment where the virtually distorted room reverts to the original physical room layout. The timeline of the transformations used in each trial for all distortion effects can be seen in Figure~\ref{fig:pic6}.

The dependent variable is the participant's reactions, which was measured through user locomotion. User locomotion is when a participant moves around the room while an assigned stimulus is being applied or returned. The Reality Distortion stimulus was measured as either a directional effect (user movement along an axis) or the Central Effect (change of the user's distance to the room center) and depends on the design of the distortion treatment.

The directional effect measures the user’s movement in relation to the axis along the room's smaller dimension (i.e. left and right in Figure~\ref{fig:pic11}) while the apply and return stimuli are in effect and consists of Elongation, Warp, and Shift Distortions. The central effect (distance to center) measures the change in the user’s positioning away from or towards the center of the room compared to before/after the stimulus segment and consists of Elevation and Enlarge Distortions. The average of the total user movement during each stimulus segment (10 seconds) was used for evaluation.

Baseline refers to the situation in which no distortion treatment is augmented. I established a baseline by running trials with particles but no visual effects or treatments. I separated the baseline data into 10-second segments for the analysis and randomly selected 15 segments from each user's trial to include all potential data segments from the trial. Along with the five distortion treatments, I conducted two additional trials: 1) no distortion treatment with no particles and 2) no distortion treatment with particles. As all trials with distortion treatment included particles, base data points were collected in trials with particles although no distortion treatment was applied.

\subsubsection{Pilot Study}
The experimental portion of this study was designed to measure the effect of distortion treatment on user locomotion responses. First, experiment design (distortion treatment) is a set of geometric distortions that affect the user's movement patterns within the room along the axes, central point, and rotation. All the distortion effects occur with respect to the physical room, regardless of the orientation of the user. I examined all ten distortion designs in the pilot study. In addition, each treatment was tested with two different segment speeds (7 seconds and 10 seconds), for a total of twenty treatments in which ten induced the directional effect while the other ten induced the central effect.

I also assessed the impact of deploying particles to encourage increased walking. I found that distortion effects characterized by excessively rapid transformations, intricate geometric alterations, and rotation effects did not conform to discernible patterns. For example, some participants noted difficulty comprehending the transformations based on the speed or geometry of their implementation, citing dizziness and confusion. Based on this feedback, I omitted certain distortion treatments, and altered the geometry and speed of the remaining transformations. Here are the five distortion designs included in the main study:

\begin{figure}[t]
\centering
\includegraphics[width=0.85\textwidth]{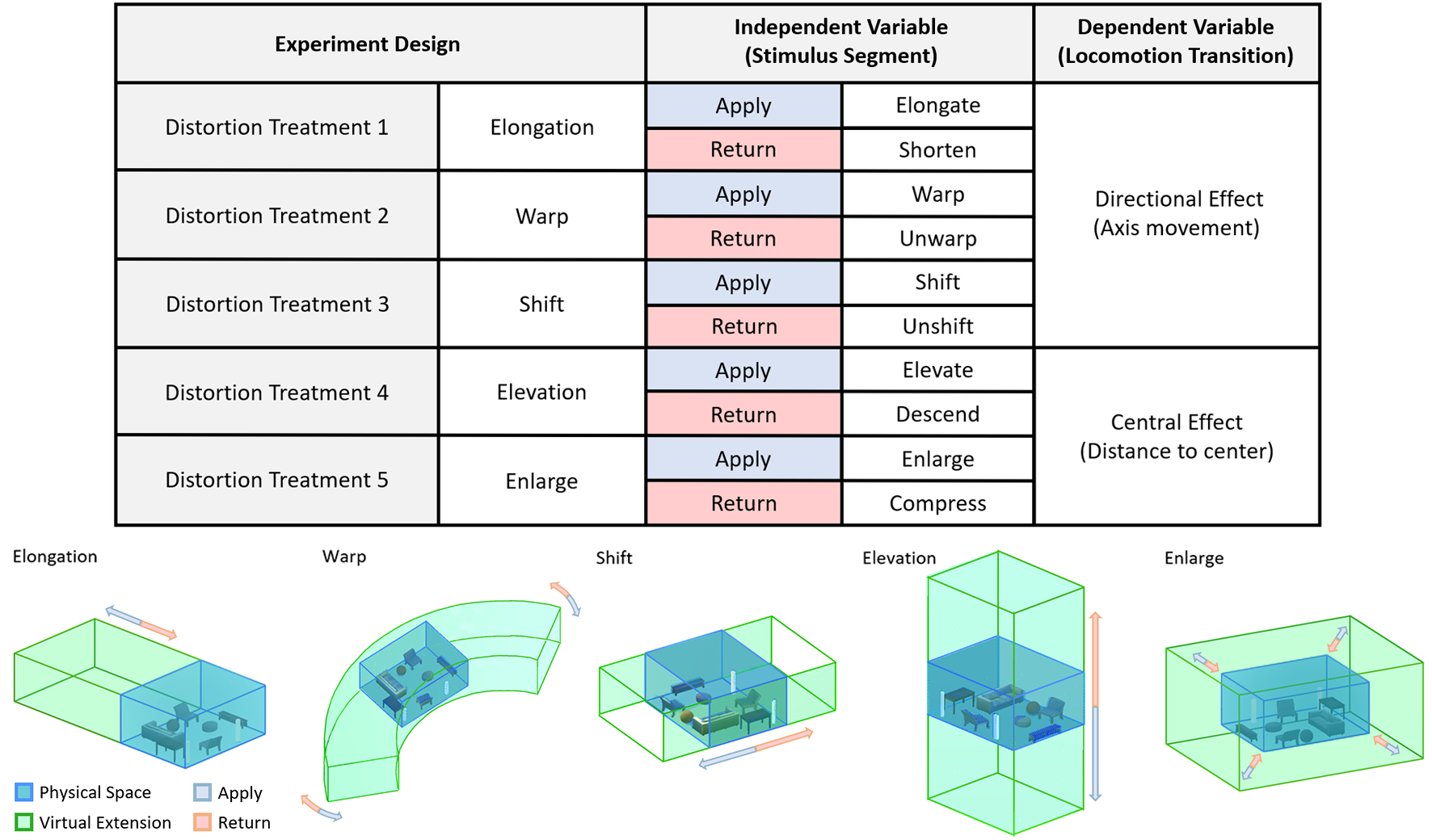}
\caption{Distortion treatments 1, 2 and 3 stimulate axis movement, while distortion treatments 4 and 5 stimulate distance to the center.}
\label{fig:pic5}

\end{figure}

\begin{itemize}
    
    \item Distortion treatment 1 (Elongation) refers to a geometric layout transformation of a room where one wall recedes into the distance and then returns (elongates and shortens). This is called Elongation Distortion and the treatment is designed to have a directional effect. The Elongation Distortion (Figure~\ref{fig:pic2}) elongates one side of the wall outward horizontally for 3.35 meters during the Apply segments of the stimulus, while in the Return segments the elongated wall shortens back to align with the physical room. 
    \item Distortion treatment 2 (Warp) refers to a geometric layout transformation of a room where the entire room warps and unwarps. This is called the Warp Distortion and the treatment is designed to have a directional effect. The Warp Distortion consisted of the continuous warping (bending) of extended versions of two opposite parallel walls of the room. The distortion treatment consists of one warp and unwarp action for each stimulus segment. The bend angle of the parallel walls consists of 160° with the bend executed along a 19.33m length of the wall. 
    \item Distortion treatment 3 (Shift) refers to a geometric layout transformation of a room where parallel walls shift and unshift (shift back) horizontally. This is called the Shift Distortion and the treatment is designed to have a directional effect. The Shift Distortion utilizes two parallel walls of the room, shifting 5.14 meters side to side in the horizontal direction. During the Apply segments the wall shifts in a set horizontal direction, while in the Return segments (unshift) the wall shifts back in the opposite horizontal direction.
    \item Distortion treatment 4 (Elevate) refers to a geometric layout transformation of a room that ascends from and descends to the ground level. The user experiences this as either being elevated above or sinking below the ground. 
    This is called the Elevation Distortion and the treatment is designed to have a central effect (changing the user's distance to the room center). The Elevation Distortion consists of the virtual room going up and down 8.07 meters vertically. During the stimulus segments, the room elevates from and descends to the base ground level.
    \item Distortion treatment 5 (Enlarge) refers to a geometric layout transformation of a room that enlarges and compresses. Users experience this as the walls move away from or come closer to them. 
    This is called the Expansion Distortion and, like for Elevate, the treatment is designed to have a central gathering or dispersion effect. The Enlarge Distortion consists of two segments: virtual room ``enlarge'' and ``compress''. The room enlarges to double the width, length and height while expanding its volume from 56.82 m$^3$ to 454.56 m$^3$.
    
\end{itemize}

\subsection{Experiment}
Twenty participants, ages 23 to 32, of which eight identified as male and twelve identified as female, were recruited for the experiment. Four participants had previously tried VR while only two previously experienced AR. On average, the study lasted approximately 30 minutes, including the time needed for the instruction, and consisted of seven trials and a 10-minute interview. I conducted one trial for each distortion treatment and two additional baseline trials: no treatment with particles and without particles. The Reality Distortion Room system described in Section~\ref{sec:system}. System Infrastructure is used in a room environment that is approximately 4.5m × 5.5m × 2.5m (Figure~\ref{fig:pic3}).
Based on dominant results from prior work in Spatial AR that highlight creative ways to interact with extended reality, and the ways that participants responded to controlled and comprehensive space, the the following hypotheses were defined:
\begin{itemize}

    \item H1: The augmented distortion treatment can induce participants to move more in some directions than others.
    \item H2: The augmented distortion treatment can induce participants to move closer to or away from the center of the room.
    \item H3: The augmented distortion treatment can induce participants to turn or move in a circular direction.


\end{itemize} 

\subsubsection{Procedure}
Participants were introduced to all five distortion treatments in each trial in addition to two trials for the baseline: static room with particles and without particles in the space. Each trial lasted a full 60 seconds, consisting of two cycles of stimulus segments: two ``Apply'' segments and two ``Return'' segments.

When participants arrived, a researcher guided them into the room equipped with the projection system. Upon entry, the room projectors remained off though ambient lighting allowed the user to see objects in the room. The ambient lights remained on through the entire trial but were outshone by the lights from the projectors.

Participants were given a brief introduction to my system and verbal instructions for the study: they are free to move around and interact naturally within the room during the trial but asked to refrain from sitting down on any surfaces. Participants were encouraged to walk and examine the room during the trial and were informed that a few questions would follow the trial. Participants were also informed that they were free to leave the room at any time if they felt discomfort (i.e. sickness, fright). A researcher remained outside the room to monitor participants through the 3D reconstructed live inspector.

After each 60-second trial, I turned off the projection and asked the participants to sit in the center seat of the couch. In between trials, I asked a few questions about their experience to ensure the participant felt okay and ready to continue. Before moving to the next trial, I asked if they could describe in a few words what they saw and experienced. After the final trial, participants were asked to reflect on their experience of each treatment and safety concern. Finally, each participant was compensated with a {\$}10 gift card.   

\begin{figure}[t]
\centering
\includegraphics[width= 1.0\textwidth]{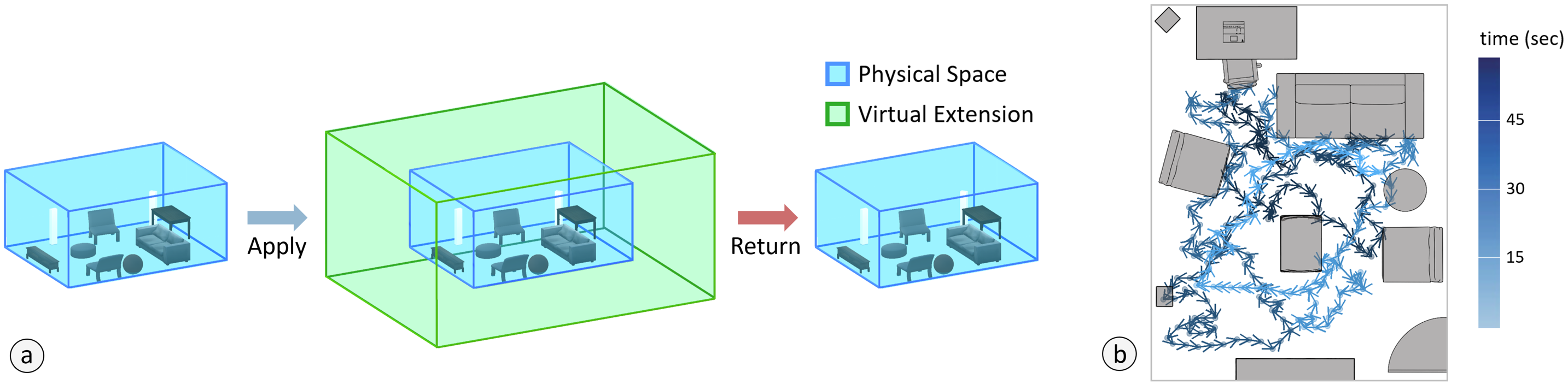}
\caption{(a) The room transformation process during active distortion treatment, using the Expansion Distortion as shown above. During the ``Apply'' segments, the virtual room enlarges for 10 seconds. This is immediately followed by the ``Return'' segments, where the virtual room is compressed for 10 seconds until the virtually extended space merges back to the original room layout. (b) Example path of a user moving around the room during a one-minute trial, demonstrating full possible utilization of the physical space without fear of bumping into objects in the dark environment.}
\label{fig:pic6}
\end{figure}

\begin{figure}[t]
\centering
\includegraphics[width= 1.0\textwidth]{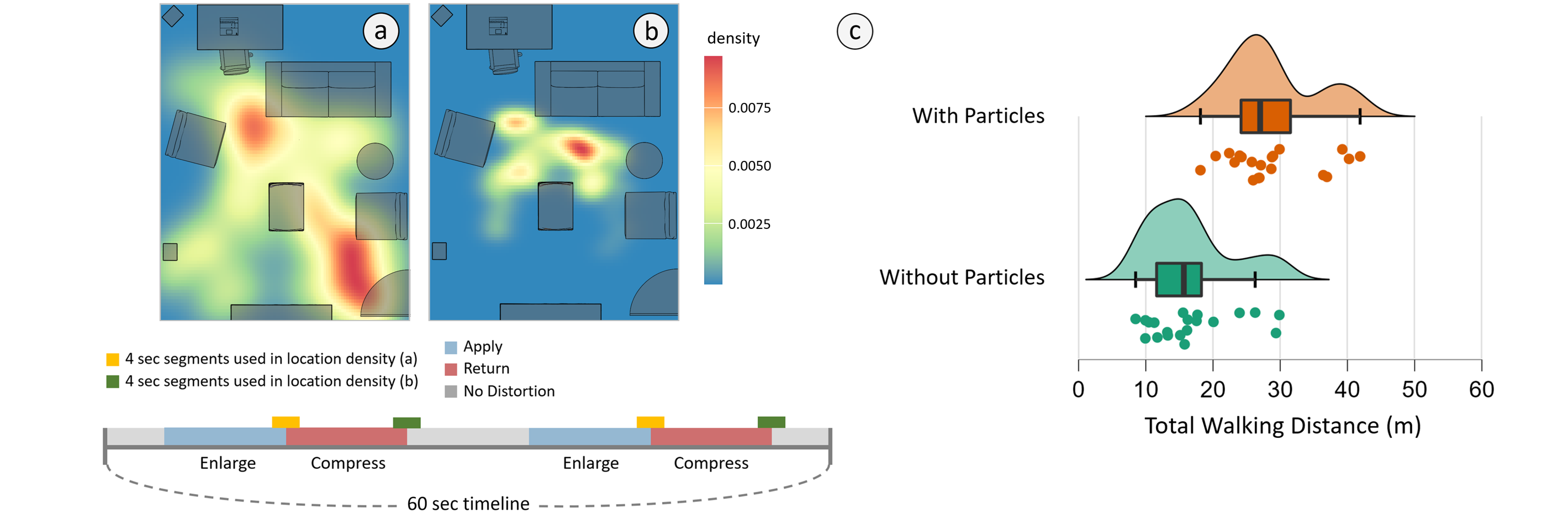}
\caption{The two location density maps, shown above, reflect data collected from twenty users during each segment: (a) Enlarge (apply phase) and (b) Compress (return phase). As shown in the timeline, collections are taken from two 4-second intervals: 2 seconds after and 2 seconds before the end of each phase of ``Apply'' and ``Return''. This shows how floor space was utilized immediately before and after the end of each phase. (c) Total walking distance (TWD) in effects with and without particles, revealing that the existence of particles increases walking distance ($p$ $<$ 0.001). The raincloud plot shows user distribution of TWD result and box indicating the median and interquartile range (IQR).} 
\label{fig:audioPerf}
 \end{figure}

\subsection{Results}
The experimental design (Figure~\ref{fig:pic5}) of this study intended to evaluate the changes in users' positions before and after the geometric transformation in the Spatial Augmented Reality room due to the application of distortion treatments. In this section, two types of distortion effects are reported, with three distortion treatments designed for directional effect, and two distortion treatments designed for central effect. 

\subsubsection{Particle Effect and Natural Locomotion}
The experiment's goal was to induce more natural walking without influencing the participant's movement in a specific direction, as there were no user tasks assigned. Here I compared the total walking distance in the room with particles, without particles, and without the distortion treatment. Without particles, the mean total walking distance was 16.70 m with a standard deviation of 6.38 m. Furniture outlines and wall grids were present in baseline trials with and without particles. Users walked an average distance of 28.80 meters, with a standard deviation of 6.75m, in studies with particles, revealing that the existence of particles significantly increased walking distance. ($p$ $<$ 0.001). As a result of an ANOVA analysis by classifying walking distance into a ``No particle'' group and a ``With particles'' group, there was a significant difference in the mean between the two groups (Figure~\ref{fig:audioPerf}c). In the ANOVA table, the $F$ value was 34.538, and the $p$-value was less than 0.001 demonstrating a statistically significant difference between the ``No particle'' group and the ``With particles'' group.

\begin{figure}[t]
\centering
\includegraphics[width= 1.0\textwidth]{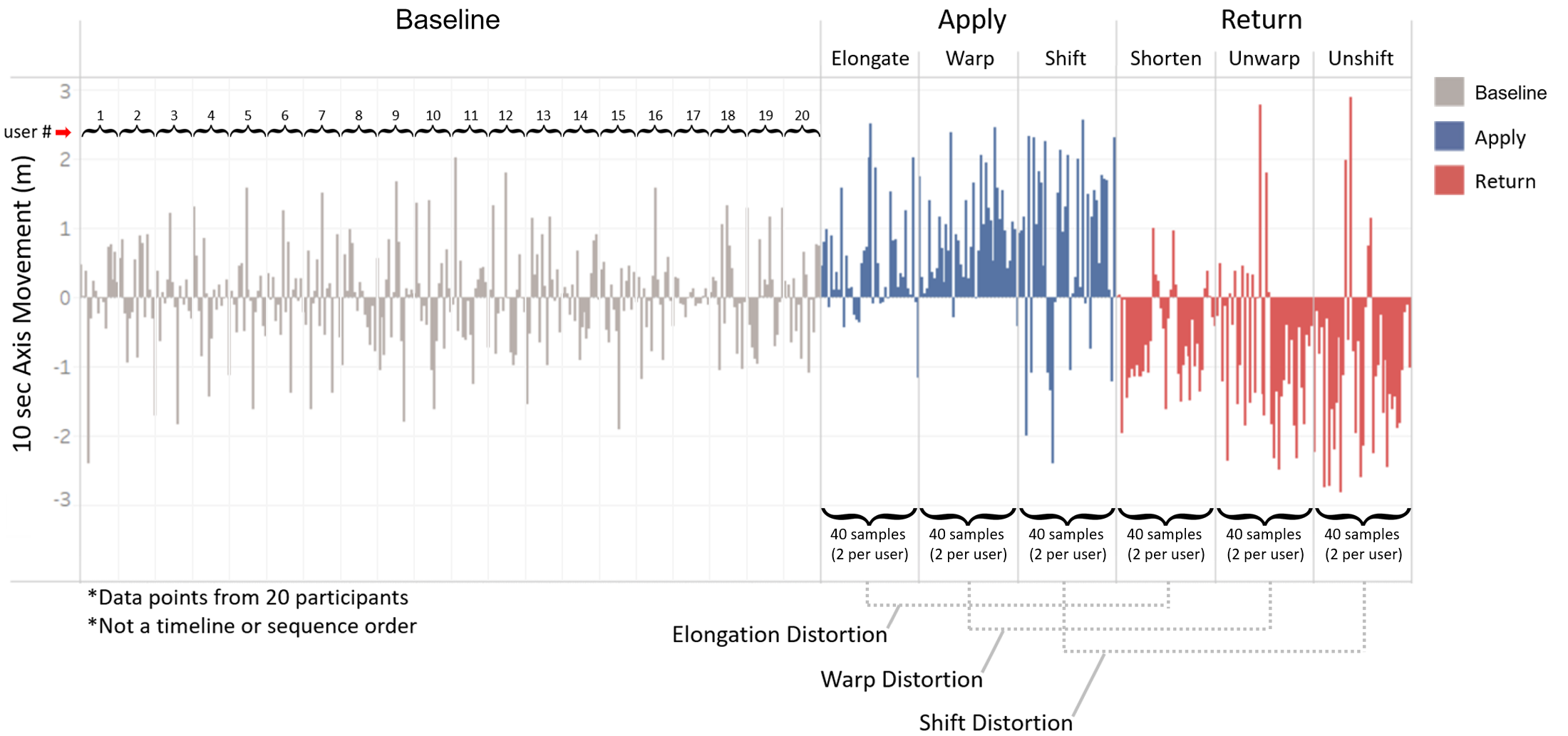}
\caption{Users’ movements on the axis of consideration corresponding to the effect applied to the room.
The chart shows data points (vertical lines) from each stimulus segment from 20 participants. Each vertical line shows data points denoting the average movement made in the axis direction over a 10-second period. We show stimulus segments among four conditions (Baseline, Elongation Distortion, Warp Distortion and Shift Distortion). Visible trends of users’ axis movement in the apply/return phase of three groups of distortion treatments become apparent by comparing with the baseline condition.
}
\label{fig:pic9}
\end{figure}

\begin{figure}[t]
\centering
\includegraphics[width= 1.0\textwidth]{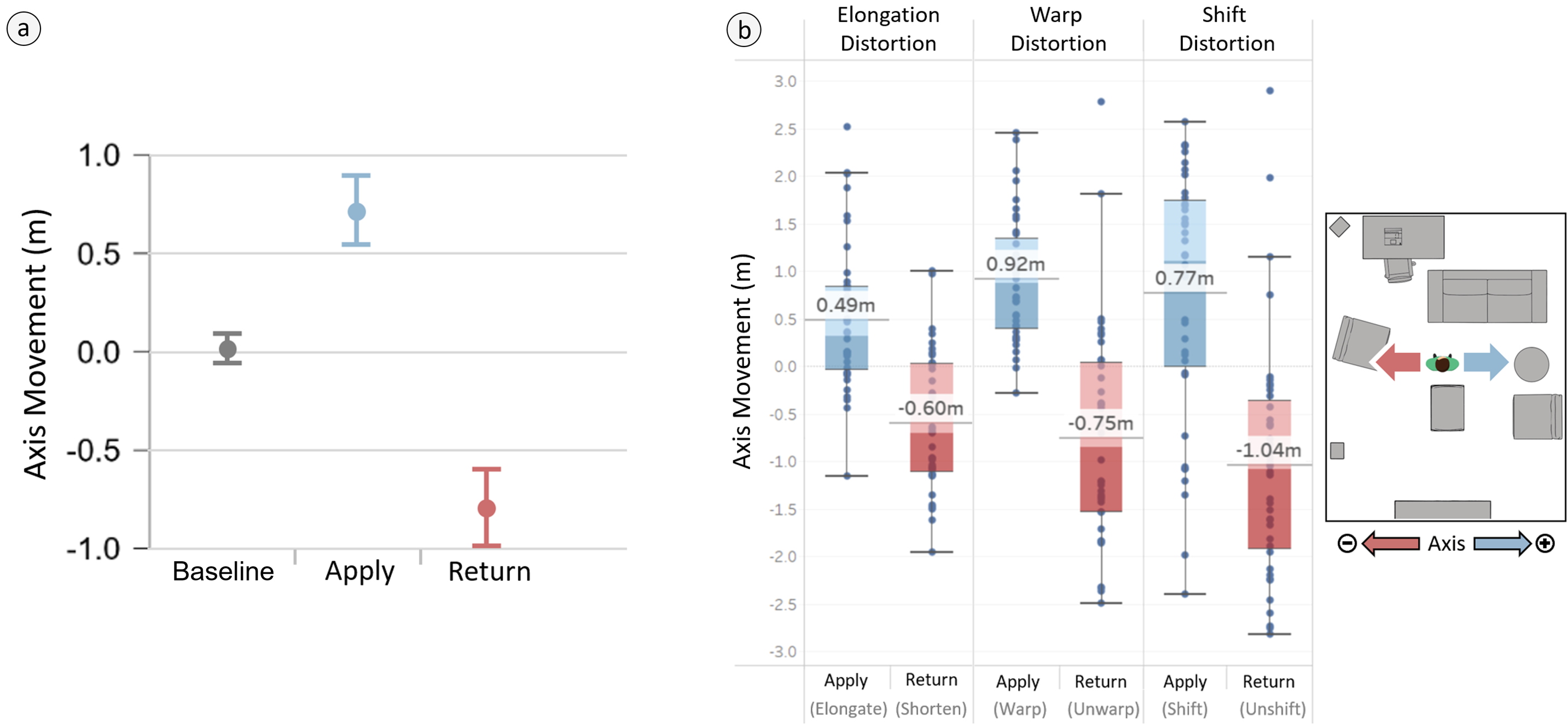}
\caption{(a) Users’ movements on the axis of consideration corresponding to the effect applied to the room. ``Baseline'' represents the movement of twenty users during randomly chosen 10-second intervals without distortion effects. During the ``Apply'' segments, users generally moved in the positive axis direction and in the opposite direction during the ``Return'' segments. Error bars indicate 95\% \emph{CI}. An ANOVA with Bonferroni-corrected post-hoc pairwise comparisons reveals significant induced user movement during the distortion treatments. (b) Magnitudes of the apply and return of axis movements compared between three different distortion effects. Based on mean distance moved, the shift effect resulted in the most user movement towards the positive and negative ends of the axis, followed by the warp effect, and the elongation effect.}
\label{fig:pic11}
\end{figure}

\subsubsection{Directional Effect: Axis Movement}
The following reports the directional effects of the three distortion treatments designed and tested. Ten-second user movements along the axis direction within each stimulus segment were evaluated. As seen from the data point of Figure~\ref{fig:pic9}, a clear signal and trend between each stimulus segment and baseline was present. ``Baseline'' shows point data generated from 20 participants' baseline trials (no distortion treatment + particles) in two 10-second intervals throughout the trial. In the apply stimulus segment, users tend to move toward the positive axis direction (refer to Figure~\ref{fig:pic3} for axis directions), while moving in the opposite direction for the return stimulus segment.

The experimental group was classified by Baseline Group, Apply Group, and Return Group and the ANOVA analysis showed significant differences between the three. In the ANOVA table between the three groups, the F value was 198.329, and the $p$-value $<$ 0.00005, showing a statistically significant difference in the size of apply segment and return segment. Post hoc analysis using Bonferroni adjustment also showed a significant difference in the mean between the three groups as follows. The difference between Baseline Group and Apply Group is $t$ value = -7.747, $p$-value $<$ 0.001. The difference between Baseline Group and Return Group is $t$ value=-8.924, $p$-value $<$ 0.01. The difference between Return Group and Apply Group is $t$ value=-13.948, $p$-value $<$ 0.001.

Each stimulus segment was tracked from apply and return, respectively. The movement of the axis for each group is analyzed as shown in the Figure~\ref{fig:pic11}. The average value of the Return and Apply segment is different for each group of the Elongation Distortion, Warp Distortion, and Shift Distortion. Based on mean distance moved, the Shift Distortion resulted in the largest directional user movement, followed by the Warp Distortion, and the Elongation Distortion.

\begin{figure}[t]
\centering
\includegraphics[width= 1.0\textwidth]{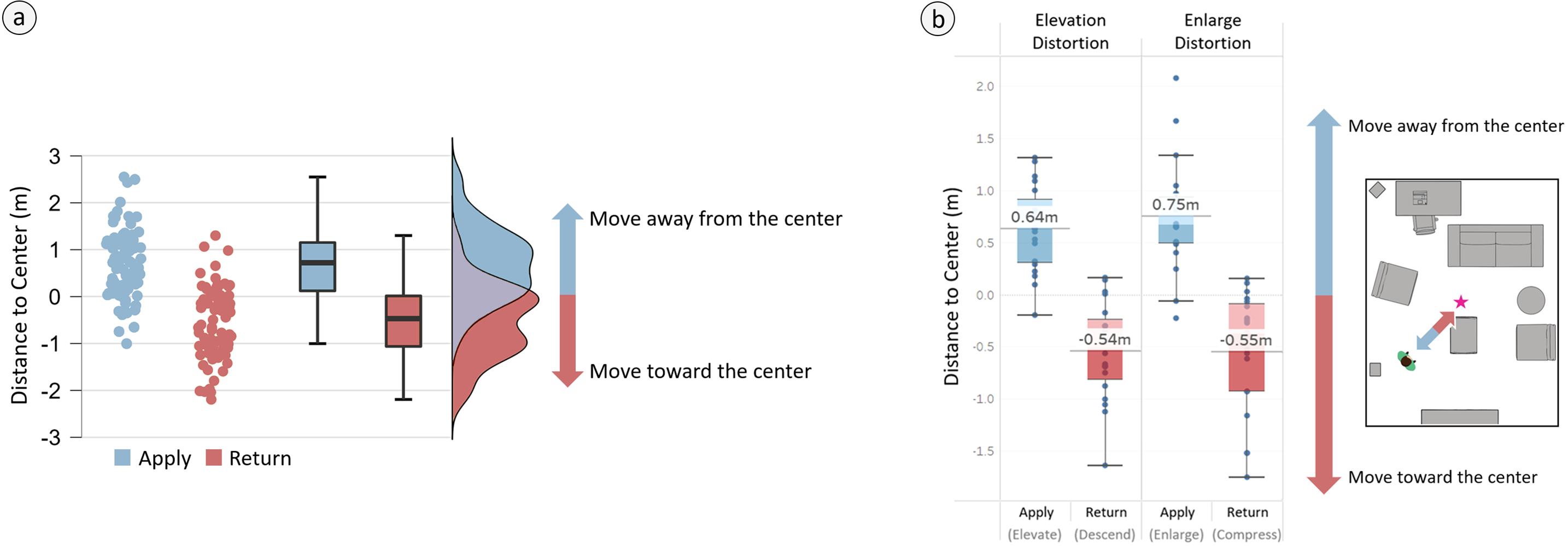}
\caption{(a) A raincloud plot depicting movement away and towards the center of the room when the elevation and expansion effects are applied. The upper whisker boundary of the box-plot is the largest data point that is within the 1.5 IQR above the third quartile. According to apply and return, respectively, the distance  to the center was tracked. (b) Magnitudes of the apply and return movements measured how far a user was from the center of the room. Based on the mean distance moved, the expansion effect resulted in slightly more movement away and towards the center, followed by the Elevation Distortion.}
\label{fig:rotation}
\end{figure}

\subsubsection{Central Effect: Distance to Center}
The average movement change of the user throughout the 10-second distortion segments was analyzed by comparing their position in the room before and after the segments. In other words, the displacement of the user during the 10-second interval was measured. Out of the five distortion treatments, two of them were designed and tested to manipulate the user's position relative to the distance from the center of the room. During the application stimulus segment, users generally moved away from the center, whereas during the return stimulus segment, users tended to move toward the center of the room. The remaining three distortion treatments had no consistent central effect on the user movement.

Experimental groups were divided into ``Apply'' and ``Return'' groups based on stimulus segments, and their distance to the center was analyzed using ANOVA. As a result, significant differences between groups are shown. In the ANOVA table, the $F$ value was 109.123 and the $p$-value $<$ 0.001, showing a statistically significant difference in the size of apply and return. The movement from the distance from the center for each group is analyzed as shown in Figure~\ref{fig:rotation}b. Based on the mean distance moved from the center, the Enlarge Distortion resulted in the most user movement. This was found by comparing both stimulus segments although both distortion effects performed comparably. 

Overall, the study aimed to investigate the influence of augmented distortion treatment on users' natural locomotion while relying on their ability to comprehend the spatial transformation of the environment. I hypothesized that H1: The augmented distortion treatment can induce participants to move more in some directions than others and H2: The augmented distortion treatment can induce participants to move closer to or away from the center of the room.
    
The user study results indicated that both H1 and H2 hold true. Additionally, the Shift Distortion showed the largest directional effect while the Enlarge Distortion showed the largest central gathering/dispersion effect. The third hypothesis (H3) aimed to test whether specific augmented distortion treatments could influence participants' natural locomotion to turn or move in a circular direction. During the pilot study, however, none of the designed treatments including those intended to induce circular movements, such as Rotation or Twist, were successful in producing this motion pattern. Participants also reported experiencing difficulty understanding and dizziness, which made it challenging to comprehend the nature of these transformations from Rotation or Twist Distortion effects. Therefore, the third hypothesis was not confirmed.


\subsection{Discussion}
As no clear objective was given, participants generally assumed that I wanted feedback on the designed space, or that I was showcasing the new AR system. Many of them shared ideas on new 3D space ideas and what I should try next. This worked in my favor as I did not want to hint that their locomotion was being examined; the goal was to capture users' natural locomotion from the distortion treatment. 

I found compelling differences between the stimulus segments apply, return, and the baseline segment in locomotion response. Among all three distortion treatments for manipulating directional effect, the Shift Distortion had the largest distance manipulated before and after the stimulus segments, while the Warp Distortion treatment showed the biggest positive axis movement when the room was being warped. This is likely due to users trying to get a better vantage point to see the end of the hallway and get a better idea of what is happening in their surroundings. The Elongation Distortion had the weakest effect, but when comparing each stimulus segment to the baseline segment, the treatment still worked with the shortest mean average distance effect to axis. Of the two central effects, the Enlarged Distortion exhibited the most prominent distance-to-center effect before and after the stimulus segments.  

Only one participant removed themselves from the study (to answer a phone call). No participant requested a break during the distortion treatment augmentation or expressed sickness during our study. Participants' responses to Elevation Distortion showed a generally negative sentiment, stating that ``it took some time for me to understand what was happening'' (P4) and ``it was apparent when the room was going up but when the room was going down I wasn’t convinced'' (P17). In contrast, many participants expressed that the Enlarge Distortion was fun and refreshing: ``The room gradually expanding in all directions was my favorite'' (P13) and ``I felt like I was floating in space'' (P14). Two participants requested to experience the Enlargement Distortion again after the study. In contrast to these many positive experiences, a few participants bumped into objects during the enlarge stimulus segment. Lastly, most participants found particles in the space to be a nice addition, stating it ``magical''and ``fun to interact with'' (P5).

\subsubsection{Limitations}

Reality Distortion Room demonstrates a visual perception locomotion aid designed for the purpose of safety, entertainment, and interaction. While the study provides insight into how visual perception can be used to manipulate a user's natural locomotion, its findings are based on a small participant sample and brief trials, underscoring the need for deeper exploration of design considerations.

I tested five distortion treatment designs in this study understanding that many more distortion treatment designs could be explored. The user study suggests interesting directions for future work and I foresee many ways to expand this concept. The challenge in using distortion treatment is that the transformation is very noticeable and, at times, intrusive. Creative solutions must be explored in using distortion treatment to deliver a cohesive user experience. I suggest distortion treatment to be adopted in catered space and situations, where the design of the geometric distortion is modified to the designated physical space. This curated experience, where aspects of the real-world deviate from reality, requires further study. 

The study utilized a full-surround augmented reality platform with projections covering the entire human field of view and beyond, achieving this setup at home is unlikely. Additionally, while the system was the best fit for my use, whenever the user came within 60 cm of the wall, the user’s shadow became visible, blocking the projection of the wall. Also, the outline projected to furniture disappeared when the user blocked the projection. 

Though there were overwhelmingly positive responses from the participants, I believe the novelty factor may have played a role in how actively the participants explored the room, inducing more walking in the room. In addition, people moved without any instruction or clear objective in our study, as I wanted to examine if I could manipulate users without using instruction and by only using distortion treatment. This raises questions about the performance of distortion treatments when a user is presented with tasks in the space and the effectiveness of the distortion effect when using parts of the projected environment rather than the grid system. I hope to tackle this question in my future work. 

\subsection{Conclusion}
Reality Distortion Room is a proof of concept that shows how visual perception of certain room distortion effects can invoke cohesive natural locomotion responses from the user. The user study demonstrated that the distortion treatments we designed were effective in influencing users’ natural locomotion in predictable ways. By relying on users' reactions to their visual perception of space, this research opens new ways to engage with familiar environments or to navigate in an enhanced, altered, or even completely virtual reality. It is especially exciting about the possibilities of instilling movement in people solely through visual deformations of the AR spaces they populate.

\begin{section}{Chapter Conclusion}

As demonstrated in the two projects discussed in this chapter, the Spatial Orchestra empowers users to control their environment for creative expression, while the Reality Distortion Room shows how the environment itself can predictably influence user behavior. Together, these prototypes highlight the vast potential for incorporating diverse elements into interactive narrative experiences using AR stages tailored to specific spaces. These works emphasize the need for further research, particularly in human-computer interaction within mixed reality, and the possibilities these interactive AR theater platforms offer. These AR stages serve not only as venues for user expression but also as environments that can significantly influence and enhance participants' experiences.

\end{section}

\chapter{Personalized AR Theater System}

In this chapter, I will discuss a two-part final project. The first part, Dynamic Theater, explores the use of AR in immersive theater as a platform for digital dance performances. This project offers a locomotion-based experience that allows users to fully explore a large indoor AR theater space, enabling them to freely navigate the augmented environment. The curated wide-area experience employs various guidance mechanisms to direct users to the main content zones. A study involving 20 participants highlighted how users interacted with the performance piece using the guidance system. The findings emphasize the importance of stage layout, guidance systems, and dancer placement in immersive theater experiences, aligning with user preferences and enhancing the reception of digital content in wide-area AR. Insights and feedback from working with dancers and choreographers are also discussed.

The second part of the project, Audience Amplified, introduces a personalized ``always-ready'' social experiment. While discussing Dynamic Theater with others, I developed the concept of an adaptive virtual audience for an AR theater play, allowing users to explore and discover diverse performances throughout the environment. In this phase, I expanded on the ideas initially discussed during the planning stages, illustrating what an always-ready social experiment might look like.

My project utilizes ML-agents trained with motion tracking data from previous users and digital twin models of physical layouts. Through imitation learning, I create virtual audiences that mimic human behavior in specific physical settings as users engage with the AR play. This effort aims to explore the future of virtual production and mixed reality spatial content, offering a glimpse into how these technologies will be presented. Additionally, I share my experience teaching an AR development course focused on creating wide-area, augmented reality theater interactive narrative apps. Over a 10-week period, students successfully utilized this open-source AR Theater Platform.

\begin{section}{AR Theater Project 1: Dynamic Theater}
Dynamic Theater showcases the future of virtual production and how people will consume digital content through spatial computing. It fully utilizes the ability to walk around familiar places, allowing users to experience personalized content and performance narratives that expand their reality as a story unfolds at their own pace. Dynamic Theater was presented at the 29th ACM Symposium on Virtual Reality Software and Technology (VRST) held in Christchurch, New Zealand, in October 2023, and premiered to the public at Kirby Crossing in November 2023 with over 50 users having the opportunity to try it out.

Dynamic Theater explores the use of AR in immersive theater as a platform for digital dance performances. The project offers a locomotion-based experience that enables full spatial exploration within a large indoor AR theater, designed to allow users to freely explore the augmented environment. A curated wide-area experience is presented, employing various guidance mechanisms to direct users to the main content zones. The results of a 20-person user study reveal how participants engaged with the dance performance while navigating the space using a guidance system. The study highlights the importance of stage layout, guidance systems, and dancer placement in immersive theater experiences, catering to user preferences and enhancing the reception of digital content in wide-area AR. Observations and feedback from working with dancers and choreographers are also discussed.

\begin{figure}[t]
\centering
\includegraphics[width= 1.0\textwidth]{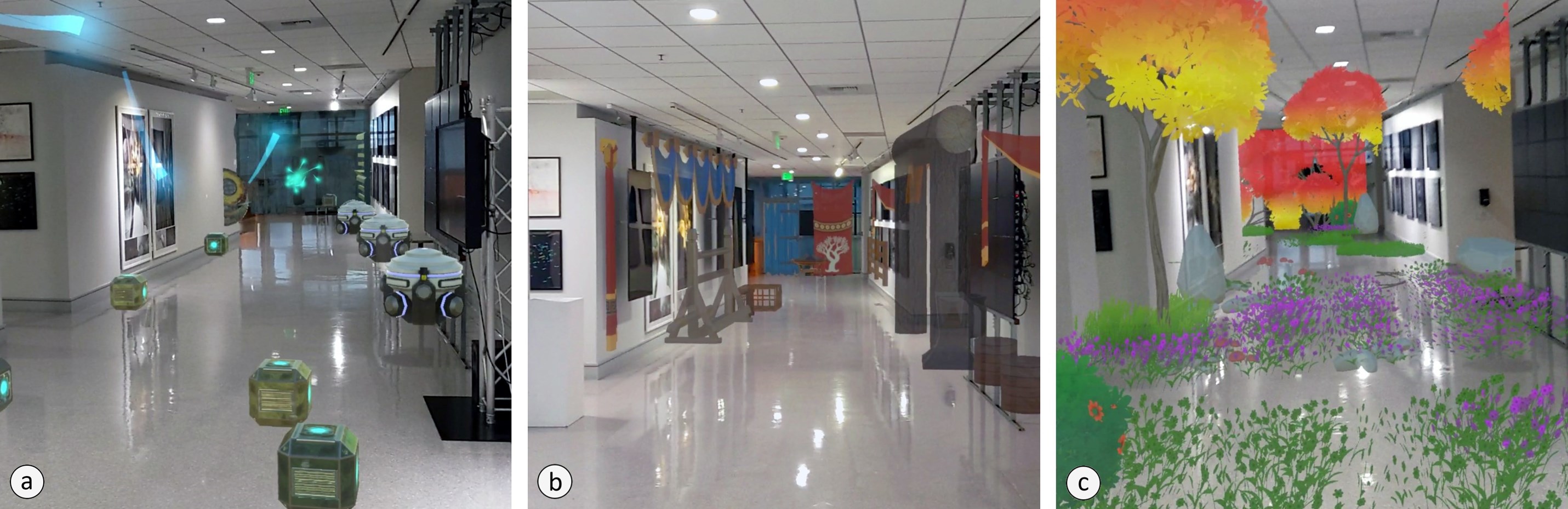}
 \caption{Three virtual stages were captured from the headset using Microsoft MR Capture. Here, both virtual objects and the physical environment are visible within the (a) future stage, (b) fantasy stage, and (c) forest stage.}
\label{fig:all-scenes}
\end{figure}

\end{section}

\subsection{Introduction}
Theater choreographer McAuley emphasizes the intricate interplay between space, performer, and narrative. Choreographers must consider the theater building's overall physical arrangement when planning dance sequences. McAuley suggests that creative teams should collectively explore the building and stage before dance rehearsals to enhance engagement~\cite{mcauley1999space}.

Dance floorwork encompasses movement throughout the entire stage, emphasizing the role of expression and movement in human bodies~\cite{diana2018overlooked}. In contrast, digital media consumption typically involves minimal physical activity. While modern gaming can offer expansive virtual environments, recent advancements in extended reality (XR) aim to replicate aspects of live theater performances~\cite{erkert2003harnessing, lella2014us, cheok2002interactive, gochfeld2018holojam, oculus2017dear, game2019presents}.

Previous studies demonstrated the potential of presenting virtual content through AR in large-scale settings~\cite{feiner1997touring,thomas2002first,cheok2004human,rompapas2019towards, kim2022investigating, kumaran2023impact}. They employ multiple tracking and registration mechanisms such as occlusion models, cloud anchor points, and user-centric design elements to ensure comfort and safety during navigation. Current XR practices, however, still underutilize available space. To optimize the utilization of a large indoor space for the application domain of immersive dance theater, I investigate user preferences on guidance systems to enhance the theatrical experience for a mobile AR audience. Focusing on choreography, which encompasses the entire floor area and involves user interaction, I explore how the effective utilization of space can showcase immersive dance pieces.

According to McAuley, theater intertwines fact and fiction, creating a nuanced and cohesive theatrical experience~\cite{mcauley1999space}. In the context of AR theater attendance, my aim is to allow users to navigate the narrative at their own pace. Therefore, a guidance system that subtly enhances the core experience while seamlessly blending into the environment is necessary.Preserving the essence of theater and dance, which has been studied and explored for years, is crucial, especially in capturing elements like footwork. Steve Paxton, a trailblazer in contact improvisation, focuses on the dynamic relationship between dancers and the forces of gravity and momentum. His floorwork approach emphasizes tactile exploration, treating the floor as a partner in the exploration of balance and movement~\cite{paxton1975contact}. Pina Bausch, renowned for her theater choreography work, utilizes floorwork to convey complex human emotions and narratives, often incorporating interactions with the floor that deepen the storytelling~\cite{gervasio2012toward}. Scott Graham of Frantic Assembly combines dynamic physicality with storytelling, using the floor to support athletic and fluid movements, thereby enhancing narrative-driven performances~\cite{evans2019performance, robinson2017practical}.

Furthermore, the placement of digital content in a vast open area poses a significant challenge in facilitating users' intended discovery of AR narrative content~\cite{cheok2002interactive, windschitl2000virtual, winn2001learning}. Ensuring freedom to navigate the room while implementing direction to specific physical locations for viewing digital content requires meticulous consideration of space arrangement.

Location-based narratives in open areas enable natural locomotion, which replicates real-world exploration and allows users to navigate at their own pace. Previous studies showed a preference for realistic modes of travel~\cite{sayyad2020walking}. Leveraging location-based narratives embraces the benefits of natural locomotion while addressing associated challenges. I build upon the foundation of interactive theater experiences in virtual environments on performance stages~\cite{gochfeld2018holojam}, expanding it to various location-based indoor environments. My work, Dynamic Theater, offers curated immersive experiences, going beyond traditional in-person theater by enabling user interaction with visual content and exploration of the environment.

\begin{figure}[t]
\centering
\includegraphics[width= 1.0\textwidth]{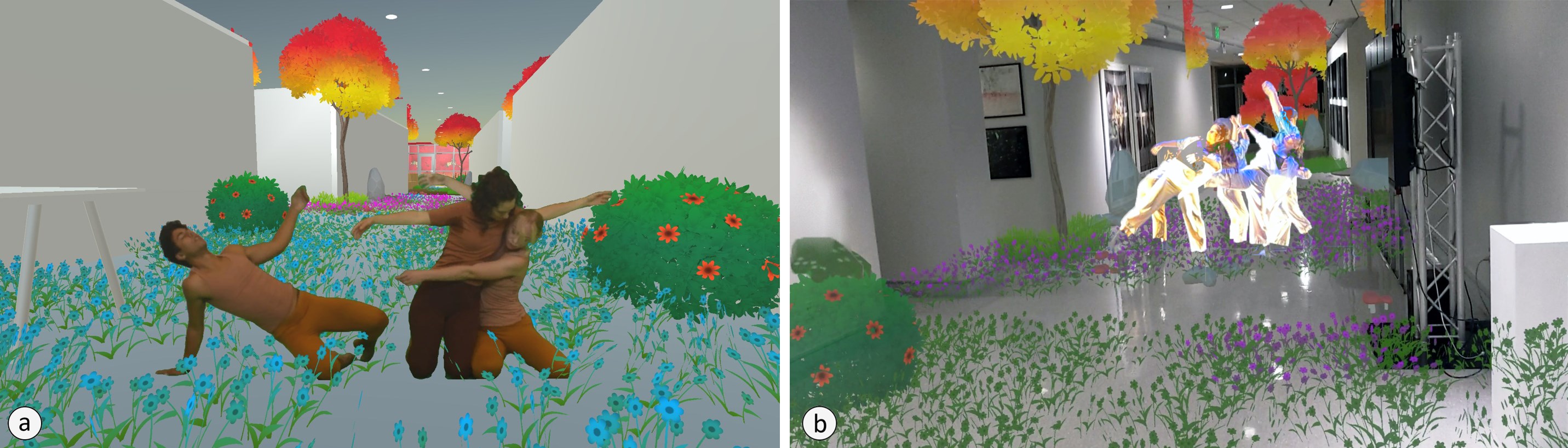}
 \caption{A digital twin of the physical space was used to situate the dancers correctly in relation to the environment to preserve their footwork. Image (a) features an illustration shown to the choreographer for precise dancer placement in Unity while (b) shows how dancers were projected through the headset.} 
 \label{fig:billboarding}
\end{figure}

Dynamic Theater presents an immersive system showcasing recorded dance performances in a large indoor space. The goal is to create an augmented dance experience supported by a visually complementary and unobtrusive user guidance system, experienced through the HoloLens 2. Using billboarding techniques, transparent dancer holograms are strategically positioned in collaboration with a professional choreographer to integrate dance movements into the physical arrangement (see Figure \ref{fig:billboarding}). Navigational guidance systems based on virtual particle systems are evaluated to enable a narrative-driven AR theater experience. Through the deployment of various guidance methods and themed virtual stages, insights are obtained towards optimizing future stagecraft and dynamic user experiences. My contributions to the immersive theater experience include the following:

\begin{itemize}
    \item Developed a Dynamic Theater system to seamlessly merge dance choreography with physical surroundings. Performance playback involves carefully placed billboarded projections of recorded dancers, enabling anyone with the headset to enjoy the play within the designated area at any time.
    \item Created a narrative-driven AR theater experience utilizing navigational guidance systems with projected content. This facilitates natural audience locomotion, guiding viewers to content zones. 
    \item Extracted valuable insights from surveys and interviews with 6 dancers and a choreographer. Their input, especially regarding dancer involvement, influenced the project direction throughout.
\end{itemize}

Altogether, my work enables the expansion of interactive theater to diverse location-based AR theater design, involving AR content tailoring to physical layouts and collaborations with dancers for floor and footwork integration.

\subsection{Related Work}

In this section, I review AR research on storytelling, interactive design, and location-based narratives.

\subsubsection{Storytelling in Mixed Reality}

A shift in recent years from traditional theater production to mixed reality theater production, which shares many of the same components, presents new opportunities and fresh obstacles\cite{coulombe2021virtual, gochfeld2022tale}. Mixed reality has been extensively studied for narrative presentation, allowing users to experience curated stories in immersive environments that enhance the conveyance of artistic works \cite{geigel2004theatrical}. Various performances, including choreographed performances, have been adapted to mixed reality, such as \emph{The Life} \cite{abramovic2020life}, \emph{Debussy3.0} \cite{clay2014integrating}, and \emph{Nautilus} \cite{fischer2016nautilus}, integrating digital elements and augmented visual effects into live dance performances. Other works involve close collaboration between the artist and technical engineers to create augmented dance experiences \cite{clay2012interactions}. My project builds upon these prior works, extending the foundations of choreography in mixed reality to create a more interactive, and mobile, user experience. 

While my focus is on dance performances, insights from other narrative theater experiences have also informed my research. Works like \emph{CAVE} \cite{layng2019cave} and \emph{CAVRN} \cite{herscher2019cavrn} offer shared cinematic experiences in large virtual environments for collective audiences. \emph{Gulliver} \cite{ARGulliver} is a play that incorporates AR headsets and augmented effects, while the use of physical props (e.g., Augmented Playbill, Prayer Wheel, and Tarot Cards) expands the theatrical experience beyond the stage \cite{nicholas2021expanding}. Building upon the insights gained from narrative theater experiences, my research aims to further explore the potential of mixed reality in the context of dance performances. 

\subsubsection{Interactive Design with Locomotion}

My understanding of human decision-making and time perception in mixed reality experiences is growing rapidly~\cite{awe, FLAVIAN2019547}. Immersion enhances focus and extends perceived time, potentially leading to increased satisfaction. An element of awe can expand time perception, influence decision-making, and enhance well-being. The perception of self within the play relates to the interactive narrative experience~\cite{self}. 

Research has explored locomotion, particularly walking, as a fundamental input for spatial computing in mixed reality designs, especially in location-based elements of AR games. This approach allows users to physically explore the environment and engage with interactive narratives. While games and performances differ, the concept of creating location-based experiences aligns closely with my research. Notably, Soul Hunter \cite{weng2011soul} demonstrates how players navigate multiple rooms in a castle and interact with translucent ghosts using an AR weapon. Other AR games, such as ARQuake \cite{thomas2000arquake} and HoloRoyale \cite{rompapas2018holoroyale}, aimed to bring first-person shooter experiences to larger physical spaces. Additionally, Madsen et al. \cite{madsen2022fear} explore integrating an exploration-based horror game into an indoor AR environment. Collaborative and multi-user experiences are facilitated through AR applications such as Human Pacman\cite{cheok2004human}, where users interact as characters in a large physical environment, and MapLens \cite{morrison2009like}, which encourages collaborative task-solving in a city-wide AR setting. Notably, the combination of natural walking experiences and sightseeing has been used to highlight and preserve important historical events using a mixed reality interactive narrative approach \cite{lehto2020augmented, fujihata2022behere, wither2010westwood}. My research expands the understanding of mixed reality in theater by exploring the possibilities and challenges of integrating dance performances into designated spaces while maintaining a strong connection with the audience and the surrounding environment through the interplay of narrative, locomotion, and location.

\subsubsection{Location-Based Narrative}

In location-based works, the relationship between the narrative and physical setting is essential for enhancing the immersive and realistic aspects of the storytelling \cite{location}. In this section, I discuss works that extend beyond dancing, focusing instead on the utilization of location-based storytelling and interactive narrative as a means to convey stories \cite{han2023architectural}.

As briefly mentioned, sites of historical significance rely on the integration of physical locations with digital information to deliver impactful experiences. For example, BeHere 1942 transports users back in time to the actual location where the forced expulsion of Japanese Americans occurred at the old Santa Fe train station in LA \cite{fujihata2022behere}. Lights On! is an augmented reality game that connects users to real cultural heritage sites \cite{lehto2020augmented}. The Westwood Experience narrates the story of a mayor and guides users to physical locations mentioned in the narrative itself \cite{wither2010westwood}. Furthermore, an AR application developed for the Oakland Cemetery guides users through the physical cemetery using audio narrations and stories of its former inhabitants \cite{dow2005exploring}.

Mixed reality theaters have also embraced innovative approaches to creating interactive experiences. Cheok et al. combined location-based, outdoor exploration with a virtual theater performance of a Shakespearean play. Users are guided through an outdoor environment with content zones that provide clues, eventually transitioning into an indoor virtual reality environment featuring a captured actor portraying Hamlet \cite{cheok2002interactive}. Other mixed reality theaters offer opportunities for users to actively participate in the theater performance itself \cite{gochfeld2018holojam, pietroszek2022meeting, pietroszek2022dill, lyons2023gumball}. For instance, Holojam in Wonderland enables audience members to follow a live actor in the form of an avatar, engaging with both the actor and the virtual environment on stage \cite{gochfeld2018holojam}. Story CreatAR helps facilitate the process of incorporating locomotive storytelling in AR, particularly when narrative elements can occur at various physical locations within large environments \cite{singh2021story}.

In addition, recent work explored a novel approach to adapting narratives to real spaces for AR experiences \cite{li2023locationaware}. This system utilizes an optimization-based algorithm that automatically assigns contextually compatible locations to story events. Other techniques offer the audience more choices in terms of where they can experience the same content, including location-specific places, such as museums, or an alternative VR environment \cite{geigel2020digital}. My approach promotes active decision-making in integrating digital content into live performances and physical spaces, curated by artists.

\begin{figure}[t]
\centering
\includegraphics[width= 1.0\textwidth]{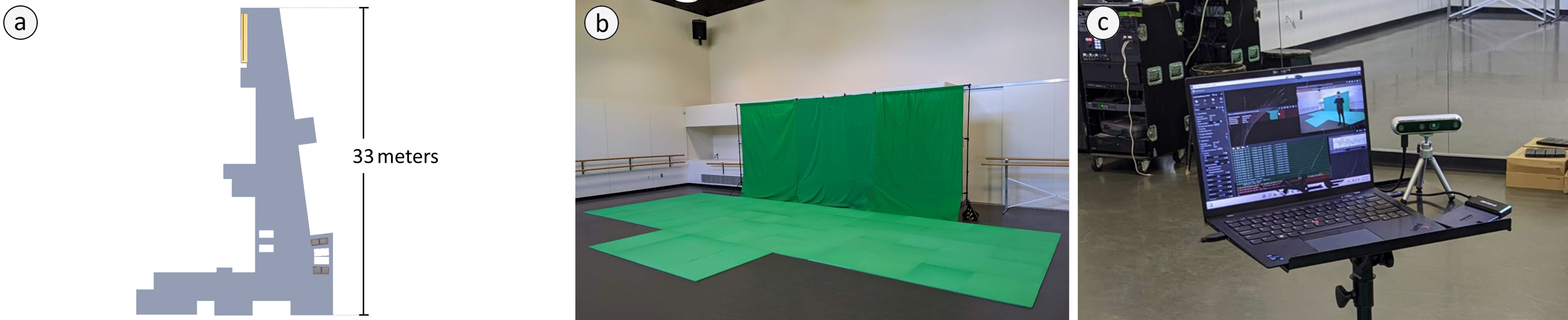}
 \caption{(a) Top view of the experiment area and its physical layout. The dimensions are 208.54$m^{2}$ (2,244 $ft^{2}$). (b) The dance capture session employed a green screen setup on a stage measuring 2.9 $m$ x 14.6 $m$, with a green screen backdrop measuring 11 $m$ x 2.6 $m$. (c) Five RealSense D455 cameras, along with a laptop and a 1.2 $m$ tall stand, were positioned between the three DSLR cameras to capture depth video frame data.}
 \label{fig:top-view}
 \end{figure}

\subsection{Dynamic Theater}
During the development of the Dynamic Theater system, I encountered technical challenges in tracking, floor plan modeling for the occlusion layer, and anchoring strategies. Collaboration with artists, dancers, and choreographers, thus, was crucial. This section explores the procedure of capturing dancer movements and presenting them in the HoloLens 2. I also discuss the design of the Virtual Stage, the guidance system, and the user experience.

The theater piece took place in a spacious indoor environment with two hallways for user traversal (Figure \ref{fig:top-view}a). The indoor space measured approximately 208.54 $m^{2}$. Virtual objects strategically restricted access to certain areas. Three distinct virtual stage designs showcased the versatility of the space.

\subsubsection{Capturing Dancer Movements}
I collaborated closely with six professional dancers and one choreographer to capture dance movements for the Dynamic Theater. Before the dance session, dancers tried out the headset and experienced the virtual stages in the physical location firsthand. Performances took place on a green-screen stage built specifically for this project. Dancers did not wear additional technology; instead, I used 5 depth cameras, 2 DSLR cameras, and 2 action cameras. In the following sections, I elaborate on equipment used, and the dance capture process.

\paragraph{Equipment and Green Screen Setup}

The dance capture process spanned 5 days, including setup, testing, and 2 days of recording with all dancers and the choreographer. Setup took 6 hours, with 3 graduate student helpers present. They operated cameras and ensured dancer safety by maintaining the cleanliness of the green screen dance floor mat.

The dance studio was equipped with 85-inch TV screens on each wall and a sound system, which played music and showcased the virtual stage during the sessions.

For video capture, I used 3 Nikon D7500 DSLR cameras. One camera was positioned at the center, 7.5 $m$ away from the screen, with a wide-angle 24 $mm$ lens to capture the entire width of the green screen. The other 2 cameras were placed at a 45-degree angle to capture different perspectives. Additionally, 5 Intel RealSense D455 depth cameras were positioned between the DSLRs but closer to the screen, approximately 5.5 $m$ away.

The dance stage featured 1.27 $cm$ thick green dance mats securely affixed using flooring adhesives and sealers. The stage measured 2.9 $m$ in width and 14.6 $m$ in length, with an additional 11 $m$ extension of the green screen backdrop. The dance mats extended 1.8 $m$ beyond each end of the screen to accommodate user footwork preparation. The screen, measuring 2.6 $m$ in height, was safely fastened to the wall structure of the dance studio.

\paragraph{Dance Capture Session}

The dance capture session and user study received the relevant ethics commission and human subject approval. Permission was obtained from the university's dance department to conduct the project. A Zoom meeting was held among the authors, tech specialists, helping crews, 6 dancers, and a choreographer to discuss dance moves, dress code, the story's overall arc, and the theme of the virtual stage. The meeting clarified the data collection process and the intended use of the captured video and data, while respecting the dancers' opinions. The choreographer and authors reviewed the story line, cue sheets, and shared their vision, considering the green screen setup possibilities. The dance capture sessions took place over 2 consecutive weekdays from 9 am to noon, totaling 6 hours. Sessions included warm-up, stretching, a review of the story board and cue sheets, and rotating dancers in groups of different sizes. Finally, consent and talent release forms were collected along with an exit interview.

\begin{figure}[t]
\centering
\includegraphics[width= 1.0\textwidth]{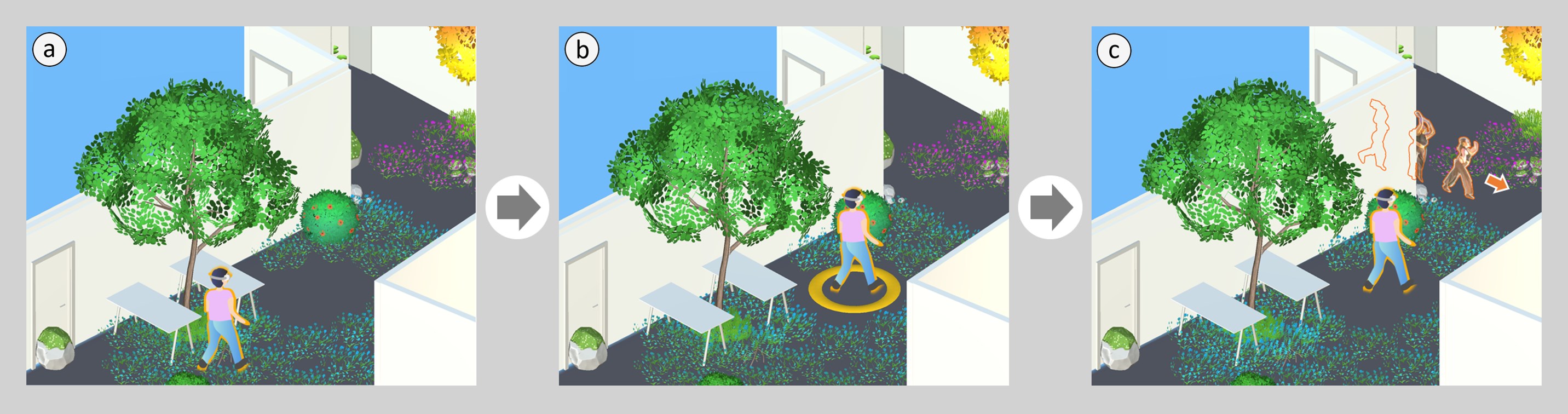}
 \caption{The location-based trigger allowed the viewer to experience the performance as the choreographer intended. The location-based trigger remains invisible and users are unaware of its presence while navigating (a). Upon entering the predetermined trigger spot (b), the dancers emerge (c), often from behind the walls, revealing themselves to the viewers.}
 \label{fig:locationtrigger}
 \end{figure}

\subsubsection{Digital Content Placement}
To enable active decision-making in dancer placement, virtual stage prompts, and location-based event triggers, an accurate floor model and a digital twin of the designated AR play area were needed. The digital twin was created using the Matterport 3D camera system, Apple's LiDAR scanning, Blender 3D, and the Unity game engine. This digital twin served multiple purposes, including visualizing the virtual stage in conjunction with projected objects and handling occlusions on the HoloLens device. Discussions with the choreographer were conducted to envision dance moves and determine dancer placement, considering the physical layout and content simultaneously. This approach allowed for strategic placement of content zones, location triggers, and stage advance triggers along the users' natural walking path (Figure \ref{fig:locationtrigger}). The nine dance content zones were positioned with distances ranging from 10 to 15 $m$ apart to facilitate exploration and navigation.

Azure Spatial Anchors were utilized to achieve precise content alignment during startup on my HoloLens 2 device~\cite{buck2022azure}. The saved spatial anchors from the server were automatically loaded to position digital content, in turn handling occlusions caused by physical walls.

Location-based triggers, invisible to the user, were strategically positioned along pathways with a radius of 1$m$. These triggers activate corresponding content when the user enters the designated collision zone, resulting in the deployment of dance performances. Dancers fade in and can be concealed behind a wall when utilizing the building structure.

\subsubsection{Dancers Placement}

To achieve the volumetric reconstruction of dancers, I initially intended to use my RealSense depth video frame data. However, this approach proved to be computationally intensive for the HoloLens 2 due to its hardware limitations, such as the Qualcomm Snapdragon 850 mobile processor and limited accessible RAM (2 GB) for external applications. Considering my goal of including detailed accessories and multiple sets of dancers on the virtual stage, which required loading full physical models, animated objects, and music, I needed to manage the additional computational costs. As a result, I opted for a 2D billboarding technique to simulate the three-dimensionality of the dance performance. This involved rendering captured dancers, a volumetric virtual stage, dynamic particle systems, and other interactive components in 3D, creating the illusion of depth when combined. Spatial audio, another crucial element for perceived 3D depth, was implemented using a separate audio source with the built-in audio spatializer, enhancing users' spatial attention in the content-rich environment.

While considering the use of avatars and skeletal animation keyframes derived from video frame data using body tracking, I prioritized maintaining the visibility of the actual dancers. Therefore, I decided to utilize 4K video footage captured from the DSLR cameras.

The captured videos were processed to extract the dancers by employing background removal techniques such as luma keys and opacity masks. This resulted in the creation of videos with a transparent background. To preserve transparency, the edited videos were exported in the {\small WebM} format, which supports an alpha channel. Within my development pipeline, these videos were integrated into materials using a fade blending mode, opaque render type, and a d3d11 vertex shader to produce the intended hologram effect.

\begin{figure}[t]
\centering
\includegraphics[width= 1.0\textwidth]{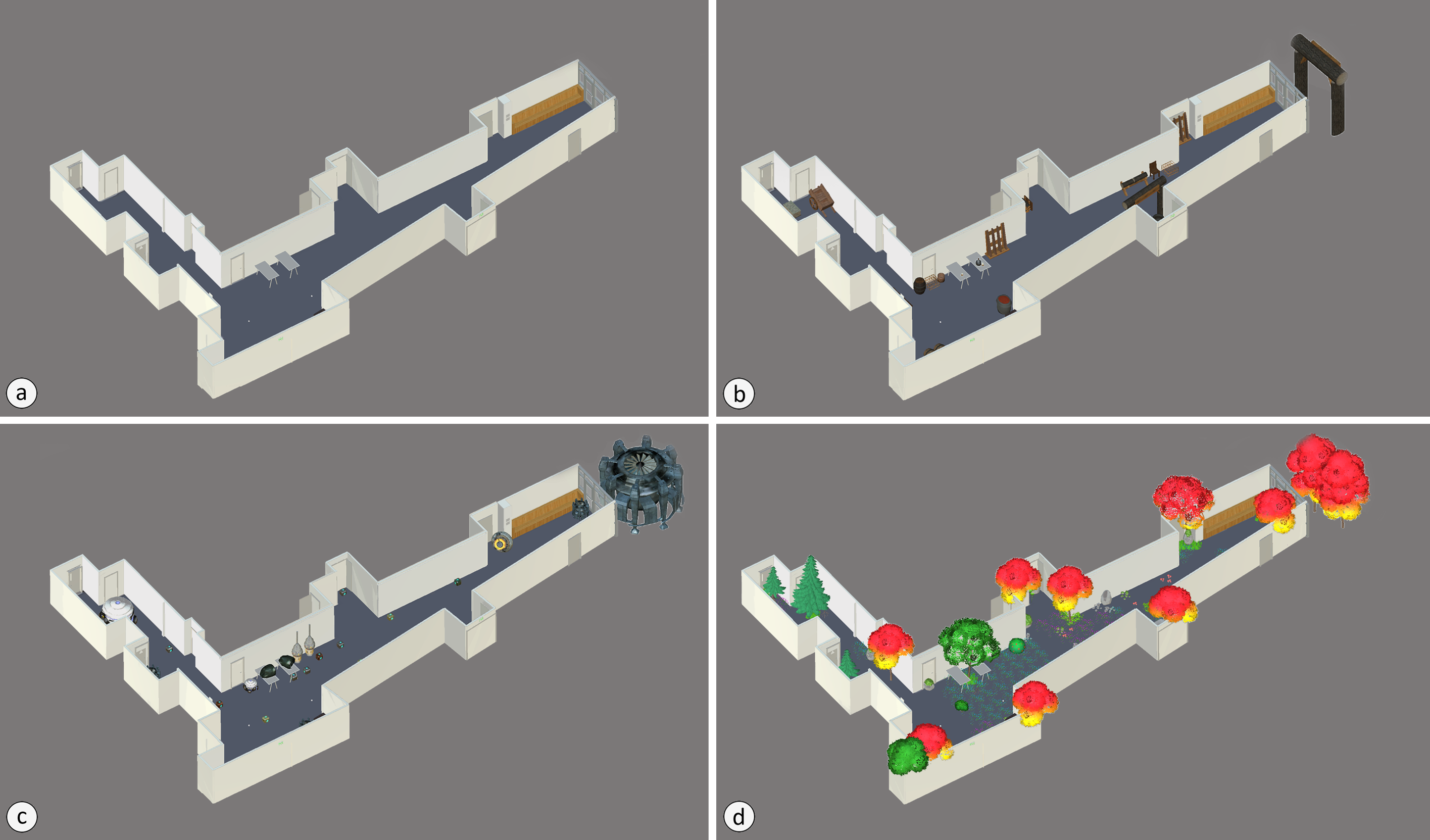}
 \caption{This figure shows an aerial view of (a) the digital twin of the physical layout and the physical layouts of (b) the ``Future”, (c) ``Fantasy”, and (d) ``Forest” stage prompts.}
 \label{fig:all-scenes}
 \end{figure}

\subsubsection{Guidance Systems}

To provide effective and non-intrusive guidance during the interactive experience, I implemented two types of guidance systems: a particle system and a traditional arrow system. These systems directed users to the content zones, blending with the overall experience and enhancing user engagement. My goal was to offer subtle cues without rushing the user, allowing them to fully enjoy the surrounding content. Both the particle system and arrow system were used as the primary navigational aid, guiding users towards the dancers or the Stage Advancement Trigger.

\begin{figure}[t]
\centering
\includegraphics[width= 1.0\textwidth]{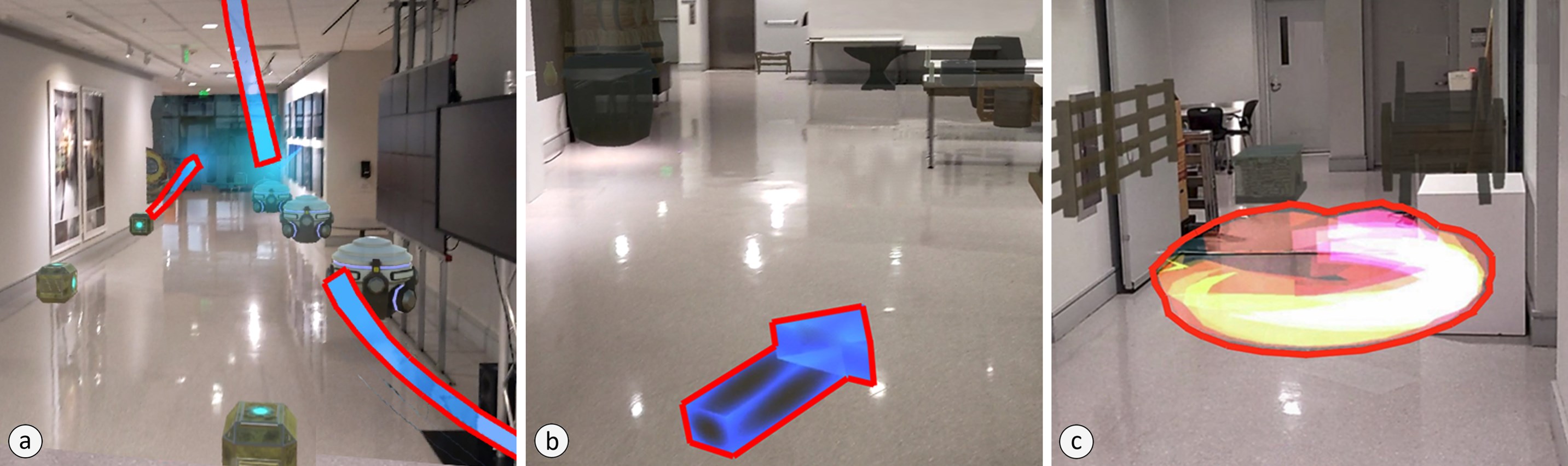}
 \caption{Visual display of my three types of navigational aids. (a) The particles: a long, visible streak of light is emitted by each particle for better view. Particles move towards the general direction of the target, with slight variations in their trajectories (b) The arrow: a typical and straightforward guidance mechanism. (c) The spiral: serves as a transition, advancing users to the next stage or ending the study.}
 \label{fig:guidance-systems}
 \end{figure}

\subsubsection{Particle Guidance}

Particle systems have enhanced the integration of visual effects in theatrical performances, particularly in choreography. For instance, Visual Flow effectively combines particle effects with dance performances, eliminating the requirement for dancers to synchronize with pre-recorded effects \cite{brockhoeft2016interactive}. My system takes a similar approach, combining particles with dancers in a digital theater within an indoor physical environment not specifically designed for performances. Users observe one to six guidance particles gracefully floating towards the target location, appearing at the outer periphery of their vision (Figure \ref{fig:guidance-systems}a). This provides a steady flow of indicators without overwhelming the screen. To ensure clear directionality, a colored streak marks the particles' trail, extending only a short distance to avoid clutter. Randomness is applied to the particle movement, simulating the appearance of naturally glowing insects or sparks to blend with the environment. The particles' spawn point is also randomized within the user's head position which introduces slight variations in their path towards the target. To counteract any deviation caused by accumulated noise, the velocity direction of the particles is reset to point directly towards the target every second.

\subsubsection{Arrow Guidance} 

The second guidance system I implemented is the arrow guidance system (Figure \ref{fig:guidance-systems}b), which uses a simple mechanism of pointing towards the next content zone. The arrow remains positioned 40$cm$ above the ground and 2$m$ away from the user's head, appearing at the center of the user's field of view (FOV) within the HoloLens display. The blue 3D arrow, measuring 30$cm$ in length and 10$cm$ in width, indicates the direction in which the user should walk. As the user reaches the content zone and a dancer appears, the arrow gradually fades away until the dance performance is completed.

\subsubsection{Other Spatial Guidance}

In addition to the arrow guidance system, I enhanced the user's guidance through other spatial cues. A spatial audio system played louder music from the content zone, creating an audio cue to direct the user's attention. Furthermore, a red light spinning spiral (Figure \ref{fig:guidance-systems}c) is positioned at the desired location, serving as a marker for the user to move to in order to transition to the next scene. This spiral system remains invisible and inactive until all the dance sequences within the content zone have been observed, ensuring that users do not transition prematurely. It functions as a cue, similar to the closing of a theater curtain before revealing a new stage.

\begin{figure}[t]
\centering
\includegraphics[width= 1.0\textwidth]{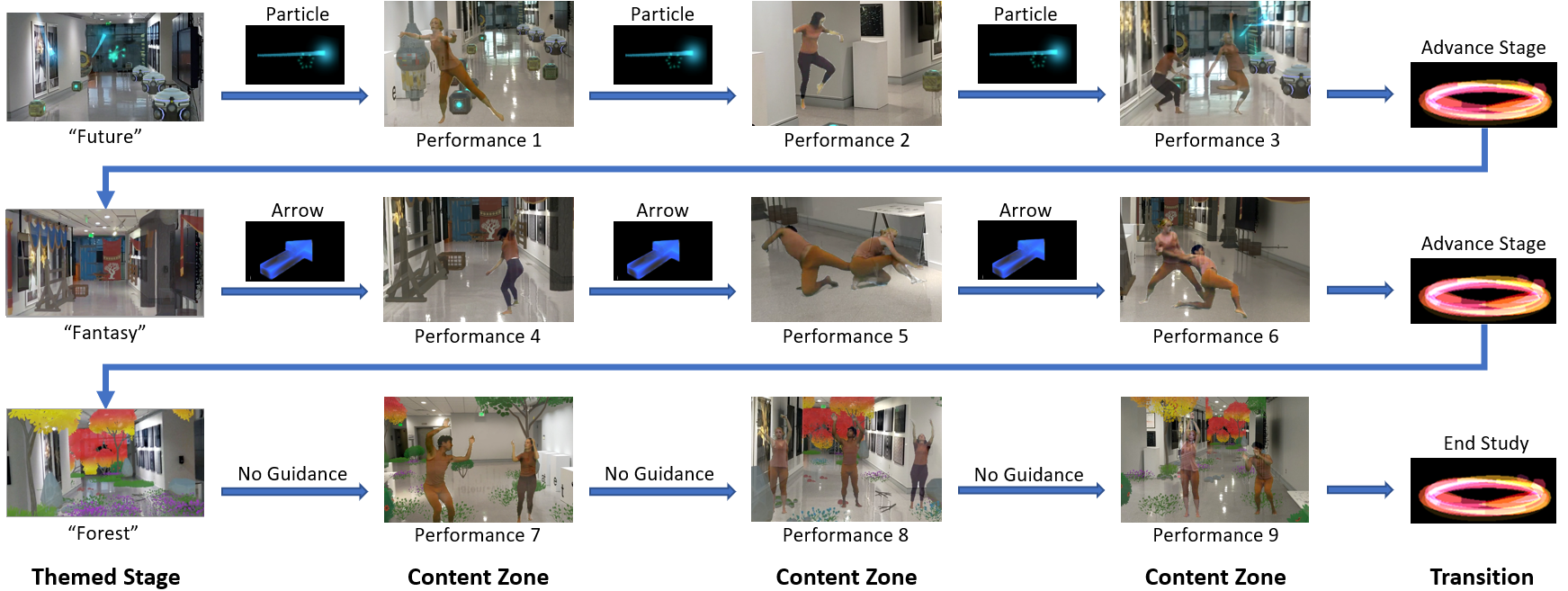}
\caption{Overview of the user experience trajectory. Within one physical space, users maneuver through three themed stages aided by a guidance system (``Particle” or ``Arrow”) for navigation during dance performances. Users are directed to the ``Content Zone” for virtual dance performances.}
\label{fig:UX}
\end{figure}

\subsubsection{User Experience Trajectory}
The play consists of three themed virtual stages: Future, Fantasy, and Forest, each with three content zones featuring 10- to 30-second dance sequences. My rationale for choosing these vastly distinct virtual stage environments was to illustrate the diverse experiences permitted in the same physical space. The design aims for a 7-minute experience, with users spending 2-3 minutes on each stage. Guidance is provided in two stages, while one stage is navigated without guidance. The order of presentation was Future, Fantasy, and Forest, with varied guidance system. After watching all three dance performances in the final stage (Forest), the spiral advance location trigger appears, concluding the trial and user study. The Interactive Narrative, displayed in Figure \ref{fig:UX}, includes spiral indicators for stage advancement and play conclusion.

Each stage was comprised of three choreographed performance segments. Users activated each segment individually by following guidance indicators. When a performance concludes, the spiral system appears, and the guidance indicators shift focus to guide the user towards it.

\subsubsection{Experiment}
This experiment examined user navigation and completion time in an AR theater play. Feedback and insights from the experiment improved the user experience of narrative interaction. Drawing on previous work in interactive narrative and AR theater projects, which explored innovative ways to interact with extended reality, as well as users' responses to spatial arrangements, I formulated the following hypotheses:
\begin{itemize} 

    \item H1: Dynamic Theater provides an enjoyable experience and creates the perception of an extended passage of time.
    \item H2: The guidance system facilitates users' natural locomotion and content viewing without compromising the AR experience.
    \item H3: The particle guidance system in Dynamic Theater outperforms a traditional arrow as a guidance mechanism, offering effective navigation and visual appeal.

\end{itemize}

\subsubsection{Design}

Users explored all three virtual stages using both guidance systems (particles and arrows) in a counterbalanced manner. To mitigate the order effect, users were randomly assigned labels A or B, in equal proportion. Label A included a future stage with a particle indicator, a fantasy stage with an arrow indicator, and a forest stage with no indicator. Label B had a future stage with an arrow indicator, a fantasy stage with a particle indicator, and a forest stage with no indicator. The trial proceeded without intervention to measure the play's length. On average, users completed the experience in 5 minutes and 5 seconds (SD = 48.27 seconds), with each stage designed for approximately 100 seconds. Typically, users took 4 to 7 minutes to finish the entire play. The walking distance per stage was approximately 53 m, with a total distance of around 150 meters for the entire play.

\subsubsection{Apparatus}
The experiment took place in a 208.54 $m^{2}$ (2,244 $ft^{2}$) indoor space, where participants were instructed to stay within a designated area. Compliance with this rule was observed. Three Philips Hue Go Bluetooth Controllable Portable LED lamps, with a diameter of 15 $cm$, were arranged in a circular pattern. The Microsoft HoloLens-2 Mixed Reality headset was used, employing low-polygon-count 3D models with varying levels of detail to manage computational load. The HoloLens utilized single-pass instanced rendering and vertex-lit shading techniques.

When designing the virtual stages, I followed guidelines to create distinct experiences that differed from the physical space. Each stage consisted of approximately fifty items, including large, medium, and small objects. The goal was to cover around 20\% of the environment in each stage with projected prompts, ensuring that the entire wall was not completely covered.

\subsubsection{Questionnaire}
Dancers completed a comprehensive questionnaire consisting of four sections and 25 questions (11 Likert-scale items and 14 open-ended responses). The questionnaire covered topics such as the green screen setup, dance schedule, assistive media and technology, and instruction. To explore the audience reception of the AR experience, a separate user study was conducted with 20 participants, involving 27 questions (17 Likert-scale items and 10 short answers), to gather insights.

\subsubsection{Participants}

For the audience study, I recruited 20 adult participants (9 men, 11 women), ranging in age from 18 to 34. Six participants used vision correction, and all reported normal or corrected-to-normal vision. 15 participants had no prior experience with augmented reality. All participants reported normal walking abilities and were able to walk normally during the trial.

\subsubsection{Procedure}

Upon arrival, participants received a guided tour of the walking area, emphasizing the virtual fence location for safety. Informed consent forms were signed, and a pre-questionnaire capturing demographic information was completed. Participants were informed about the guidance indicators leading to dance performance segments and encouraged to walk continuously at a normal pace. Portable LED lamps emitting distinct colors had been placed randomly along the hallway but were not mentioned.

The HoloLens 2 was activated, and Azure Spatial Anchors ensured accurate alignment between the virtual scene and physical environment. Participants wore the headset while a researcher remained nearby for safety and observation. 

At the end of the performance, an exit interview and questionnaire were conducted to gather feedback and insights. No instances of discomfort or dizziness were reported.
Participants completed a questionnaire assessing the guidance system's visual and functional qualities and their perception of scene duration. Additional questions covered the dance performance and participants' recall of the floor lamps. The entire procedure lasted approximately 25 minutes per participant.

\subsection{Results}
Data was collected on dance performance, perceived play time, user ratings of the AR experience, and guidance systems. Feedback from the dancers and the choreographer was also gathered. Paired t-tests and ANOVA tests were conducted to compare the results between the different guidance indicator types using a significance level of $\alpha=0.05$.

\begin{figure}[t]
\centering
\includegraphics[width= 1.0\textwidth]{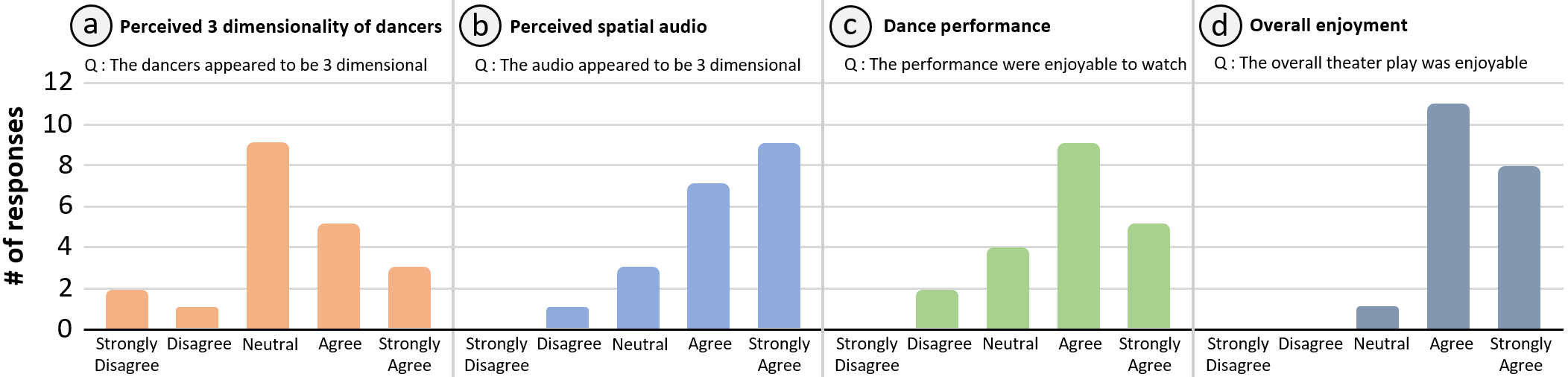}
 \caption{User responses to questions regarding their evaluation of the environment on a five-point Likert scale (1 = Strongly disagree, 5 = Strongly agree). (a) Q: ``The dancers appeared to be 3D.'' The mean response was 3.3. (b) Q: ``The audio appeared to be 3D.'' The mean response was 4.2. (c) Q: ``The performances were enjoyable to watch.'' The mean response was 3.85. (d) Q: ``The overall theater play was enjoyable.'' The mean response was 4.35.}
 \label{fig:user-eval}
 \end{figure}

\subsubsection{User Evaluation of the Dynamic Theater}
Figure \ref{fig:user-eval} presents plots of user responses on a 5-point Likert scale regarding their experience of viewing the AR dance theater performance. The question about the visual appeal of my virtual dancers yielded a mean response of 3.85, with 70\% of respondents agreeing or strongly agreeing that the performance was interesting and enjoyable to watch (see Figure \ref{fig:user-eval}c). The mean response to the question ``The dancers appeared 3D'' was 3.3 (Figure \ref{fig:user-eval}a), and for ``The audio source seemed 3D'', the mean response was 4.2 (Figure \ref{fig:user-eval}b).

The overall rating users assigned to the theater experience was 4.35 (\ref{fig:user-eval}d). One user expressed, ``This experience has proven to be surprisingly refreshing. The innovative approach, to me, ranks much higher than existing and conventional performances.'' Another user's favorite aspect was ``exploring a virtual world while having the real world as a palette.''

The project was well received by users, who voiced their individual tastes and enjoyment of various features of the play, which aligned with the Dynamic Theater project's concept of catering to individual experiences without rushing. As one user stated, ``I enjoyed walking through the forest scene due to the lack of restrictions.'' Another user found joy in ``going into the teleportation spiral,'' despite its relatively smaller role in the overall experience. The combination of virtual flowers and dancers left a lasting impression, with one user noting, ``Seeing the virtual flowers and dancers together was something else, it was great.'' The dynamic nature of the music, changing with the dancers' entrance into scenes, was particularly enjoyable for one user, while another appreciated ``the music and its ability to change based on location.'' The simple act of following the particles provided a sense of exploration, and the familiarity of virtual stages in known places was an intriguing aspect noted by one user. Additionally, the use of stereo audio added a layer of fun and immersion, as described by a user who delighted in ``spinning my head listening to the sound move'' and likening it to ``listening to 3D audio.''

\subsubsection{Exploration Time}

The average play time duration per stage was 100.54 seconds, with a perceived time of 165.83 seconds, demonstrating a significant overestimation of time (p-value = $2\times10^{-9}$). Users overestimated the experience duration in 85\% of trials, often perceiving nearly twice the actual duration. Remarkably, users estimated a perceived duration over double the actual time in 41.66\% of cases. These results support previous findings that increased immersion leads to an overestimation of elapsed time \cite{awe, FLAVIAN2019547}.

\begin{figure}[t]
\centering
\includegraphics[width= 1.0\textwidth]{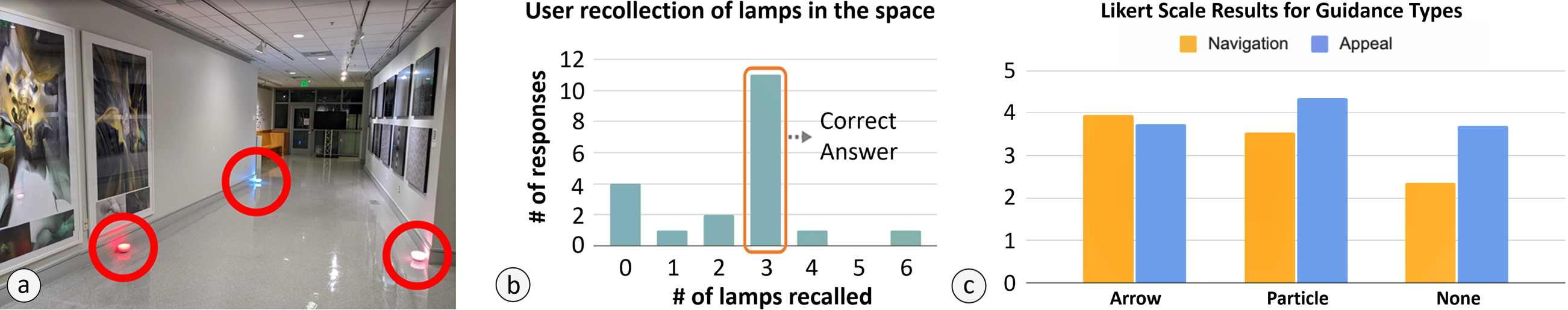}
 \caption{(a) A view of the physical environment with the three lamps on the ground. For each user, lamps were arranged and placed in random locations. These lamps were used to test users' ability to recall the physical layout, as the users were not given any instruction regarding the floor lamps. (b) Bar graph of users’ answers to how many lamps they noticed on the ground without having been alerted to them beforehand. The average answer is 2.4, with 55\% recalling the correct number. (c) User ratings for the guidance systems. The averages for navigation and appeal, shown here, are based on the users' ratings of the arrow, the particles, and the no guidance conditions. Based on these ratings, users enjoyed the experience (consistent appeal) despite their fluctuating opinions with navigation.}
 \label{fig:lamp-setup}
\end{figure}

\subsubsection{User Evaluation of Guidance Systems}

Figure~\ref{fig:lamp-setup}c shows average user ratings for each guidance system regarding navigation utility and appeal. For navigation, an ANOVA analysis showed significant differences across the arrow group, the particle group, and the no guidance group (F-value = 9.002, p-value $<$ 0.001). Bonferroni post-hoc analysis indicated significant differences between the arrow and no guidance groups (t-value = 3.448, p-value = 0.002), and the no guidance and particle groups (t-value = -3.866, p-value $<$ 0.001). No significant difference was observed between the arrow and particle groups ($t$-value = -0.418, p-value = 1.000).

Another ANOVA analysis categorized by appeal also showed significant differences across the three groups (F-value = 3.509, p-value = 0.037). However, Bonferroni post-hoc analysis did not reveal any significant differences. Between the arrow group and no guidance group, the difference was $t$-value = 0.183, $p$-value = 1.000. Between the arrow group and the particle group, the difference was $t$-value = -2.197, $p$-value = 0.096. Finally, the difference between the no guidance group and the particle group was $t$-value = -2.380, $p$-value = 0.062. These results suggest that both guidance systems had aspects that supported natural user interaction with the performances throughout the various locations in the experiment area.

My findings confirmed hypotheses 1 and 2, suggesting users enjoyed the experience and perceived a greater amount of time as having passed. Additionally, the guidance systems proved helpful in locating digital content. However, hypothesis 3 was not confirmed, as both the particle system and the arrow may have presented distinct advantages.

\subsubsection{Feedback from the Dancers}

I gathered data from dancers using a 25-question survey, comprising 11 Likert-scale items and 14 open-ended queries. 

\paragraph{Choreographer}

Regarding the use of AR headsets, the choreographer emphasized the importance of allowing the dancers to explore and discover the possibilities and limitations themselves in AR headsets, which is something I did not do enough. A few weeks before the dance session, the choreographer conducted a one-hour wide-scale AR demo to see the virtual stage. This provided them with a better understanding of the augmented reality environment and what could be achieved. Without this first-hand experience, the dancers felt it would have been challenging to contribute to the choreography. The choreographer noted a desire to have a dance studio TV showing storyboards, sketches, and cue sheets so that the dancers and choreographer could maintain the session's flow while enabling them to review and plan their footwork and dance sequence.

\paragraph{Dancers}
When asked whether background music or virtual stage graphics on big TV screens in the studio would assist in their engagement, the dancers' responses were evenly divided. Three dancers strongly favoring background music, one dancer strongly favoring visual stage imagery, and two dancers preferring visual stage graphics over background music, enabling them to maintain creative engagement in their dance.

The question posed to the dancers regarding the input they received during the dance session was: ``In the dance session, dancers were given input from both the choreographer and the tech specialist. Input from which source was more helpful in allowing you to dance for the imaginary environment?'' Based on the findings, it was revealed that the dancers perceived the instructions from the choreographer as considerably more helpful in facilitating their dance performance for the imaginary environment compared to the input received from the tech specialist. Furthermore, two dancers specifically highlighted the potential benefit of having access to the virtual stage imagery earlier, allowing them adequate time for thoughtful contemplation and strategic planning of their movements within the virtual space.

Despite this, the presence of both the choreographer and tech specialist was deemed necessary by the dancers, as the choreographer assisted in translating the desired mood and tempo for synchronized group performance, while the tech specialist provided the overarching objectives and goals. The dancers expressed appreciation for the provided landscape images, as they greatly aided in creating an imaginary world to dance and choreograph within. 

All six dancers agreed that despite initial concerns about its spatial limitations, the area provided for the project was indeed ideal, with four dancers expressing that it even exceeded their expectations. The dancers acknowledged the value of adapting to spatial limitations, emphasizing the opportunities for creativity and exploration that arise from working within given constraints. Special limitations can actually be beneficial to the choreographic process, presenting an opportunity to explore and play with the dynamics of space. One dancer aptly expressed this sentiment, stating, ``Our job as dancers is to accommodate to our space and choreograph, improvise while working with the given space. The spatial limitations can actually be beneficial to the choreographic process and serve as a creative play of space.'' Furthermore, the inclusion of a padded floor was well-received by the dancers as it offered a balance between sufficient softness for floor movements and adequate traction for more energetic choreography. 

All six dancers expressed their desire to participate in future AR dance projects, as they found the experience inspiring and believed it positively influenced their own dance practices. One dancer specifically mentioned the unique aspect of focusing solely on the dance itself and being removed from the physical space where the performance would ultimately take place. Another dancer highlighted the joy of creating a new world to dance in and the rewarding nature of the experience. 

\subsubsection{Discussion}

The study's purpose was to investigate how audiences consume AR dance performances while walking around in the virtual stage at their own pace. By timing user experience duration in each stage and comparing it with user perception of time, I was able see a noticeable greater perceived passage of time. Detailed pre-and post-study questionnaires also assessed subjective ratings of the experience using each guidance system to lead the integrative narrative. 

\subsubsection{Impact on Dance Theater}
My design successfully captivated user engagement. The significant discrepancy between perceived and actual time revealed during the exit interview suggests users' engagement and immersion in the Dynamic Theater system. Using location-based triggers, I created diverse environments and sequentially offered immersive dance experiences within several virtual scenes in a single indoor space.

My system provides a unique level of interactivity by letting users naturally explore different content zones and interact with virtual performers and the stage, which is not typically enabled by traditional stages. My AR theater, incorporating the physical layout into the virtual stage design, leveraged the existing infrastructure to create a seamless transition between physical and virtual content. 

\subsubsection{Use of Guidance System in AR Theater}
The user study data and exit interviews suggest that the particle system, though visually appealing and effective, isn't as intuitive as the arrow for guidance. The arrow had a slight edge in providing clear directions. Its direct, accurate navigation path may be essential for user safety in virtual environments like DreamWalker \cite{yang2019dreamwalker}. However, I aimed to create an experience that compliments both content and usability. In the exit interview, five users expressed difficulty in locating the dancers without any guidance system, confirming that using a guidance system while watching a performance does not diminish enjoyment or appeal when implemented correctly.

\subsubsection{Supporting Dancers}

Significant creative research supporting and involving actors and dancers as active participants in immersive media has appeared in recent years~\cite{support1, support2, kyan2015approach}. My AR theater project benefited greatly from the direct contributions of dancers, leading to valuable learning experiences for both the artists and the development team. One noteworthy example is when the dancers proposed a dance routine that involved gesturally creating a tree in response to a virtual forest scene. This approach resonated with the dancers, who expressed a desire for more opportunities to create and explore similar ideas in future opportunities.

Dancers and choreographers emphasized desire for more involvement in the development of immersive technology projects. They highlighted the importance of not underestimating dancers' comprehension abilities. As one choreographer stated, dancers possess problem-solving skills and an intuitive understanding of space and timing. Their ability to perceive subtle movements and changes, as well as envision how the audience will experience the performance and utilize the space and objects, stems from years of training. Therefore, it is crucial to allow dancers to engage with the main technology, such as AR, and provide them with opportunities to explore and take leading roles in these projects.

\subsection{Conclusion}
My research fuses wide-area augmented reality, dance, and theater, offering a framework for immersive, interactive dance experiences. I created a location-based theater where users navigate indoor space, interacting with dancers in the augmented stages. My study emphasizes a guidance system and location-centric approach, curating user movement, engagement, and focus for a seamless and enjoyable experience. Thus, my immersive theater exemplifies an innovative AR application, presenting a novel interactive platform for exploring performing arts and spatial computing.

To overcome hardware limitations and deploy several dancers at once, I employed the billboarding technique. However, this method introduces potential issues as users naturally move around during interaction. Abrupt shifts in user positioning can compromise spatial awareness, momentarily distorting the three-dimensionality. Furthermore, my experiments took place in a corridor-shaped physical environment, which suggests particular walking directions. As such, my findings may be applicable to similar environments, with wider space settings yet to be explored.

I explored location-based AR dance theater, emphasizing guidance systems and content placement to craft curated user experiences. I assessed the guidance system in tandem with my main content and performance as users navigated the space. This system potentially serves as a blueprint for open-world AR applications, recognizing that guidance system selection may vary based on the designers' goals.

\begin{section}{AR Theater Project 2: Audience Amplify}
In this second part of the project, Audience Amplified~\cite{kim2024audience}, I introduced a personalized ``always-ready'' social experiment. This work was presented at the 23rd IEEE International Symposium on Mixed and Augmented Reality (ISMAR), held in the Greater Seattle Area, USA, in October 2024. During my discussions, I developed the concept of an adaptive virtual audience for an AR theater play, allowing users to explore and discover diverse performances throughout the environment. In this section, I expand on the ideas initially discussed during the planning stages, illustrating what an always-ready social experiment might look like with spatially aware and trained AI agents playing a supporting role to enhance user experience and the feeling of a live event in an interactive narrative.

Audience reactions can considerably enhance live experiences; conversely, in anytime anywhere AR experiences, large crowds of people might not always be available to congregate. To simulate live events with large audiences more closely, I created a mobile AR experience where users can wander around naturally and engage in AR theater with virtual audiences trained from real audiences using imitation learning. This approach allows me to carefully capture the essence of human imperfections and behavior in artificial intelligence (AI) audiences. The result is a novel mobile AR experience in which solitary AR users experience an augmented performance in a physical space with a virtual audience. Virtual dancers emerge from the surroundings, accompanied by a digitally simulated audience, to provide a community experience akin to immersive theater.

In a pilot study, simulated human avatars were vastly preferred over just audience audio commentary. I subsequently engaged 20 participants as attendees of an AR dance performance, comparing a no-audience condition with a simulated audience of six onlookers. Through questionnaires and experience reports, I investigated user reactions and behavior. My results demonstrate that the presence of virtual audience members caused attendees to perceive the performance as a social experience, increasing their interest and involvement in the event. On the other hand, for some attendees, the dance performances without the virtual audience evoked a stronger positive sentiment.

Mixed reality (MR) has gained significant traction due to its ability to offer a flexible and convenient way to engage with content and interact with others~\cite{feiner1997touring, yang2019dreamwalker}. Social VR gatherings and concerts (such as in VRChat) have demonstrated a desire for virtual meetups around artistic events. The COVID-19 pandemic, with its demand for isolation and personal separation, triggered the need for new creative ways to address connection. Technological advances in AR and machine learning opened pathways to experience previously recorded performances at the convenience of the participant's time and space while recreating an aura of a social gathering through AI-enhanced audience simulation. My work demonstrates and evaluates steps in this exciting new direction of socially amplified asynchronously performed AR theater.

In-person immersive theater has influenced the design of virtual theatrical experiences, where participants collectively view content online in a shared virtual space~\cite{game2021murder, game2021shrine}. Many of these experiences allow users to pursue various activities concurrently, all while enjoying the distributed content. This contrasts current trends in immersive theater with a directed narrative, such as productions like \emph{Dear Angelica}~\cite{oculus2017dear}, which requires full immersion, undivided attention, and active interaction from the audience.

While having multiple users in the same space (real or virtual) is the best way to create shared participatory experiences, it could also be challenging to coordinate if users are not all participating concurrently. Including virtual characters in place of other users offers a possible solution. The usage of avatars to represent remote users or Non-Player Characters (NPCs) in mixed reality has been a prominent topic of research in recent years~\cite{piumsomboon2018minime, ho2022perspective}. Other research explored how users interact with crowds of people, walking together to understand the effects of virtual crowds on natural locomotion~\cite{trivedi2023human, trivedi2023human}. Additionally, researchers have sought to enhance the authenticity and engagement of mixed reality narratives by incorporating AI-trained characters~\cite{merrick2007reinforcement, merrick2006motivated, campo2023assessment}. Recent virtual reality games like Modbox, allow users to interact with GPT-3 powered NPCs~\cite{article2021npc}. While AI research in Human-Computer Interaction (HCI) has seen increased efforts, its practical application in generating real-time adaptive narratives, automating character behaviors, and dynamically adjusting virtual environments based on user interactions remains relatively unexplored. Height is another intriguing factor, given that the experience of attending live concerts can be strongly impacted by the height of viewers amidst the crowd. While height manipulation in virtual reality (VR) can change stress levels in task-oriented situations~\cite{macey2023feeling}, relatively little research has been done on the relationship between user height and interacting avatar heights. As height can directly influence task-based experiences, I felt it was important to account for how this affects the desire to watch an experience.

My work adopts a participatory open-world exploration format to grant users the freedom to pursue the narrative at their own pace. This aligns with the shared VR experiences in widely used applications such as VR Chat and Rec Room. 

I employed AI audiences trained on human audience data to simulate an immersive AR theater/concert experience where viewers can stroll and watch dance performances with AI crowds. 

While previous works investigated how crowds can be simulated and affect our experience~\cite{mel2022sentiment, yang2019dreamwalker, trivedi2023human}, I aim to bring live entertainment experiences one step closer to being ready to be experienced anytime, anywhere. I present a novel mobile AR dance performance showcase that adapts as users move through a designated area while featuring virtual dancers. This supports previous research which established that the movement and behavior of virtual audience avatars and crowds directly affect the participatory experience~\cite{yakura2020enhancing, trivedi2023human}. 

I conducted two pilot studies to determine the best appearance and behavior of the audience avatars for this experience. The findings from these studies were used to design the main AR dance performance. The main user study explored how navigation and virtual audiences affect engagement, enjoyment, and perceived realism. Users experienced the dance performance under two conditions: with and without virtual audiences. Semi-structured interviews, the SentimentAnalysis package~\cite{feuerriegel2018package}, 150-word experience essays, free-form essay responses, and analysis of questionnaire responses were employed to understand user experiences and reactions to the audience avatars. My findings indicate that virtual AI audiences trained for specific tasks can enhance MR theatrical encounters, fostering a more socially engaging experience with increased participant interest and involvement.

\begin{itemize}
    \item Participants had a more positive experience when watching the performance with a virtual audience and also spent more time in this condition. This suggests increased engagement when virtual audiences are present.
    \item Participant experience differed depending on which of the two conditions they saw first, with a more positive reaction overall when virtual audiences were included after first watching the performance alone. This indicates the importance of careful introduction of characters into the interactive narrative.
    \item The user's height emerged as a factor influencing their overall experience. This finding highlights the importance of considering physical attributes and user perspective in the design of immersive MR experiences.
\end{itemize}

I see great possibilities for AR performances that can be experienced via mobile AR by individual audience members at their leisure, while still experiencing some form of audience connection. In the future, such an audience member's social viewing behavior could influence the performance experience of subsequent users.

\begin{figure}[t]
\centering
\includegraphics[width= 1.0\textwidth]{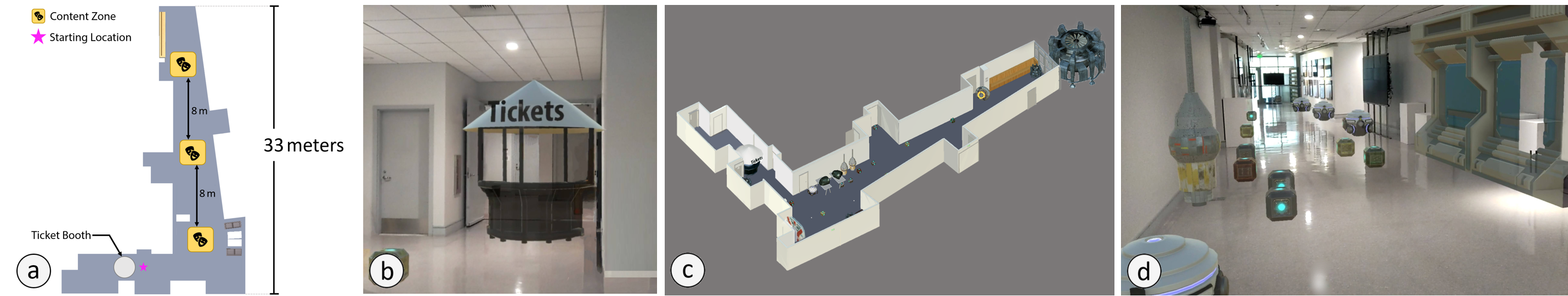}
 \caption{(a) Top view of the experiment area and its physical layout. The dimensions of the space are 208.54$m^{2}$ (2,244 $ft^{2}$). Also indicated are the three Content Zones, location of the ticket booth, and the viewer's starting location. (b) In the trial with audiences, virtual audiences would spawn in the vicinity of the ticket booth. (c) This figure depicts an aerial view of the spaceship-themed stage's physical and augmented virtual stage layout. (d) The physical surroundings and projected stage props are visible within the stage from the viewpoint of the viewer in the experimental area, as can be seen in this image.}
 \label{fig:2top-view2}
\end{figure}

\subsection{Related Work}
This section reviews immersive theater, the usage of NPC and ML-trained avatars, and user sentiment.

\subsubsection{AR Immersive Theater}
Recently, there has been a shift toward bringing theater experience to the mixed reality. Despite some overlap with traditional theater production, MR theater development presents novel challenges and opportunities~\cite{coulombe2021virtual, VR_SharedSocialSpaces}. My research draws inspiration from narrative theater experiences, such as \emph{CAVE}~\cite{CAVE_Leon}, \emph{CAVRN}~\cite{herscher2019cavrn}, \emph{Gulliver}~\cite{ARGulliver}, and the integration of physical props with AR features~\cite{nicholas2021expanding}.
My project builds upon these foundational works to create an interactive user-centered dance performance experience through walking. This aligns with my broader research focus on crafting experiences for specific locations and spaces~\cite{wang2022exploring, kim2023reality}. Notable examples of location-based experiences include Soul Hunter~\cite{weng2011soul}, ARQuake~\cite{thomas2000arquake}, HoloRoyale~\cite{rompapas2018holoroyale}, and exploration-based horror games~\cite{madsen2022fear}. The use of walking experiences in mixed reality extends to the engagement with historical events~\cite{fujihata2022behere, dow2005exploring}.

Since the research on outdoor exploration in wide-area virtual theater performances by Cheok et al. in 2002~\cite{cheok2002interactive}, various mixed reality theaters, such as \emph{Holojam in Wonderland}~\cite{gochfeld2018holojam}, \emph{The Meeting}~\cite{pietroszek2022meeting}, \emph{Dill Pickle}~\cite{pietroszek2022dill}, \emph{Samuel Beckett’s Play,}~\cite{Play_Leon} and \emph{Gumball Dreams}~\cite{lyons2023gumball}, have provided unique opportunities for active user engagement within the theater performance itself. These examples indicate the increasing importance of a player's involvement in production and development. Moreover, there is a growing trend of collective experiences where users gather to participate in interactive narratives, live concerts, and other events, sharing virtual spaces and moments~\cite{game2023japan, game2020capsule, CAVE_Leon}.

Within open-world platforms like VR Chat, users generate novel virtual spaces that provide theater-like experiences with immersive narratives~\cite{game2021murder, game2021shrine, game2023flow}. These user-generated spaces and experiences are shared through various platforms such as Reddit, YouTube, and VR infusers highlighting the growing popularity of user-generated content in the VR space~\cite{web2023reddit, web2023YT2}.

The Under Presents~\cite{game2019presents}, a multiplayer VR game and performance space, enables users to collectively view live or recorded performances, encouraging a shared virtual experience in a participatory theater experience. Although experienced entirely on the user's own, these shared spaces where audiences build experiences together are on the rise. The deployment of AI-trained audiences in my work, intended to capture the essence of live events within my interactive AR theater experience, is said to have been directly influenced by user engagement.

\subsubsection{The Usage and Effect of Virtual Agents}

In mixed reality environments, NPC audiences have traditionally found their primary use in video games, with some research dedicated to understanding their influence on user experiences~\cite{10.1145/2967934.2968092}. Emmerich and Masuch extended their research~\cite{watchmeplay} to examine the impact of real observers and virtual agents on gaming experiences, while Yee et al. \cite{10.1111/j.1468-2958.2007.00299.x} conducted two studies to investigate the effect of transformed self-representation on user behaviors. Similarly, Leyrer's study \cite{10.1145/2077451.2077464} showed that avatars could significantly influence the verbal estimates of egocentric distances during the tasks and the eye height had a significant effect on the verbal estimates of both egocentric distances and the dimensions of the room. Zhu et al. \cite{zhu2023} also determined that the appearance of interactive conversational avatars impacts user experience, comfort, and the ability to recall information from the AR experience.

A recent study by Kao~\cite{kao_effects_2021} further contributed to the exploration of virtual audiences by proposing the intentional incorporation of observation and surveillance based on text phrases within games. This approach demonstrated the potential to enhance players' performance, overall gaming experience, and motivation. Notably, Haller et al. \cite{10.1145/3290688.3290752} revealed that NPC audiences who cheered and applauded led to improved player performance. In addition, Xu et al.\cite{10.3389/fpsyg.2023.1079132} conducted research to investigate the effect of the size of NPC audiences and their feedback on user performance and gameplay experience. Their results illustrated that as the size of NPC audiences grew larger, the user's performance and gameplay was enhanced. Similarly, Yu et al. \cite{Yu2023} found that the presence of NPC audiences and their feedback can enhance elderly users' performance and gameplay experience, like competence, immersion, and intuitive controls. 

While the impact of virtual audiences in some video games and MR contents has garnered increasing attention, it remains relatively understudied in virtual environments and AR theater experiences. Consequently, I would like to understand the role and effect of virtual audiences on users' performance and experience in the AR theater environment.

\subsubsection{Evaluation of User Experience}
Assessment is important to understanding user experiences within virtual environments and AR settings. Here, I present an evaluation of methodologies crucial to my experiment.

There are different definitions of presence from previous research \cite{10.1162/105474698565686, 10.1162/pres.1992.1.4.482}. My study aligns most with Witmer's definition \cite{10.1162/105474698565686}, which defines presence as a subjective experience of being in one place or environment. He also stated that the user's self-reported sense of presence is an important metric to evaluate the effectiveness of virtual environments. Presence is evaluated by questionnaire \cite{lessiter2001cross, usoh2000using, 10.1162/105474698565686}, physiological responses~\cite{meehan2002physiological}, and in some cases using behavioral indicators~\cite{kisker2021behavioral}. In my study, I use a combination of post-trial questionnaires, with questions from the Game User Experience Satisfaction Scale (GUESS) \cite{doi:10.1177/0018720816669646}, the Immersive Virtual Environment Questionnaire \cite{10.1145/2927929.2927955}, the Augmented Reality Immersion (ARI) Questionnaire \cite{GEORGIOU201724}, and the Game Engagement Questionnaire (GEQ)~\cite{geq2013}, to assess user interest and boredom. Additionally, I utilize the Presence Questionnaire (PQ) \cite{10.1162/105474698565686} to collect quantitative feedback on the sense of presence experienced by participants in each trial letting them write 150 words or more about the experience and talk to interviewer freely for additional 5 minutes; gathering insights and reasoning behind the user experience.

The concept of immersion is often employed in VR narrative games~\cite{GEORGIOU201724}. Brown et al. \cite{10.1145/985921.986048} proposed a division into three levels, including engagement, engrossment, and total immersion. Georgiou et al. \cite{GEORGIOU201724} applied the same divisions to develop and validate the Augmented Reality Immersion (ARI) questionnaire, which is one of the main sources for building my questionnaire. This notion of immersion is not to be confused with Slater's definition of immersion, which refers to objectively measurable characteristics of technology, but instead aligns more with his definition of Presence~\cite{slater2003note, slater2023sentiment}. User sentiment is another important assessment.

\subsection{Immersive Theater Design}
My immersive theater concept transforms a 208.54$m^{2}$ corridor into a spaceship's hallway where users are free to walk about while watching a performance of a dancer through the Microsoft HoloLens-2 Mixed Reality headset. This section outlines the user experience in my study and explains the recording process of the dancer's performance, as well as the training of the audience using a machine learning-trained deep neural network (DNN).

\subsubsection{Location-based Immersive Theater}
I utilized the Unity game engine to develop my immersive AR theater experience featuring three content zones that showcase prerecorded contemporary dance performances through volumetric avatars. The experiment was conducted in an indoor space where users could move among the evenly distributed content zones with the help of a navigation aid. Upon reaching each zone, the corresponding recorded performance was played. Each session was designed to allow users to explore for approximately 4 minutes. 

\paragraph{{Navigational Aid}}
I added a guidance system to help users locate the next content zone (dance performance) and navigate the environment at their own pace. To streamline the experience, a guiding system with an arrow pointed toward the next destination (Figure \ref{fig:guidance-systems}b).

\begin{figure}[t]
\centering
\includegraphics[width= 1.0\textwidth]{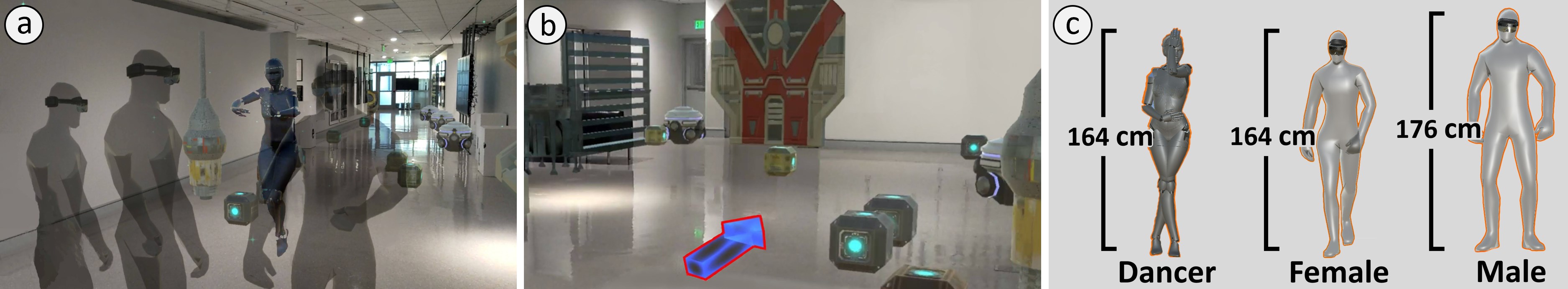}
\caption{(a) The perspective of an augmented reality headset user witnessing a dance performance alongside virtual audience members in a physical layout with augmented stage props. (b) In every trial, a navigational aid (blue arrow) was present and guided users to the next Content Zone. The arrow vanished at the beginning of the performance and reappeared at the end of the dance. If user got lost, they could look down to find the arrow, which would lead them to their next content zone. (c) Avatars in AR Theater.}
\label{fig:aa}
\end{figure}

\paragraph{{Dancer Performance}}
In a traditional theater setting, spotlights draw attention to performers. However, in an open-world performance such as mine, I needed to find new ways to draw attention to the dancer and alert viewers that the performance was in progress. 

The dancer's performances utilized motion capture data collected by Morro Motion technology~\cite{web2023mocap} using a Vicon motion capture system, consisting of six strategically positioned Vicon T160 cameras. The dancer was fitted with 36 reflective ball markers, allowing for the recording of joint trajectories and muscle activities~\cite{wojtusch2015humod}. All dancers are standardized to a height of 164$cm$. I used 34-bone avatar for the dancers in the pilot study and 102-bone avatar for the main study (Figure \ref{fig:aa}c) as the pilot studies indicated that the dancers' movements appeared rigid. I used the same dance motion animation in both studies. Yet, the avatar with more bones could convey expressive movements, which enabled me to eliminate ``stiffness.''

\paragraph{{Physical and Virtual Layout}}
A digital twin of the corridor was used to handle occlusions on the headset and to train the DNN audiences. To ensure precise content alignment on HoloLens 2 devices, I used Azure Spatial Anchors~\cite{buck2022azure}. These spatial anchors from the server were automatically loaded to position digital content, addressing occlusions caused by physical walls. I strategically placed three 2.8 $m^{2}$ content zones to act as location-based triggers along pathways from the user. These zones triggered dance performances when users entered specific areas on the floor layout. The zones were eight meters apart. 

\paragraph{{User Experience and Instruction}} \label{instruction}
My user experience was crafted to promote exploration, enabling the user to navigate and watch dance performances for approximately 4 minutes as they wished. After watching all three dance performances in the trial, users could continue exploring the space or exit the hallway to conclude their experience. Participants were asked not to run, and to wait for each dance performance to end before moving on. They were told about the arrows that could guide them if they were lost, and that they were free to exit the hallway when they were ready to end the experience. They were not given any details of the virtual audience(s), to prevent participant bias or preconceived notions.

\begin{figure}[t]
\centering
\includegraphics[width= 1.0\textwidth]{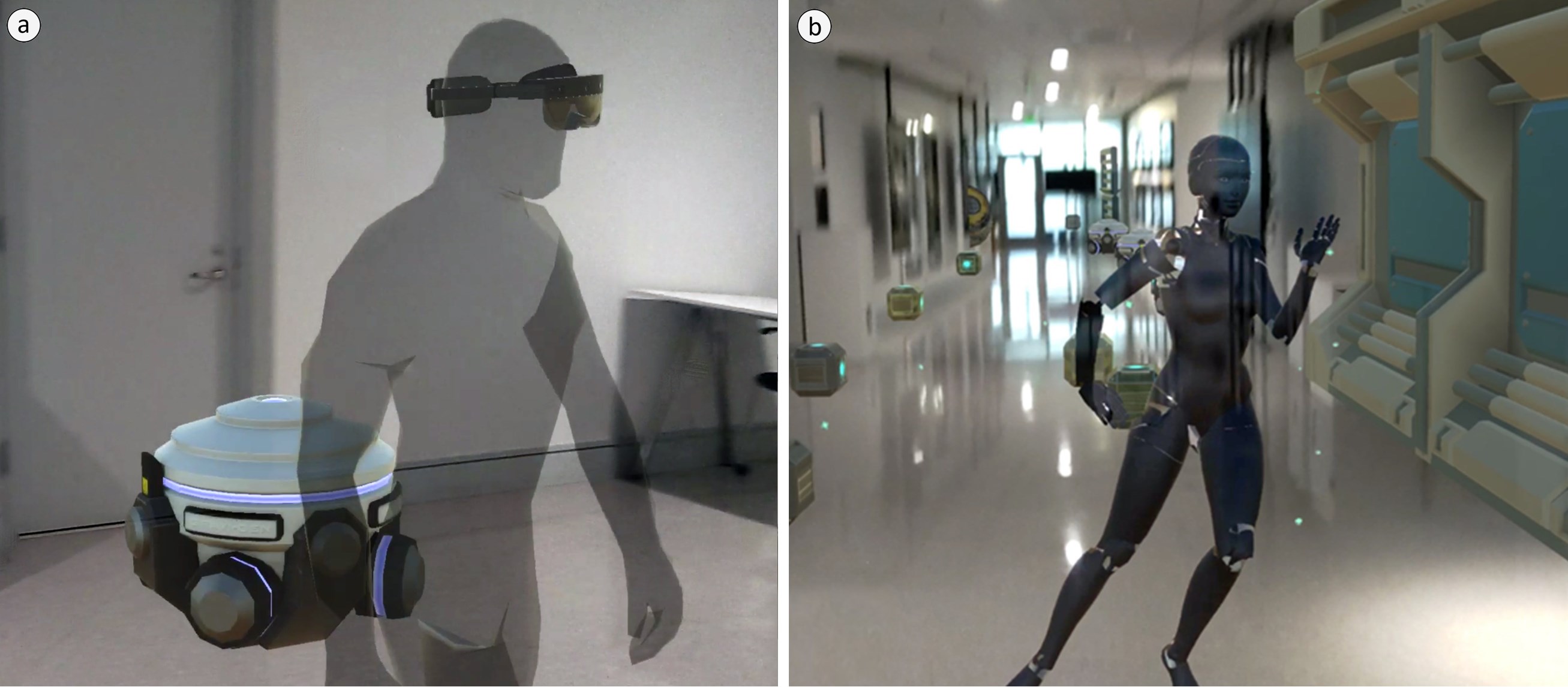}
 \caption{Shows audiences and dancer avatar in the main study. (a) HoloLens 2. The audience avatar is shaded with 50\% transparency so that viewers may feel the audience's presence in their periphery while still being able to see the performance without occlusion. In the DNN trial, three female avatars measuring 164 cm in height and three male avatars measuring 176 cm in height were deployed. (b) Shows 102-bone avatar. The performer's avatar was replaced to a 102-bone avatar from the 34-bone avatar utilized in the pilot studies to ensure that dancers could adequately convey the intricacy of the same chirography performance.}
 \label{fig:audiences-dancer}
\end{figure}

\subsubsection{Virtual Audience}
With an understanding of the physical layout, my virtual audiences imitated the behavior of a real audience. Virtual audience agents are spatially aware, walking and observing performers with similar curiosity and attention as human users. I trained the audience agents to blend into the user's environment and experience to ensure the focus remained on the dance performers. I designed all six virtual agents to spawn close to the ticket booth, which indicated they were fellow audience members. 

\paragraph{{Training Workflow}}
I employed a combination of imitation and reinforcement learning for my virtual audience. First, I collected tracking data of users navigating the AR theater environment while watching performances. This data served as the basis for imitation learning, where a teacher agent performed the task and a student agent imitated it. My goal was to achieve human-like behavior rather than machine-like perfection. Each deep neural network (DNN) model underwent 2 million training steps using the same demonstration data. I identified the top-performing 6 models by selecting those that exceeded the threshold set at the top 30 percent of cumulative reward points.
\end{section}

I utilized a deep neural network with three hidden layers, each consisting of 128 neurons. The neural network was trained using the proximal policy optimization (PPO) algorithm, implemented through Unity's open-source project, Machine Learning Agents Toolkit (ML-Agents).

\begin{figure}[t]
\centering
\includegraphics[width= 1.0\textwidth]{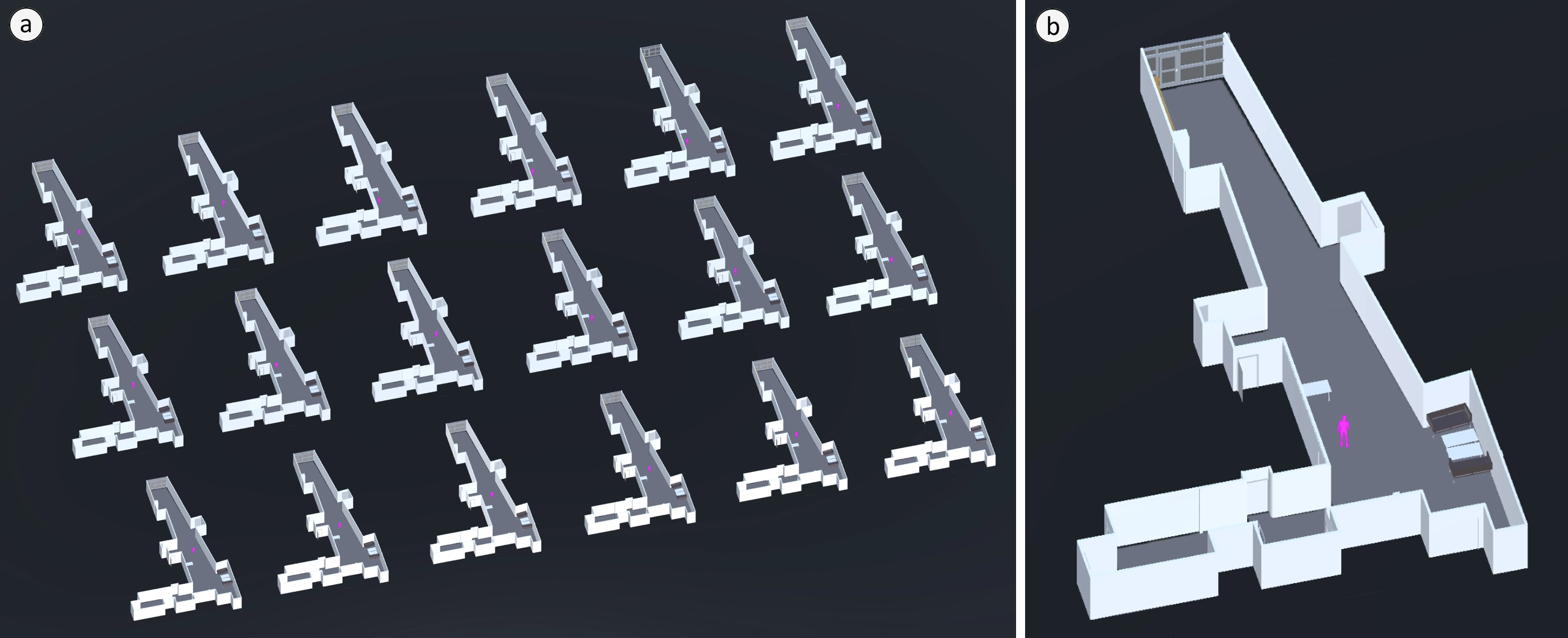}
 \caption{(a) I produced 6 neural network models for each virtual audience, utilizing concurrent training sessions with 18 training AI environments at once to speed up the training session using data from prior participants who participated AR theater play in the space. (b) Zoomed into one of the AI environments, displaying its distinctiveness and fidelity to the physical layout. Training sessions were conducted in this environment to run imitation learning with the goal of having it behave in the environment way a real human would.}
 \label{fig:training-environment}
\end{figure}

The virtual audience avatars were trained to move within the hallway, exploring content zones and mimicking real human behavior. I trained six neural network models using position data from previous users, which included 60 trials and 3 hours and 21 minutes of tracking data. To ensure accuracy, I used a digital twin model of the hallway at the precise scale and matched it with the demonstration data, creating a realistic training environment (Figure~\ref{fig:training-environment}). These agents operated independently but shared the same model, facilitating parallel training in 18 AI environments. Using the digital twin allowed me to capture nuanced behaviors specific to my scenario, rather than relying on a generic model. 

\paragraph{{Reward System}}

To enhance my results from imitation learning, I fine-tuned the reward system to incentivize behavior aligned with my task of watching performances while walking. AI agents received rewards for various actions related to content zones, such as moving towards them, entering them for the first time, and remaining within them. The longer an agent stayed within a content zone, the more rewards it accrued, up to a maximum duration matching the performance duration in each zone.

To optimize the timing of engagement with the virtual audience, I introduced additional rewards. Entering the first content zone resulted in a reward, with increasing rewards for subsequent zones. Completing entry into all three zones earned the agent an extra reward. Additionally, agents received a small incentive for proximity to the content zone but were penalized for touching the wall.

Agents received positive rewards for various actions, including moving towards content zones, entering them for the first time, and staying within them. The longer they stayed in a content zone, the more rewards they earned, up to a maximum of 17 seconds, which matched the duration of the performance occur in each content zone. To refine the timing of virtual audience, I introduced additional rewards. Entering the first content zone was rewarded with 48.2f, the second with 63.7f, and the third with 85.5f. Completing entry into all three zones earned the agent an extra 41.0f. I also provided a small incentive for getting closer to the content zone (0.03f/ sec) and penalized the agent for touching the wall (0.01f/ sec).

\paragraph{{Avatar Motion}}
\label{sec:logic}
I placed particular emphasis on improving avatar motion for mixed reality interactions, recognizing the significance of avatar representation in such scenarios based on existing research findings in recent years~\cite{yakura2020enhancing, doi:10.1080/10447318.2022.2121038, wang2020effect}. I leveraged Unity ML-Agents' ActionBuffers function to handle continuous and discrete actions. This approach significantly smoothed the motion during initial training sessions and prevented overlapping actions.

To provide a more realistic representation of audience movement, I took an additional step by incorporating motion-matching technology, specifically Kinematica. This technology enabled seamless transitions between different behavioral motions based on the path and speed of the avatar's movement. My avatar animations were controlled through State Machines, encompassing gestures like walking, idle, turning around, and subtle expressive motions, thereby enhancing the overall realism of the avatar's movements.

\paragraph{{Virtual Audience Voice Responses}}
I collected voice recordings from participants who had previously experienced the same virtual dance performance. To encourage users to express themselves more naturally, I assured them that speaking loudly during the performance was acceptable. To allow participants to react and express themselves freely while being voice recorded, the chatter of four planted audience members was played in the background. These recorded voice reactions were then played back as virtual audience response voices. This decision was motivated by real-world experiences and the way users of VR social experiences can always hear the people around them. 

\subsection{Experiments}
To optimize my study's ability to gauge the impact of virtual audiences on user experience, I conducted two pilot studies. In the first, 20 participants ranked preferred avatar and behavior combinations. In the second, based on the insights gained, 10 participants evaluated four conditions. Finally, my main study, involving 20 participants, focused on two trials: dance performance only and performance with a DNN avatar as the virtual audience. The participants for the three experiments (two pilots, one main) were drawn from separate populations. No participant took part in more than one experiment. I recruited 50 participants for my experiments.

\begin{figure}[t]
\centering
\includegraphics[width= 1.0\textwidth]{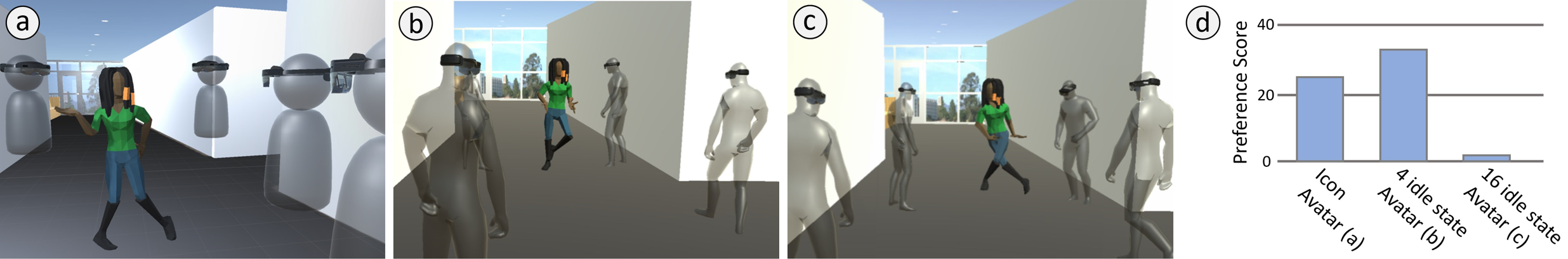}
 \caption{Users' perception of watching a dance performance with each of the three virtual audiences in the first pilot study: (a) A human figure made of basic geometric shapes represents the virtual audiences I named Icon Avatar. (b) A human figure avatar with 4 idle states represents virtual audiences. (c) A human figure avatar with 16 idle states, that enable the avatar to be more expressive and receptive, represents virtual audiences. (d) Accumulated preference ranking for each of the 3 avatars as virtual audiences (2 for the most preferred avatar, 1 for the next, and 0 for the least preferred, for a total of 20 participants). The avatar with 4 idle states was most preferred, followed by the basic geometric avatar, Icon Avatar. }
 \label{fig:pilot-study1}
\end{figure}

\subsubsection{Pilot Study 1}
The pilot study was used to select the best avatar models to use in the second pilot study and the formal user study. For this purpose, I let the 20 participants experience different avatar types to participate in viewing the performance. My designed AR theater environment included (1) an icon avatar, (2) a 4-idle-state avatar, and (3) a 16-idle-state avatar, which appeared more expressive than the 4-idle-state avatar (Figure \ref{fig:pilot-study1}). 

As I briefly explain in the Avatar Motion section, in the first pilot study, condition 1, my virtual audience was represented as human figures without hands or legs. As seen in Figure \ref{fig:pilot-study1}a, this was to give the viewer an idea that someone is watching nearby and nothing more. For conditions 2 and 3, I produced a human figure with arms and legs, three male avatars, 176$cm$ in height, and three female avatars of the same height as the dancer (164$cm$). For condition 2, I deployed four idle animation states on top of basic navigational motion as seen in Figure \ref{fig:pilot-study1}b. For condition 3, I deployed sixteen idle animation states on top of basic navigational motion and demonstrated much more expressive reactions, as illustrated in Figure \ref{fig:pilot-study1}c.

\paragraph{{Pilot Study 1 Procedure}}
I recruited 20 participants from a university campus, ranging in age from 18 to 51. The group comprised 10 individuals who identified as male and 10 who identified as female. These individuals took part in a survey involving the evaluation of a 1-minute performance under three distinct conditions within a physical environment while using an AR headset. The sequencing of the conditions was counterbalanced across participants.

Participants ranked their preferred audience from most to least desirable and participated in brief exit interviews. They discussed their ranking preferences for watching performances, the types of audiences, and which experience most resembled a live concert, explaining their favorite and least favorite choices. During the survey, participants experienced three trials in which they were presented with distinct virtual audience types while the dance performance remained consistent. Additionally, I gathered subjective feedback and suggestions from participants to enhance my future studies.

For ranking, a preference score system was employed: the first choice received 2 points, the second choice was awarded 1 point, and no points were allocated for the last choice (Figure \ref{fig:pilot-study1}d).

\paragraph{{Insight and Qualitative Feedback}}
My initial pilot study outcomes indicated that viewers exhibited a preference for the 4-idle-state avatar model as their virtual audience, followed by the icon avatar. Conversely, the expressive 16-idle-state avatar model was generally not favored, as observed in Figure \ref{fig:pilot-study1}d. The main reason cited for this preference was that fellow audience members in the expressive avatar model appeared too distracted and moved around too much, detracting from their ability to focus on the dancer's performance. In a direct quote, participant P17 expressed a preference for the 4-idle-state avatar, stating, ``I am much better at maneuvering around crowds of people and need to be able to see how they are walking, so I prefer the avatar with arms and legs''. P4 remarked, ``It makes more sense for the audience to be more like the performers, as it seems odd to me that the two should not match. Additionally, a more intricate human figure simply fits in with the surroundings''.

In light of these findings, I opted to use the 4-idle-state avatar model in my NPC and DNN trials in the second pilot study.

In addition to these findings, the subjective feedback gathered from all participants after the study indicated that many participants described the 16-idle-state avatar model as ``creepy.'' Consequently, based on these results, I decided to utilize an audience avatar with arms and legs, the 4-idle-state (less expressive) avatar model for my main user study.

\begin{figure}[t]
\centering
\includegraphics[width= 1.0\textwidth]{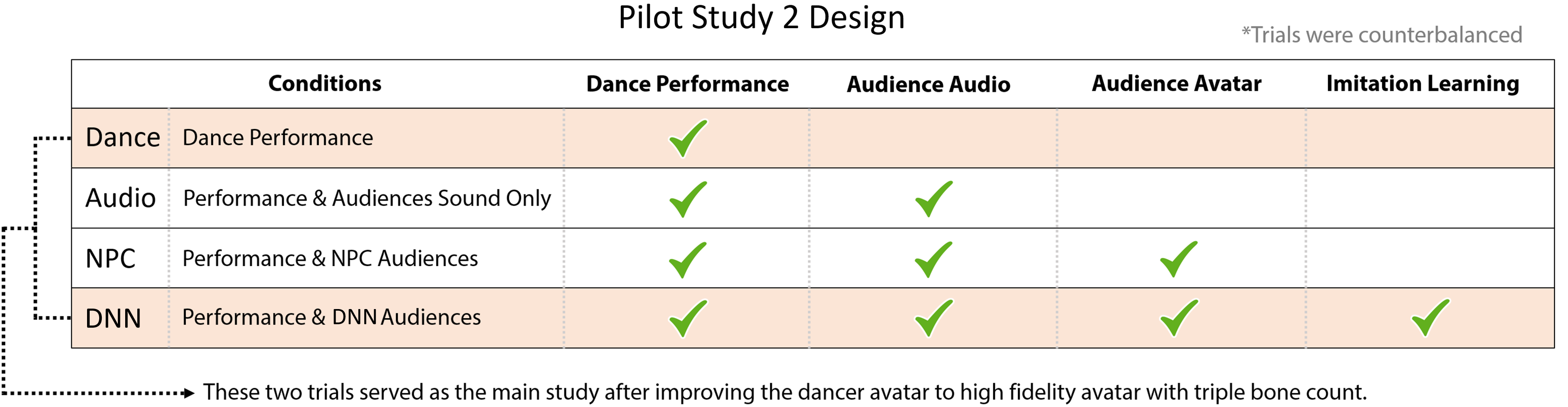}
\caption{Design chart for Pilot Study 2, showing the four trial conditions and their components. }
 \label{fig:pilot-study2}
\end{figure}

\subsubsection{Pilot Study 2}

Pilot study 2 was conducted to determine which of three different implementations of the virtual audience to use in the main study. The control condition was the dance performance with no audience (hereafter referred to as \textit{Dance}). All trials included audio, except for the Dance condition.  The three different audience implementations are detailed below (Figure \ref{fig:pilot-study2}):

\begin{itemize}
\item\textit{Audio}: In addition to the dance performance, users also heard recordings of six real user reactions to the performance, implemented as spatial sound.

\item\textit{NPC}: In addition to the dance performance, users also saw six humanoid avatars as virtual audience members, along with the same audio reactions as \textit{Audio}. These avatars always stood close to a content zone and directly faced the performer (Figure \ref{fig:virtual-audiences}a).

\item\textit{DNN}: This condition was similar to the NPC condition, except that the audience avatars were now trained to emulate human-like movement and behaviors (see Section \ref{sec:logic}). The difference between the NPC and DNN audiences was that the NPC audience tended to spend their time looking directly at the performance and paying attention to it, while the DNN audience typically displayed more individual behaviors, and acted more participatory and dispersed (Figure \ref{fig:virtual-audiences}b).
\end{itemize}

Upon closer observation and interaction with the NPC and DNN models throughout many development sessions, the differences became clear to me. A clear distinction emerged between NPC and DNN audiences in how they interacted. In terms of behavior and interaction, the AI agents positioned themselves in the hallway in a way that mirrors how real human crowds naturally gather, particularly in the hallway we tested. For example, the AI agents demonstrated how congestion occurs when many people are present in that hallway, moving to sensible areas rather than arranging themselves randomly, much like real people would. These AI agents appeared to learn and adapt to using space respectfully, ensuring they maintain an appropriate distance while observing dancers, similar to human behavior. Additionally, the AI agents learned to incorporate human-like imperfections in their movements, such as small, non-essential motions that occur when humans are indecisive or wandering.

\begin{figure}[t]
\centering
\includegraphics[width= 1.0\textwidth]{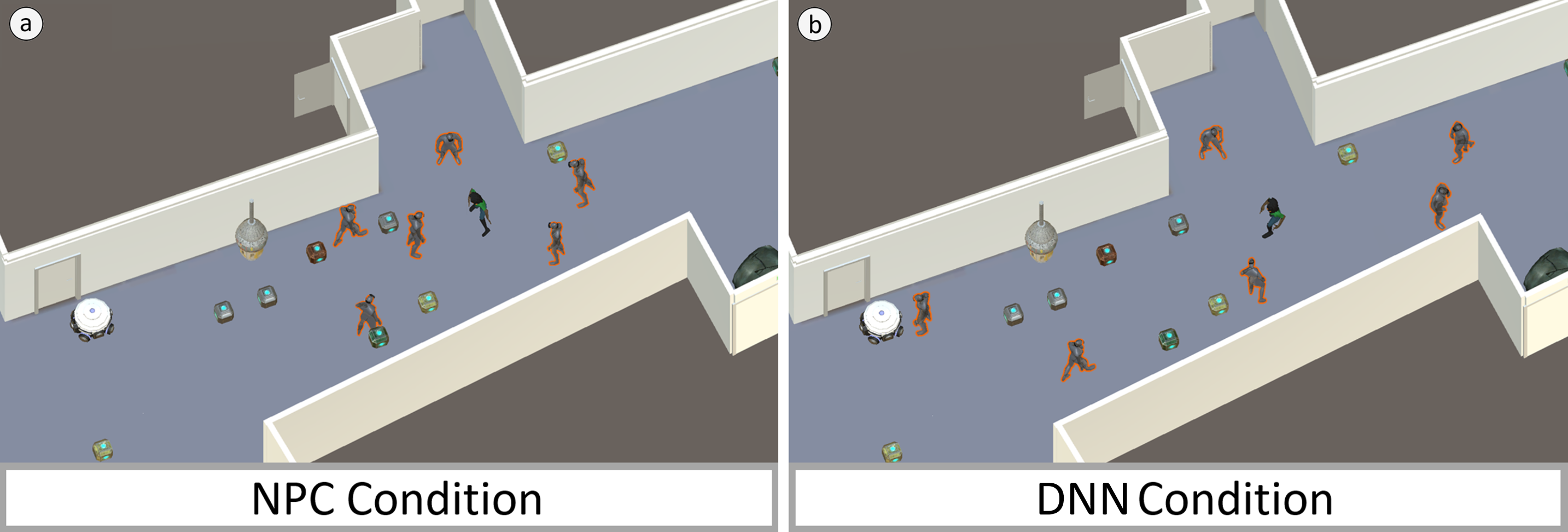}
 \caption{The positioning and dispersion of virtual audiences and the differences between the DNN and NPC models. (a) Demonstrates how audiences are positioned close to the dancers and are looking in the dancer's direction. (b) Illustrates how the crowd is dispersed and spaced out, with some people watching the dancers and others moving independently.}
 \label{fig:virtual-audiences}
\end{figure}

\paragraph{{Pilot Study 2 Procedure}}\label{procedure-pilot2}
Ten participants (5 self-identified as male, and 5 as female) experienced the four conditions in a randomized order (counterbalanced between subjects) and gave feedback on their experience and preference. Each condition lasted approximately four minutes, though participants were free to explore the space for longer if they wanted to. Demographic information was collected before they experienced any of the conditions, and qualitative feedback was collected after each condition (specific to that experience) as well as after the entire study (regarding the overall experience).

The demographic questionnaire had questions related to previous experience with immersive theater, narrative games, and VR games. The post-trial questionnaire, which was administered after each condition, included a five-question sub-scale of the Game User Experience Satisfaction Scale (GUESS) \cite{doi:10.1177/0018720816669646} to measure ``Enjoyment'', eleven questions from a sub-scale of Immersive Virtual Environment Questionnaire \cite{10.1145/2927929.2927955} to measure \textit{presence}, a twelve-question sub-scale of the Augmented Reality Immersion (ARI) Questionnaire \cite{GEORGIOU201724} to measure \textit{immersion}, and a ten-question sub-scale of the ARI Questionnaire \cite{GEORGIOU201724} to measure \textit{engagement}, three questions from Game Engagement Questionnaire (GEQ) to measure interest and boredom~\cite{geq2013}. Participants also completed a free-form short essay (150 words or more) within 10 minutes about their experience, along with four 7-point Likert scale responses mentioned above. At the end, they engaged in a 5 minute voice recorded free talk for analysis and insight.

\paragraph{{Insight and Qualitative Feedback}}
Using the SentimentAnalysis package, I generated a word cloud visualization to identify trends in user experience~\cite{ALAMOODI2021114155}. Additionally, I employed Grounded Theory for my analysis, involving validation distributed among different team members~\cite{corbin2014basics}. This process allowed me to review the user interviews, discuss my findings collaboratively, and compile an overview of the general trends, which I report here. The DNN condition emerged as the most preferred condition, with the NPC condition being second-most preferred. Both conditions were significantly more entertaining and enjoyable than the Audio and Dance conditions. The DNN condition also fostered better presence and engagement compared to NPC, with participants saying that ``I liked the second one (DNN) because I can walk behind them, feeling like I am watching with a cloud of people and not by myself'' (P2) and ``The audience naturally guided me to the dancer without needing to rely on an arrow which was nice.'' (P5). Seven out of 10 participants reported feeling ``distracted'' by NPC avatars, with some expressing that they felt ``crowded'' and that their view was ``blocked,'' leading to occasional overwhelm. The DNN avatars, on the other hand, helped to improve the experience by their tendency to disperse in the environment (``I like that people are there but I prefer the one (where) user(s) are more spread out, I can focus on dance performance without feeling like I have to fight for the best view while still feeling like I am in (a) concert,'' (P6)).

\subsubsection{Main Study}
For the main study, I recruited 20 adult participants (N=20, 9 who self-identify as female, 11 who self-identify as male, ranging in age from 18 to 53, an average age of 26.25 (SD=8.37)) from a local university campus. They were compensated at a rate of \$15 per hour. All participants were able-bodied without hearing and vision issues and were able to move around during the trials.

\paragraph{{Experimental Design}}
Based on the results of my two pilot studies, I decided to use the 4-idle-state avatar model in the main study (as shown in Figure \ref{fig:aa}b). I also opted to utilize machine-learning-trained virtual audiences, as I believe they closely mimicked a live audience experience. This main study followed a within-subject design with one factor--the presence of NPC audiences, which is evaluated with two conditions: \textbf{Dance} is the baseline condition where no audience was provided during trials and \textbf{DNN} with six Imitation Learning-Trained NPC audience avatars presented during the AR theater play. Trials with virtual audiences included audio, whereas no audio was present in the Dance condition as there was no virtual audience in those trials. The order of these two conditions was counterbalanced in the experiment.

\paragraph{{Measurement}}
Despite eliminating two trials compared to pilot study 2, I utilized the same set of questions and interview process. As for outcome assessment, I collected the following data to evaluate users' overall experience during trials:

\begin{itemize}
    \item User Performance: I collected the completion time of each trial for each user.

    \item User Experience: As in pilot study 2, I measured and recorded participants' experience via my designed questionnaire (see Section \ref{procedure-pilot2}). I focused on measuring four main metrics of user experience, including  \textit{Enjoyment}, \textit{Presence}, \textit{Immersion}, and \textit{Engagement}.

    \item User Sentiment: Users were required to complete a brief text-written essay with open-ended questions targeting their overall experience and feelings in each trial. For the user sentiment analysis, I chose to use the R package sentimentr \cite{ALAMOODI2021114155}, the VADER system, the syuzhet package, and SentimentAnalysis package \cite{feuerriegel2018package} to analyze user sentiment level of each trial based on what I collected from participants' answer.
    
\end{itemize}

Users also responded to questions regarding their subjective preferences and feedback about the two trials, with some open-ended questions from Presence Questionnaire (PQ) \cite{10.1162/105474698565686}, such as, ``Which trial did you enjoy more?'', ``Which trial felt more like a live performance?''. Semi-structured interviews were conducted after the entire experiment, and users were allowed to share any thoughts in a free-form interview post trial to understand the data. 

\paragraph{{Main Study Procedure}}

The procedure of the main study mirrors the pilot study 2 (detailed description in Section \ref{procedure-pilot2}), with the only difference being that there are only two conditions in the main study. Thus, the entire procedure of the main study lasted approximately 60-80 minutes per participant. 

\paragraph{{Analysis}}

The Shapiro-Wilk test was used to check for violations of normality in the data. The Aligned Rank Transform (ART) \cite{Jacob_ART} was applied to all data that violated normality. 

One-way repeated measures ANOVAs were used to assess the impact of the two experiment conditions (Dance and DNN) on user experience ratings and metrics such as completion time, presence, immersion and engagement, with experiment condition as the only independent variable. Bonferroni corrections were used for all pairwise comparisons. Independent-sample t-tests were used to examine the impact of other independent variables (trial order and height) on user experience ratings and metrics.

\subsection{Main Study Results}

I analyzed data to compare the two experiment conditions and investigated the effect of trial order and participant height on reactions to the experience.

\subsubsection{Audience Type}
I compared the two main experiment conditions--virtual audience present (DNN) and no virtual audience present (Dance)--using both qualitative metrics of user experience (interest and involvement) as well as a quantitative metric of engagement (time spent in the experience). Results are highlighted in Figure \ref{fig:audience-type}.

\begin{figure}[t]
\centering
\includegraphics[width= 1.0\textwidth]{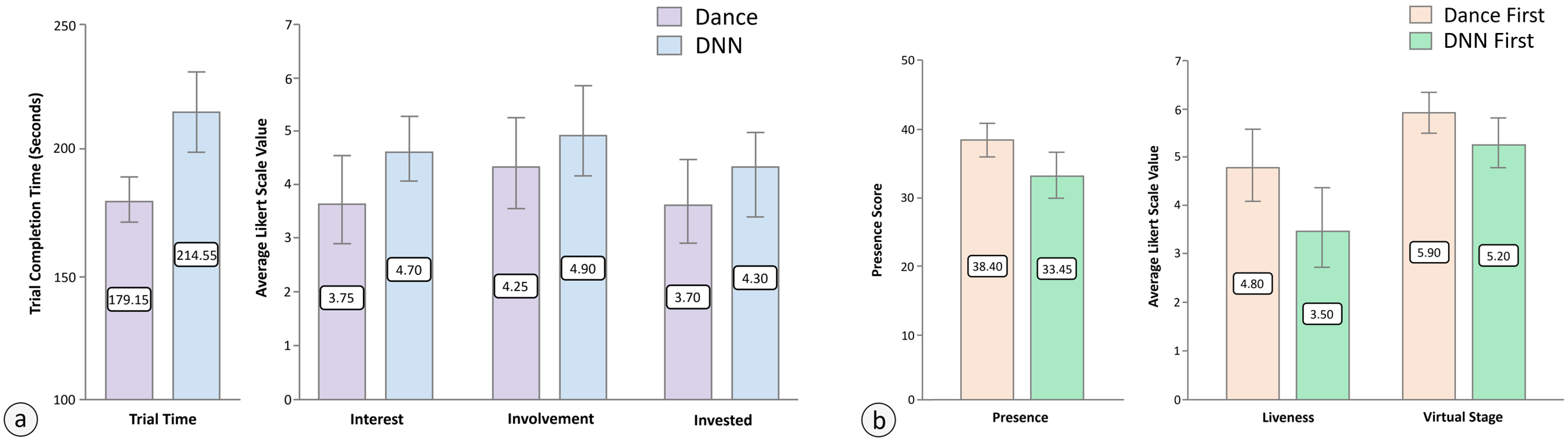}
 \caption{(a) Trial completion time and user experience ratings for the two experiment conditions. Participants spent more time in the experience in the DNN condition and also reported higher interest, involvement (``How much did the auditory aspects of the environment involve you?''), and investment(``I was so involved that I felt that my actions could affect the activity''). For this and all following charts: Error bars represent 95\% confidence interval. User experience ratings were on a 7-point Likert scale (7=Strongly agree/Very much, 1=Strongly disagree/Not at all). (b) Presence and user experience ratings for the two groups of participants based on trial order (which condition they experienced first). Participants who experienced the Dance (no audience) condition first had a higher presence score, higher liveness (feeling of actually being in a live performance), and appreciated the virtual stage more.}
 \label{fig:audience-type}
\end{figure}

A one-way repeated-measures ANOVA indicated a significant difference in interest between the two conditions ($F_{(1,19)}=4.524, p=.047, \eta_{p}^{2}=.192$), with the post-hoc test showing that participants reported more interest in the experience during the DNN condition compared to Dance ($p<.05$). There was also a significant difference in involvement ($F_{(1,19)}=6.011, p=.024, \eta_{p}^{2}=.240$), with participants reporting more involvement with the auditory aspects of the experience (``How much did the auditory aspects of the environment involve you?'') in the DNN condition ($p<.05$). Participants felt that their actions could influence the experience differently ($F_{(1,19)}=4.557, p=.046, \eta_{p}^{2}=.193$), with a stronger perceived influence in DNN compared to Dance ($p<.05$).

Time spent in the experience changed as a function of the experiment condition ($F_{(1,19)}=28.672, p<.001, \eta_{p}^{2}=.601$). Participants spent more time watching the performance in the DNN condition compared to Dance ($p<.001$), which tracks with the increased interest score from GEQ questions~\cite{geq2013} in this condition. These results suggest that virtual audiences had a positive impact on user experience and engagement in shared immersive performances.

\subsubsection{Trial Order}

When comparing participants' reactions to their entire study experience (recorded after both sessions, and regarding their impressions of both sessions combined), participants who saw the Dance session first, followed by DNN, had a more positive reaction to the experience than participants who saw the DNN session first (Figure~\ref{fig:audience-type}b). This was demonstrated by a higher presence ($t(38)=2.145, p<.05$), higher appreciation of the virtual stage design ($t(38)=1.807, p<.05$), and a stronger feeling of actually being in a live performance ($t(38)=2.145, p<.05$). These results suggest that the order of introduction to the virtual audience could alter user experience and appreciation of the stage and performance.

\begin{figure}[t]
\centering
\includegraphics[width= 1.0\textwidth]{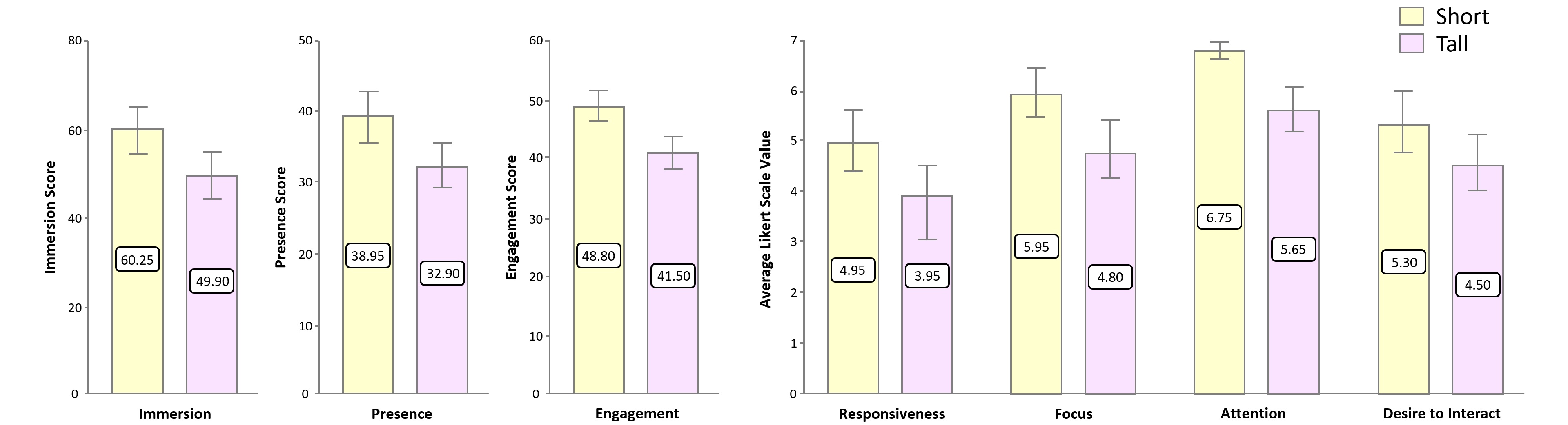}
\caption{Immersion, presence, engagement, and user experience ratings for the two height groups. Shorter participants had higher immersion, presence and engagement. They also reported more focus, more attention, more of a desire to interact with the virtual characters and felt that the environment was more responsive to their actions.}
 \label{fig:height}
\end{figure}

\subsubsection{Height}
Considering how height influences a person's live concert experience, I investigated the role of height and how it affects the desire to watch an AR play catered to a freely walking user. Therefore, I divided the 20 participants into two groups of equal size based on their height--the 10 tallest participants (Tall) in one group and the 10 shortest participants in another (Short). Results are highlighted in Figure \ref{fig:height}. An independent-samples t-test indicated that the shorter participants had significantly higher immersion ($t(38)=-3.113, p<.005$), overall presence $t(38)=-3.234, p<.005$) and engagement ($t(38)=-3.799, p<.001$) than taller participants during the study. Shorter participants also felt that the environment was more responsive to their actions ($t(38)=-2.380, p=.022$), were more focused on the activity ($t(38)=-3.598, p=.001$), and wanted to interact with the virtual characters and objects more ($t(38)=-2.312, p=.026$) than taller participants. 

On the other hand, taller participants felt like they had more external distractions than shorter participants ($t(38)=-2.427, p=.020$), and also felt that the virtual content captured their attention less than the shorter participants did($t(38)=-4.341, p<.001$). To verify my results, I compared the 5 tallest participants' and the 5 shortest participants' experiences. My results held, with even stronger significance being demonstrated. An examination of the participant's gender did not result in significant variations. The observation that shorter participants enjoyed the experience more, irrespective of gender, indicates that this effect cannot be attributed to gender differences alone. Additionally, I did not find any other correlations that could explain this effect.

\subsection{Discussion}

Audience Amplified explored the inclusion of virtual audiences in AR theater applications and their impact on user experience. I introduce a training combining imitation and reinforcement learning to create virtual audiences for target space layout that accurately capture the nuances of human imperfections and behaviors.

A novel aspect of this work involves capturing audience behavior by collecting user motion and behavior data to train AI agents. This approach effectively represents audiences from a specific era, time, or demographic, creating a "time capsule" of audience behavior while engaging with specific content from 2023. This enables capturing, archiving, and preserving the entire audience experience of live events, such as concerts. This also presents numerous opportunities such as re-experiencing, or even reliving, events and environments in a fully immersive and accurate way. 

Participants had increased involvement, interest, and engagement with the experience when virtual audiences were present. There appear to be a number of reasons for this, including the experience being more realistic (``It definitely brought more of a real experience compared to having no audience at all.'', P7) and social (``I think that the social aspects are the best part, because ... digesting (the experience)) together makes the experience meaningful...'', P3; ``it truly felt like you were interacting with others; more so than an online video game'', P10). Some participants mentioned, however, that the virtual audience sometimes blocked their view of the dancers. This might have happened as a result of the virtual audience being trained using data gathered from an earlier study \cite{kim2023dynamic}, which made use of user data from encounters with AR theater alone and without audiences. 

Additionally, participants enjoyed the experience more when interacting with the ML-trained virtual audience, feeling higher levels of presence and appreciation for the content. The auditory aspect enhanced their sense of environment and engagement, especially when they first experienced the play without the virtual audience and were introduced to it later.

Participants who experienced the virtual audience after initially watching the performance with no audience expressed a greater appreciation for the dance play and stage environment, as they were able to enjoy the play without any interruptions. While virtual audiences can enhance presence and enjoyment, there is a tradeoff to consider. This insight is crucial for making design choices in AR theater, as adding an audience does not always enhance the experience; it depends on the director's objectives. I believe that using a combination of scenes, sometimes with an audience and other times in solitude, can be a powerful tool for AR theater.

Shorter participants reported a more positive experience overall than taller participants, with higher reported presence, immersion, and engagement. Taller participants also reported more external distractions and less focus on the activity, which could be driven by the fact that all of the virtual characters were much shorter than the taller participant group on average while the shorter participant group was about the same height as, or much shorter than, the virtual avatars. For example, a participant from the taller group said ``What drew me out were ... the different sizes of the avatars'' (P19). Further exploration is needed to determine the impact of scale and content placement on user experience in immersive storytelling, but my results suggest that it is important to consider user height and perspective when designing such experiences.

\subsection{Limitations and Future Work}

While Audience Amplified offers the sensation of being in the company of others, it is important to note that users are, in reality, engaging with these virtual audiences alone. Not directly comparing the experience with a real audience is a limitation. Future research should examine how virtual audiences perform in group settings and how this alters the user experience. Despite their realistic appearance and behavior, my virtual audiences are not interactive. The DNN model I presented contributed to this non-interactive nature of AI agents. Furthermore, compared to the NPC model used, my virtual audience exhibited more distributed attention and did not provide full focus on the performance. Training was not conducted for more delicate gestures such as hand gestures and facial expressions, as this information was not part of the training data. In the future, when I have a clear vision for these gestures and how they should be trained, I aim to refine my avatar model to enhance real-time content engagement in future projects. Surprisingly, some participants spontaneously danced along with the avatar dancer, suggesting that virtual audiences perhaps should at times react to and participate in unexpected situations. Finally, while my virtual audience can be implemented in various environments, their behavior was most effective in the corridor layout I tested. Future projects will investigate training virtual avatars to adapt to building data and user context, using deep neural networks to automatically adjust to floor layouts.

\subsection{Conclusion}
I present Audience Amplified, a novel augmented reality dance performance that includes trained virtual audience avatars that perform human-like behaviors and movements using imitation learning. Two pilot studies were held to determine the optimal appearance and behavior of the audience avatars. A following 20-subject user study was conducted to understand differences in user experience when viewing the performance with and without the virtual audience. I found that participants reacted positively to the experience that included virtual audiences and appreciated the social aspect of watching a performance with other virtual audiences, especially as it provided live-event like experience. 

Some important design considerations for mixed reality theatrical experiences are highlighted in my analysis, including the presence and absence of virtual audiences in the interactive narrative, as well as optimizing content placement and scale based on user height and perspective. 

In this project, I focused on experiences that followed the participatory open-world exploration style, which provided users freedom to personalize their experience as a viewer. I see great possibilities for AR performances that can be experienced via mobile AR by individual audience members at their leisure, while still experiencing some semblance of audience connection. In the future, the AR spectator’s social viewing behavior could influence the amplified audience experience of subsequent users, leading to a new kind of asynchronous, yet still, socially connected experience.

\end{section}

\begin{section}{Teaching an AR Theater System}
I had the opportunity to teach and serve as the instructor of record for a course using the AR Theater platform I developed, which allowed students to create their own AR theater experiences. The course, titled ART 22: Programming for Arts: Wide-Area Augmented Reality Theater Experience, uniquely blended engineering and artistry by focusing on the development of AR applications. It attracted a diverse group of students from disciplines such as fine arts, computer science, dance and theater, and humanities, fostering a dynamic environment where the intersection of technology and creativity could be thoroughly explored.

The course adopted a project-based learning approach, where students worked in teams to develop AR projects. Each team consisted of three students, assuming roles such as Lead Developer, Version Controller, or Project Manager, to ensure active engagement and skill development in both technical and creative domains. Using Unity as the primary platform for AR development, students created five-minute AR theater plays for HoloLens 2, utilizing the open-sourced Immersive AR Theater platform I developed during the Dynamic Theater and Audience Amplify projects. The platform was made open-source before the course began, and students accessed it throughout the class, supported by a comprehensive seven-hour YouTube tutorial that lowered the entry barrier for non-technical users. This allowed participants to focus on creativity and experimental interaction design in AR stage settings.

\begin{figure}[t]
\centering
\includegraphics[width= 1.0\textwidth]{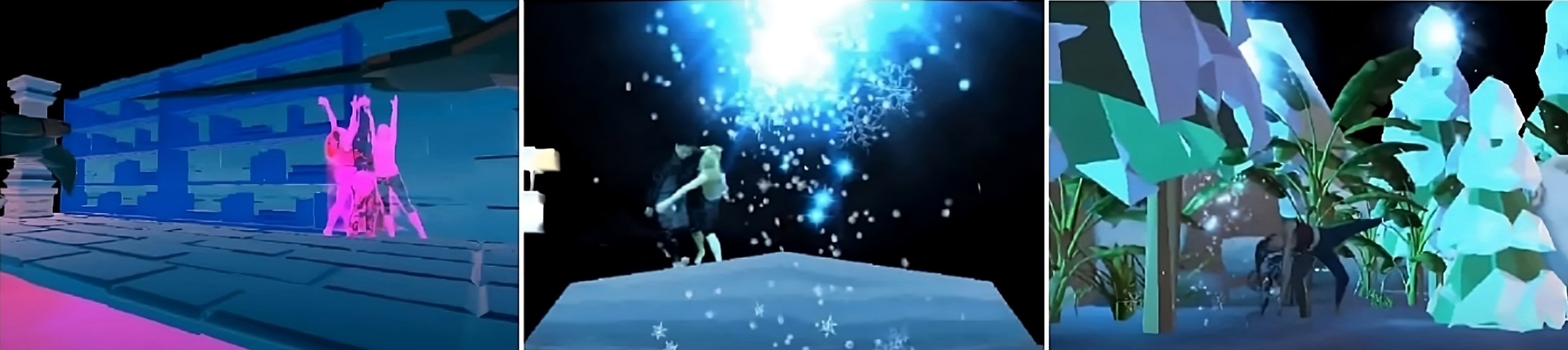}
\caption{An example of an AR theater play created for the class final project is "The Passage of Time," with a duration of 3 minutes and 31 seconds. This image serves as an example of student projects and is shared courtesy of Section 2, Team 2.}
 \label{fig:teachingAR}
\end{figure}

Collaboration was a fundamental aspect of the course, facilitated by weekly surveys for student feedback and course adjustments. Discord served as a collaborative platform, encouraging peer support and discussion, which helped students manage technical difficulties and foster a strong learning community. This setup incentivized students to be active community members, allowing them to troubleshoot issues with peers before seeking help, providing quick answers outside of TA office hours.

Students found the course both challenging and rewarding, appreciating how it demystified complex AR topics and offered practical technology understanding. Despite technical challenges, such as version control on GitHub, these obstacles became valuable lessons in teamwork and problem-solving. Using GitHub for issue reporting and ticket support aligned with practices from the game development community, offering real-time solutions.

The course's emphasis on collaboration and feedback resulted in high student satisfaction, especially in terms of teamwork and project outcomes. Many students expressed interest in continuing their collaborations beyond the course, highlighting the strong community bonds that were formed and demonstrating the need for an easy-to-use platform for creative, interactive narratives in mixed reality. The course successfully combined technical and artistic elements, underscoring the educational value of interdisciplinary learning.

While my earlier projects, Dynamic Theater and Audience Amplify, focused on user experience, this course demonstrated how the platform could effectively support both artists and engineers. Choreographers, dancers, and theater artists found the AR platform particularly beneficial for dynamic storytelling through cue-triggered spatial interactions, reminiscent of traditional theater practices. This innovative approach allowed performances to unfold in an interactive narrative, where user choices influenced the story's progression.
The modular design of the user experience enabled students to easily envision lighting design changes when entering ``Content Zones,'' adding dynamic elements to their performances. Many students specifically appreciated the cue scene system provided by my AR platform, which mirrored traditional theater cue sheets, helping them transition their ideas into mixed reality more seamlessly. Additionally, the GitHub repository, equipped with project settings for both Mac and Windows users, simplified the setup process, allowing students to quickly initiate their projects using Unity's automatic asset deployment and begin collaboration from day one.

I was pleasantly surprised by the production level and uniqueness of the 10 AR theater plays presented at the end of the course, and how much students could achieve with minimal technical support. The platform provided step-by-step video guides and instructions made by myself, empowering them to push their creative boundaries. \textit{The Passage of Time} is of the student projects created for this class as seen in Figure~\ref{fig:teachingAR}. This AR play offers an immersive theater experience set in an ice world, with a duration of 3 minutes and 31 seconds~\cite{section2team22022}.

Despite using the same platform and targeting the same physical location (UCSB campus, Kirby Crossing area), each five-minute play was distinct, with different themes and innovative uses of AR lighting that uniquely set the mood for each performance. It was incredibly rewarding to see artists using my system to create immersive AR theater plays, and students demonstrated strong ownership and pride in their creations. The platform's ease of use significantly lowered the entry barrier for creative artists and theater students without technical skills, empowering them to create AR theater in mixed reality outdoor settings without being constrained by technical limitations.

\end{section}

\chapter{Conclusion}

Tom Stoppard, a British playwright, once said, ``Look on every exit as being an entrance somewhere else.''~\cite{stoppard2013rosencrantz} In the complete openness of virtual production and its endless possibilities, it is crucial to reflect the director's vision of the play, while being intentional to user experiences. With more resources, creators must consider the many possibilities for human interaction to curate user experiences that align with the vision of what we are trying to capture and present to the user. The main goal of this dissertation is to propose a new way of immersive media consumption in spatial computing and predict how mainstream media will evolve in the future.

Where no physical constraints dictate how things are done, human-computer interaction (HCI) design and user experience become key to presenting virtual environments. Considering the design of the stage and how users experience the space is crucial to these virtual productions, especially when there is unlimited potential. An experimental approach is key to presenting a lasting experience for the user; however, this also could generate confusion for the user if they are unsure of what to do. Designing an intuitive experience that guides users through various methods of clues and visual cues becomes essential, especially in open-world experiences.

The underutilization of physical layouts in today’s immersive experiences presents a significant challenge. Through this dissertation, I present and examine methodologies for optimizing space utilization and user interaction in these spaces for interactive narratives. By incorporating large-scale 3D digital twins, thoughtfully placed spatial objects, and AI audience simulations, this research demonstrates how AR technology can optimize physical space to enhance user and audience engagement. Layering the AR stage and objects within the physical space provides a more full user experience. Moreover, the ability to play back how each user interacted in the space and with spatial content can be used to design future interactions and improve upon them. By systematically exploring spatial arrangements from the inception of the immersive play with stage designers and employing guidance systems, this work offers insights into integrating digital spatial content into physical environments. 

Furthermore, this dissertation provides a glimpse of how virtual productions may look for open-world experiences, where the story unfolds through walking and exploration. Following all the experiments and technical discoveries through my artistic HCI projects presented in this dissertation, such as Reality Distortion Room, Spatial Orchestra, Dynamic Theater, and Audience Amplify, I aim to show how user experiences can be expanded through simple interactions. The system I present leverages everyday user actions, such as walking, to expand the scale of interactions. Designed for wide-area and large-scale interactions, my vision is simple--to provide users with dynamic open-world experiences.

The research also introduces AI-enhanced audience simulations to replicate the social aspect of theater. By creating AI-trained virtual audiences, the dissertation highlights the potential of AI to foster socially enriched experiences, even in the absence of live spectators. This is particularly relevant for digital performances, where maintaining a social connection among audience members is challenging. The use of virtual audiences enhances the immersion of the performance, creating a more real, interactive, and engaging experience for users that is truly an always-ready social experience.

Finally, this work calls for greater nurturing and active involvement with the artistic community to expand the excitement of theater to all through AR. A creator's vision of a narrative can be meticulously scheduled and organized using mechanisms that trigger the next content or narrative. Doing so allows users to experience the creator's vision similarly to their intention. Through collaboration with artists and choreographers, I discovered how their points of view and visions can be transported into an AR platform, which confirmed to me that AR holds more potential and excitement in interactive storytelling. There is also so much beauty in capturing and presenting in physical spaces that correlate with human art forms, dance, and movement specific to those spaces. Connecting and translating these components into AR, and doing so in a way that aligns with how creators and artists see their performances happening in their imaginative spaces, allows viewers to see it through these virtual productions. Alternatively, it can be a more open experience, where users navigate the same environment but find completely different storylines. My exploration into creating interactive narratives and designs suggests that some of the techniques proposed in this dissertation can be further adapted to shape these experiences.

\begin{section}{Future work}
Looking ahead, the potential for AR in performing arts and virtual production is vast and varied. The development of open-source platforms that make AR theater experiences accessible to a wider audience and allow more creative artists and like-minded individuals to express freely without being limited by technology presents a promising avenue. By democratizing access to AR technology, we can encourage broader participation and innovation within the artistic community. Additionally, adapting performances for various devices and platforms, such as Meta Quest, could expand the reach and impact of AR experiences. Another exciting prospect is the integration of volumetric data to create more lifelike virtual performers. This technology could significantly enhance the realism and fidelity of AR performances, offering audiences a richer and more engaging experience. Furthermore, the incorporation of machine learning techniques to train virtual audiences and performers could lead to more adaptive and responsive environments, tailored to the preferences and behaviors of individual users. 

As demonstrated in Chapter 5, I explored methods to expand user interactivity in augmented reality environments by expanding user expression and predictability of movement through experimental visual and spatial techniques. This involves the search for more creative methods to interact and express ourselves, ultimately enriching our experiences in AR. In future work, I would like to utilize these methods to craft and design user experiences in AR theater storytelling, specifically focusing on interactive narrative in mixed reality.

AI agents trained in digital twins, designed to interact within physical spaces, will become more common. As more motion-tracking data from human participants becomes available, these AI entities will exhibit more human-like behaviors. Their presence enriches the immersive social experience by acting as extra actors who enhance the overall experience with minimal effort. I envision first-person technologies, such as augmented and mixed reality, as gateways to interacting with these virtual agents on a personal level. As these entities advance rapidly, bringing them to human form to interact or assigning them roles in the experience can enrich the already personal first-person, mixed reality experience. This allows for a new relationship between people and these virtual entities in a more personal manner. Ultimately, this experience, where immersive technologies meet users ``face''-to-face, can make complex interactions more understandable. The collaboration between human users and AI in the immersive AR platform, I argue, can make large and unexplainable concepts more digestible, thus providing a more natural and inviting experience for users.

Future research must delve into projects that replicate realistic audience behaviors through AI agents, such as dancing, singing, or interacting with virtual environments. While virtual audiences can simulate a social environment, they cannot fully replicate the spontaneity and energy of a real audience due to their lack of interactive nature. Current models often lack the full scope of human-like behaviors and attentiveness, which detracts from the immersion and authenticity of the experience. While this may be acceptable for audience-simulating ML-agents, which involve simpler human interactions as they are often idle, more complex interactions require further development. This exploration raises important questions regarding the goals and methodologies for training AI agents to authentically mimic human responses. For instance, what specific data and algorithms are needed to ensure these AI agents can convincingly participate in events without breaking the illusion of reality? Additionally, researchers must consider the impact on human participants: in a scenario where an audience comprises 92 AI agents and 8 real humans, does the presence of AI enhance or detract from the experience of the real attendees? Future research should focus on refining avatar models to enhance real-time engagement and responsiveness, such as those in more delicate gestures, such as hand gestures and facial expressions. Additionally, it is essential to acknowledge the limitations of virtual interactions, particularly when compared to live, in-person performances. Understanding these dynamics can facilitate the creation of immersive, hybrid experiences where AI contributes to the atmosphere, making events more engaging and memorable for human participants. 

At the same time, as I proposed in this dissertation, capturing and preserving human body movements, actors' subtle expressions, and the intensity of mood remains an important challenge in immersive virtual productions. This also corresponds to the idea that, as one of the key aspects of the Dynamic Theater project, the goal is not to alter the dancer's performance but to make it the center of the user experience. Despite the ability to expand art in AR worlds, this work does not advocate for a changing of human art forms. Rather it calls to enhance human performance through AR. Especially as AI-trained, human-like supporting roles become more easily deployable, the human form of art and the artist themselves are worth capturing and should be the center of the user's experience. Directors in this virtual production should aim to present an experience that is curated and controlled, even if that ``controlled'' experience is for the user to wander around and be lost or simply experience the beautiful scene without tasks or objectives. These open-world-like interaction modalities will soon come to virtual production, where users seek these plays for certain interaction modalities. Designing the experience around human forms of art is what makes these experiences special. 

Further, the use of AR in large-scale open environments introduces the possibility of enhancing real-time maps. Using high-fidelity digital twin maps, any location (e.g. city, town, etc.) could employ a comprehensive virtual layer onto their place-based maps to feature interactive content. Similar to experiences like Pokemon Go, where users travel to different locations to collect special items, open AR environments can introduce and encourage visitors to explore the places they visit in unique ways. With the use of spatial anchors, for example, a city could employ location-specific puzzles or artwork for tourists to engage with. This allows tourists and residents alike to extend their experience with the place they are in and creates opportunities to introduce narratives and experiences that are not readily or physically available. In this way, large-scale AR experiences also generate opportunities for potential funding and community cultural development.  

The foundational interactions and experimental approaches we are pioneering today will become the norm in the future. Enabled by our deepening understanding of space and the limitless possibilities of human imagination, I am excited by the future possibilities of large-scale AR environments and look forward to seeing how our present-day small-scale projects and basic effects have evolved into our daily lives. 
\end{section}

\begin{section}{Discussion}
The exploration of AR in the context of spatial content and interactive narrative through the interactive theater system I introduced in this dissertation shows its potential to present immersive and personalized experiences. By using wide-area augmented reality settings to present performing arts, my projects aim to create immersive experiences that transcend the physical limitations of conventional stages into re-skinned environments that appear to be new places. This allows users to revisit and feel present in new spaces and content. While this dissertation presents journey to creating an immersive media platform using today's best technology to imagine how people might use it, it concludes with my vision for virtual production. 

Unlike conventional stages, where there is a strict order and a centered performance zone, AR theater offers multiple performances throughout the stage, allowing users to find their narrative in their preferred order as they encounter prepared performances in their trajectory. Utilizing the physical layout, virtual stages are created and expanded to include multiple stages that change to the next themed stage. For example, in Dynamic Theater, users hop into a spiral that appears once all content zones are experienced, which provides a virtual stage change without the physical changing of sets. The Reality Distortion Room also highlights how much there is to explore and the potential future of virtual production in stage use.

The design of AR-enhanced theatrical experiences requires careful consideration of several factors, including user guidance systems, content placement, and the physical layout of performance spaces. Effective guidance systems are crucial for curating user movement and focus, ensuring that audiences can navigate the virtual environment intuitively and seamlessly. The selection of guidance systems should be aligned with the designer's objectives and the narrative structure of the performance. The key experience I wanted to highlight in my AR theater projects was bringing dance performances into wide-area, open-world experiences. By engaging how the performances would look, as well as identifying some of the challenges in creating these experiences, I designed virtual stages that encouraged users to navigate the space at their own pace. In this way, AR facilitates a personalized experience that adapts to individual preferences and interests. In making Dynamic Theater, it became clear that some form of guidance helped users stay on track with the experience even in open-world settings. It also revealed that more creative ways to guide users are preferred, such as butterflies leading them to the next content zone where spatial contents are prepared.

The integration of virtual audiences further amplifies this anytime, anywhere experience. My work illustrated how an ``always-ready'' experience where AI-trained audiences are spatially aware and good at playing audience roles enhances these physical layouts. These virtual audiences allow live-event simulation and can provide a social dimension that enriches the overall experience in a very personalized way. The presence of virtual spectators can also create a sense of shared experience, even when users are physically isolated, thereby enhancing the realism and emotional impact of the performance. Findings suggest that the height of the virtual audience can also impact user experience greatly, as viewing angles can be very important when seeing computational content. Also, the timing of introducing an audience can impact the overall user experience. Therefore, we should consider this factor in user experience design. This consideration is particularly relevant when deploying AR in open-world or location-based settings, where environmental variables vary widely.

An important achievement in this virtual production was the crucial collaboration between dancers, choreographers, and technologists, which showed how conventional performers and choreographers were essential in designing this AR theater system. This synergy is pivotal in crafting performances that are not only technologically innovative but also artistically compelling. Dancers' intuitive understanding of space and movement plays a vital role in the development of AR experiences. Their insights into spatial dynamics and audience perception contribute significantly to the design of immersive environments that are both engaging and aesthetically pleasing. Encouraging greater involvement of performers in the technological development process is essential. By allowing dancers and choreographers to actively participate in the design and implementation of AR projects, we can leverage their creative problem-solving skills and artistic vision into the AR stage, where users get to see and experience them. This collaboration not only enhances the quality of the final product but also fosters a sense of ownership and investment among performers, leading to more authentic and impactful performances where both engineers and artists are on the same page in production.

“There are as many applications for VR as you can think of; it’s restricted by your imagination,” once said Jon Goddard of HTC Vive~\cite{viarium2018teleport}. My research into AR dance and theater represents a significant step toward reimagining the possibilities of interactive performance art. By blending technological innovation with artistic creativity, we can create immersive experiences that capture the imagination and engage audiences in novel ways. As we continue to refine and expand these projects, the insights gained will pave the way for new forms of expression and storytelling in the digital age. Through ongoing collaboration, experimentation, and exploration, the future of AR in performing arts holds the promise of transformative experiences that connect and inspire audiences worldwide.

\end{section}

\appendix

\dsp

\end{mainmatter}

\ssp
\bibliographystyle{JHEP3}
\bibliography{dissertation}

\end{document}